\documentstyle[12pt,epsfig]{article}

\catcode`@=11
\def\citer{\@ifnextchar [{\@tempswatrue\@citexr}{\@tempswafalse\@citexr[]}}
 
\def\@citexr[#1]#2{\if@filesw\immediate\write\@auxout{\string\citation{#2}}\fi
  \def\@citea{}\@cite{\@for\@citeb:=#2\do
    {\@citea\def\@citea{--\penalty\@m}\@ifundefined
       {b@\@citeb}{{\bf ?}\@warning
       {Citation `\@citeb' on page \thepage \space undefined}}%
\hbox{\csname b@\@citeb\endcsname}}}{#1}}
\catcode`@=12

\def\refeq#1{\mbox{eq.~(\ref{#1})}}
\def\refeqs#1{\mbox{eqs.~(\ref{#1})}}
\def\reffi#1{\mbox{Fig.~\ref{#1}}}
\def\reffis#1{\mbox{Figs.~\ref{#1}}}

\def\refse#1{\mbox{Sect.~\ref{#1}}}

\def\citere#1{\mbox{Ref.~\cite{#1}}}

\newcommand{\mst}{m_{\tilde{t}}}
\newcommand{\delmst}{\Delta\mst}

\newcommand{\mste}{m_{\tilde{t}_1}}
\newcommand{\mstz}{m_{\tilde{t}_2}}

\newcommand{\msbe}{m_{\tilde{b}_1}}
\newcommand{\msbz}{m_{\tilde{b}_2}}

\newcommand{\MstL}{M_{\tilde{t}_L}}
\newcommand{\MstR}{M_{\tilde{t}_R}}

\newcommand{\Mtlr}{M_{t}^{LR}}

\newcommand{\mf}{m_f}

\newcommand{\msfi}{m_{\tilde{f}_i}}
\newcommand{\msfe}{m_{\tilde{f}_1}}
\newcommand{\msfz}{m_{\tilde{f}_2}}

\newcommand{\msq}{m_{\tilde{q}}}

\newcommand{\sfl}{\tilde{f}_L}
\newcommand{\sfr}{\tilde{f}_R}
\newcommand{\sfe}{\tilde{f}_1}
\newcommand{\sfz}{\tilde{f}_2}

\newcommand{\sfez}{\tilde{f}_{12}}
\newcommand{\sfze}{\tilde{f}_{21}}

\newcommand{\Pe}{\phi_1}
\newcommand{\Pz}{\phi_2}
\newcommand{\Pez}{\phi_{1,2}}
\newcommand{\PePz}{\phi_1\phi_2}
\newcommand{\mpe}{m_{\Pe}}
\newcommand{\mpz}{m_{\Pz}}
\newcommand{\mpez}{m_{\PePz}}

\newcommand{\oas}{{\cal O}(\alpha_s)}
\newcommand{\oaas}{{\cal O}(\alpha\alpha_s)}
\newcommand{\cp}{{\cal CP}}
\newcommand{\wz}{\sqrt{2}}
\newcommand{\edz}{\frac{1}{2}}
\newcommand{\twol}{two-loop}
\newcommand{\onel}{one-loop}

\newcommand{\tc}{{\em TwoCalc}}
\newcommand{\fa}{{\em FeynArts}}
\newcommand{\fh}{{\em FeynHiggs}}
\newcommand{\rp}{$\rho\,$-parameter}

\newcommand{\MW}{M_W}
\newcommand{\MZ}{M_Z}
\newcommand{\MA}{M_A}
\newcommand{\mh}{m_h}
\newcommand{\mH}{m_H}

\newcommand{\mt}{m_{t}}
\newcommand{\mtms}{\overline{m}_t}
\newcommand{\mq}{m_{q}}
\newcommand{\mb}{m_{b}}

\newcommand{\mgl}{m_{\tilde{g}}}

\newcommand{\Sferm}{\tilde{f}}
\newcommand{\Stop}{\tilde{t}}

\newcommand{\Sbot}{\tilde{b}}

\newcommand{\tst}{\theta_{\tilde{t}}}

\newcommand{\tsf}{\theta\kern-.20em_{\tilde{f}}}
\newcommand{\tsfp}{\theta\kern-.20em_{\tilde{f}\prime}}
\newcommand{\tsq}{\theta\kern-.15em_{\tilde{q}}}
\newcommand{\sw}{s_W}
\newcommand{\cw}{c_W}

\newcommand{\sinQZtt}{\sin^2 2\tst}

\newcommand{\sintf}{\sin\tsf}

\newcommand{\sinQtf}{\sin^2\tsf}

\newcommand{\costf}{\cos\tsf}

\newcommand{\cosQtf}{\cos^2\tsf}

\newcommand{\KL}{\left(}
\newcommand{\KR}{\right)}
\newcommand{\KKL}{\left[}
\newcommand{\KKR}{\right]}
\newcommand{\KKKL}{\left\{}

\newcommand{\VL}{\left( \begin{array}{c}}
\newcommand{\VR}{\end{array} \right)}
\newcommand{\ML}{\left( \begin{array}{cc}}
\newcommand{\MLd}{\left( \begin{array}{ccc}}
\newcommand{\MLv}{\left( \begin{array}{cccc}}
\newcommand{\MR}{\end{array} \right)}
\newcommand{\dd}{\partial}

\newcommand{\hc}{\mbox {h.c.}}
\newcommand{\re}{\mbox {Re}\,}

\newcommand{\Tb}{\tan \beta\hspace{1mm}}

\newcommand{\CTb}{\cot \beta\hspace{1mm}}

\newcommand{\Sb}{\sin \beta\hspace{1mm}}
\newcommand{\SQb}{\sin^2\beta\hspace{1mm}}
\newcommand{\SDb}{\sin^3\beta\hspace{1mm}}
\newcommand{\Cb}{\cos \beta\hspace{1mm}}
\newcommand{\CQb}{\cos^2\beta\hspace{1mm}}
\newcommand{\CDb}{\cos^3\beta\hspace{1mm}}
\newcommand{\Sa}{\sin \alpha\hspace{1mm}}
\newcommand{\SQa}{\sin^2\alpha\hspace{1mm}}
\newcommand{\Ca}{\cos \alpha\hspace{1mm}}
\newcommand{\CQa}{\cos^2\alpha\hspace{1mm}}

\newcommand{\SZb}{\sin 2\beta\hspace{1mm}}

\newcommand{\CZb}{\cos 2\beta\hspace{1mm}}

\newcommand{\tev}{\,\, {\mathrm TeV}}
\newcommand{\gev}{\,\, {\mathrm GeV}}

\newcommand{\BC}{\begin{center}}
\newcommand{\EC}{\end{center}}
\newcommand{\BE}{\begin{equation}}
\newcommand{\EE}{\end{equation}}
\newcommand{\BEA}{\begin{eqnarray}}
\newcommand{\BEAnn}{\begin{eqnarray*}}
\newcommand{\EEA}{\end{eqnarray}}
\newcommand{\EEAnn}{\end{eqnarray*}}
\newcommand{\non}{\nonumber}
\newcommand{\id}{{\rm 1\kern-.12em
\rule{0.3pt}{1.5ex}\raisebox{0.0ex}{\rule{0.1em}{0.3pt}}}}
\newcommand{\lsim}
{\;\raisebox{-.3em}{$\stackrel{\displaystyle <}{\sim}$}\;}
\newcommand{\gsim}
{\;\raisebox{-.3em}{$\stackrel{\displaystyle >}{\sim}$}\;}

\def\al{\alpha}
\def\als{\alpha_s}

\def\be{\beta}
\def\ga{\gamma}
\def\de{\delta}

\def\si{\sigma}

\def\De{\Delta}
\def\Si{\Sigma}
\def\Sip{\Sigma'}
\def\Sie{\Sigma^{(1)}}
\def\Siz{\Sigma^{(2)}}

\def\hSi{\hat{\Sigma}}
\def\hSip{\hat{\Sigma}'}
\def\hSie{\hat{\Sigma}^{(1)}}
\def\hSiz{\hat{\Sigma}^{(2)}}

\newcommand{\htad}{\hat{t}}
\newcommand{\tade}{t^{(1)}}
\newcommand{\tadz}{t^{(2)}}

\newcommand{\deMZe}{\de \MZ^{2\,(1)}}
\newcommand{\deMZz}{\de \MZ^{2\,(2)}}

\newcommand{\deMAe}{\de \MA^{2\,(1)}}
\newcommand{\deMAz}{\de \MA^{2\,(2)}}

\newcommand{\deZHie}{\de Z_{H_i}^{(1)}}
\newcommand{\deZHiz}{\de Z_{H_i}^{(2)}}

\newcommand{\deZHez}{\de Z_{H_1}^{(2)}}

\newcommand{\deZHzz}{\de Z_{H_2}^{(2)}}
\newcommand{\deTbe}{\de\Tb^{(1)}}
\newcommand{\deTbz}{\de\Tb^{(2)}}
\newcommand{\detie}{\de t_i^{(1)}}
\newcommand{\detiz}{\de t_i^{(2)}}

\newcommand{\detez}{\de t_1^{(2)}}

\newcommand{\detzz}{\de t_2^{(2)}}

\newcommand{\deVez}{\de V_{\Pe}^{(2)}}

\newcommand{\deVzz}{\de V_{\Pz}^{(2)}}

\newcommand{\deVezz}{\de V_{\PePz}^{(2)}}

\newcommand{\SLASH}[2]{\makebox[#2ex][l]{$#1$}/}

\newcommand{\pslash}{\SLASH{p}{.2}}

\def\draftdate{\relax}
\def\mda{\relax}
\def\mua{\relax}
\def\mla{\relax}
\def\draft{
\def\thtystars{******************************}
\def\sixtystars{\thtystars\thtystars}
\typeout{}
\typeout{\sixtystars**}
\typeout{* Draft mode!
         For final version remove \protect\draft\space in source file *}
\typeout{\sixtystars**}
\typeout{}
\def\draftdate{\today}
\def\mua{\marginpar[\boldmath\hfil$\uparrow$]%
                   {\boldmath$\uparrow$\hfil}%
                    \typeout{marginpar: $\uparrow$}\ignorespaces}
\def\mda{\marginpar[\boldmath\hfil$\downarrow$]%
                   {\boldmath$\downarrow$\hfil}%
                    \typeout{marginpar: $\downarrow$}\ignorespaces}
\def\mla{\marginpar[\boldmath\hfil$\rightarrow$]%
                   {\boldmath$\leftarrow $\hfil}%
                    \typeout{marginpar: $\leftrightarrow$}\ignorespaces}
\def\Mua{\marginpar[\boldmath\hfil$\Uparrow$]%
                   {\boldmath$\Uparrow$\hfil}%
                    \typeout{marginpar: $\Uparrow$}\ignorespaces}
\def\Mda{\marginpar[\boldmath\hfil$\Downarrow$]%
                   {\boldmath$\Downarrow$\hfil}%
                    \typeout{marginpar: $\Downarrow$}\ignorespaces}
\def\Mla{\marginpar[\boldmath\hfil$\Rightarrow$]%
                   {\boldmath$\Leftarrow $\hfil}%
                    \typeout{marginpar: $\Leftrightarrow$}\ignorespaces}
\overfullrule 5pt
\oddsidemargin -15mm
\marginparwidth 29mm
}

\begin{document}
\thispagestyle{empty}

\def\thefootnote{\fnsymbol{footnote}}

\begin{flushright}
KA-TP-17-1998\\
DESY 98-194\\
CERN-TH/98-405\\
hep-ph/9812472 \\
\end{flushright}

\vspace{1cm}

\begin{center}

{\large\sc {\bf The Masses of the Neutral $\cp$-even Higgs Bosons in 
the MSSM:}}

\vspace*{0.4cm} 

{\large\sc {\bf Accurate Analysis at the Two-Loop Level}}

\vspace{1cm}

{\sc 
S.~Heinemeyer$^{1}$%
\footnote{email: Sven.Heinemeyer@desy.de}
, W.~Hollik$^{2,3}$%
\footnote{email: Wolfgang.Hollik@physik.uni-karlsruhe.de}
, G.~Weiglein$^{3}$%
\footnote{email: georg@particle.physik.uni-karlsruhe.de}
}

\vspace*{1cm}

{\sl
$^1$ DESY Theorie, Notkestr. 85, 22603 Hamburg, Germany

\vspace*{0.4cm}

$^2$ Theoretical Physics Division, CERN, CH-1211 Geneva 23, Switzerland

\vspace*{0.4cm}

$^3$ Institut f\"ur Theoretische Physik, Universit\"at Karlsruhe, \\
D--76128 Karlsruhe, Germany
}

\end{center}

\vspace*{1cm}

\begin{abstract}
We present detailed results of a diagrammatic calculation of
the leading two-loop QCD corrections to the masses of the
neutral $\cp$-even
Higgs bosons in the Minimal Supersymmetric Standard Model (MSSM).
The two-loop corrections
are incorporated into the full diagrammatic \onel\ result and
supplemented with refinement terms 
that take into account
leading electroweak \twol\ and higher-order QCD contributions. 
The dependence of the results for the Higgs-boson masses
on the various MSSM parameters is analyzed in detail, with a particular 
focus on the part of the parameter space accessible at LEP2 and the
upgraded Tevatron. For the mass of the lightest Higgs boson, $\mh$, a
parameter scan has been performed, yielding
an upper limit on $\mh$ which depends only on $\Tb$\hspace{-.2em}.
The results for the Higgs-boson masses are compared 
with results obtained by renormalization group methods.
Good agreement is found in the case of vanishing
mixing in the scalar quark sector, while sizable deviations occur if 
squark mixing is taken into account.
\end{abstract}

\def\thefootnote{\arabic{footnote}}
\setcounter{page}{0}
\setcounter{footnote}{0}

\newpage


\section{Introduction}

The search for the lightest Higgs boson is a crucial test of 
Supersymmetry (SUSY) which can be performed with the present and the
next generation of accelerators. The prediction of a relatively
light Higgs boson is common to all supersymmetric models whose
couplings remain in the perturbative regime up to a very high energy
scale~\cite{susylighthiggs}.
A precise prediction for the mass of the lightest Higgs boson in terms
of the relevant SUSY parameters is necessary in order to determine the
discovery and exclusion potential of LEP2 and the upgraded Tevatron, and
for physics at the LHC and future linear colliders, where eventually a
high-precision measurement of the mass of this particle might be possible.
A precise knowledge of the mass of the heavier $\cp$-even Higgs boson,
$\mH$, is important for
resolving the mass splitting between the $\cp$-even and -odd
Higgs-boson masses.

In the Minimal Supersymmetric Standard Model (MSSM)~\cite{mssm} at
the tree level the mass $\mh$ of the lightest Higgs boson is restricted 
to be smaller than the $Z$-boson mass. However, this bound is strongly 
affected by the inclusion of radiative corrections: the dominant \onel\ 
corrections arise from the top and scalar-top loops which yield 
terms of the form $G_F \mt^4 \ln (\mste \mstz/\mt^2)$~\cite{mhiggs1l}. 
These results have 
been improved by performing a complete \onel\ calculation in the
on-shell scheme, which takes into account the contributions of all 
sectors of the MSSM~\cite{mhoneloop,mhiggsf1l,pierce}. 
Beyond \onel\ order, renormalization group (RG) methods have been applied 
in order to include leading logarithmic higher-order
contributions~\cite{mhiggsRG1,mhiggsRG1a,mhiggsRG1b,mhiggsRG2}. 
In the effective potential approach diagrammatic results for the
dominant \twol\ contributions have been obtained in
the limiting case of vanishing $\Stop$-mixing and infinitely large
$\MA$ and $\Tb$~\cite{hoanghempfling}. 
The calculation of the leading QCD corrections in this approach has
recently been generalized to the case of arbitrary $\Tb$ and
non-vanishing $\Stop$-mixing~\cite{zhang}.

Up to now phenomenological analyses have been based either on the
RG results~\cite{mhiggsRG1,mhiggsRG1a,mhiggsRG1b,mhiggsRG2}, or on the
complete \onel\ on-shell results~\cite{mhoneloop,mhiggsf1l,pierce}.
These results differ by large leading logarithmic higher-order contributions,
which are not included in the \onel\ on-shell results, but also by 
non-leading \onel\ contributions, which are neglected in the RG
approach. The numerical difference in the Higgs-mass predictions 
between the two approaches reaches up to 20~GeV.

Recently a Feynman-diagrammatic calculation of the leading \twol\
corrections of $\oaas$ to the masses of the neutral $\cp$-even Higgs
bosons has been performed~\cite{mhiggsletter,mhiggsletter2}. Compared
to the leading 
\onel\ result the \twol\ contribution was found to give rise to a 
considerable reduction of the $\mh$ value. The leading \twol\
corrections have been combined with the full diagrammatic \onel\
on-shell result~\cite{mhiggsf1l} and further refinements 
have been included concerning the
leading \twol\ Yukawa corrections of ${\cal O}(G_F^2
\mt^6)$~\cite{mhiggsRG1a,ccpw} and leading QCD corrections beyond
\twol\ order.

In this paper we present in detail the steps of this calculation. The
results for the masses of the neutral $\cp$-even Higgs bosons are 
analyzed in terms of the relevant parameters of the MSSM. A parameter
scan for the lightest Higgs-boson mass is performed yielding an upper
bound for $\mh$ within the MSSM (apart from certain threshold regions
which correspond to very specific configurations of the MSSM
parameters) given exclusively in terms of $\Tb$. 
This upper bound is discussed in view of the discovery potential of
LEP2 and the upgraded Tevatron.
The results for $\mh$ are compared with the corresponding results
obtained by RG methods. The comparison is performed both in terms of
the (unobservable) parameters of the scalar top mass matrix and in
terms of the physical stop masses and the stop mixing angle.

The paper is organized as follows: Section~2 contains our notations
and a description fo the renormalization procedure as required 
for the corrections in the MSSM Higgs sector in $\oaas$.
The main features of the calculation are discussed in
section~3. In section~4 we present a detailed numerical analysis
of the results for the neutral $\cp$-even Higgs-boson masses as
functions of the different SUSY parameters. 
We perform a scan for $\mh$ over the parameters $\mgl, \MA, M, \mu$
and the $\Stop$-mixing parameter and 
determine the maximal possible values of $\mh$ as a function of $\Tb$.
Finally numerical comparisons are shown 
with results obtained by renormalization group
(RG) methods.
In section~5 we give our conclusions.


\section{Renormalization}

\subsection{The Higgs sector of the MSSM}

The Higgs sector of the MSSM consists of two Higgs doublets $H_1$, $H_2$ 
with opposite hypercharges
$Y_1=-1$ and $Y_2=+1$ and non-vanishing vacuum expectation
values $v_1$ and $v_2$.
The Higgs doublets can be decomposed according to
\BEA
H_1 &=& \VL H_1^0 \\[0.5ex] H_1^- \VR \; = \; \VL v_1 
        + \frac{1}{\sqrt2}(\phi_1^0 + i\chi_1^0) \\[0.5ex] -\phi_1^- \VR  
        \non \\
H_2 &=& \VL H_2^+ \\[0.5ex] H_2^0 \VR \; = \; \VL \phi_2^+ \\[0.5ex] 
        v_2 + \frac{1}{\sqrt2}(\phi_2^0 + i\chi_2^0) \VR~.
\label{higgsfeldunrot}
\EEA
The vacuum expectation values define the angle $\be$ via
\BE
\Tb \equiv \frac{v_2}{v_1}  ; \qquad 0 < \be < \pi/2~.
\label{tanbeta}
\EE
The Higgs potential, including all soft SUSY breaking terms
reads~\cite{hhg} ($\epsilon_{12}=-1$): 
\BEA
V &=& m_{1}^2 |H_1|^2 + m_{2}^2 |H_2|^2 
      - m_{12}^2 (\epsilon_{ab} H_1^aH_2^b + \hc)  \non \\
  & & + \frac{1}{8}({g_1}^2+{g_2}^2) \left[ |H_1|^2 - |H_2|^2 \right]^2
        + \frac{1}{2} {g_2}^2|H_1^{\dag} H_2|^2~,
\label{higgspot}
\EEA
where $m_{i}^2\equiv |\mu|^2+\tilde{m}_i^2\; (i=1,2)$;
$\tilde{m}_1$, $\tilde{m}_2$, $m_{12}$ are the soft SUSY breaking terms, and 
$\mu$ denotes the mixing between $H_1$ and $H_2$.
The coupling constants of the Higgs self-interaction are, contrary to
the SM, determined through the gauge coupling constants $g_1$ and $g_2$.
Besides $g_1$, $g_2$ two independent parameters are required to fix the
potential (\ref{higgspot}) at the tree level. Conventionally they are
chosen as $\Tb$ and $M_A^2 = -m_{12}^2(\Tb+\CTb)$,
where $M_A$ is the mass of the $\cp$-odd $A$ boson.

The diagonalization of the bilinear part of the Higgs potential,
i.e.\ the Higgs mass matrices, is performed via the orthogonal
transformations 
\BEA
\label{hHdiag}
\VL H^0 \\[0.5ex] h^0 \VR &=& \ML \Ca & \Sa \\[0.5ex] -\Sa & \Ca \MR 
\VL \phi_1^0 \\[0.5ex] \phi_2^0 \VR  \\
\label{AGdiag}
\VL G^0 \\[0.5ex] A^0 \VR &=& \ML \Cb & \Sb \\[0.5ex] -\Sb & \Cb \MR 
\VL \chi_1^0 \\[0.5ex] \chi_2^0 \VR  \\
\label{Hpmdiag}
\VL G^{\pm} \\[0.5ex] H^{\pm} \VR &=& \ML \Cb & \Sb \\[0.5ex] -\Sb & 
\Cb \MR \VL \phi_1^{\pm} \\[0.5ex] \phi_2^{\pm} \VR~, 
\EEA
with $\be$ from \refeq{tanbeta}. 
The mixing angle $\al$ is determined through
\BE
\tan 2\al = \tan 2\be \; \frac{\MA^2 + M_Z^2}{\MA^2 - M_Z^2} ;
\qquad  -\frac{\pi}{2} < \al < 0~.
\label{alphaborn}
\EE

One gets the following Higgs spectrum:
\BEA
\mbox{2 neutral bosons},\, {\cal CP} = +1 &:& h^0, H^0 \non \\
\mbox{1 neutral boson},\, {\cal CP} = -1  &:& A^0 \non \\
\mbox{2 charged bosons}                   &:& H^+, H^- \non \\
\mbox{3 unphysical Goldstone bosons}      &:& G^0, G^+, G^- .
\EEA


The masses of the gauge bosons are given in analogy to the SM:
\BE
M_W^2 = \frac{1}{2} g_2^2 (v_1^2+v_2^2) ;\qquad
M_Z^2 = \frac{1}{2}(g_1^2+g_2^2)(v_1^2+v_2^2) ;\qquad M_\gamma=0.
\EE

\bigskip
At tree level the mass matrix of the neutral $\cp$-even Higgs bosons
is given in the $\Pe$-$\Pz$-basis 
in terms of $\MZ$, $\MA$, and $\Tb$ by
\BEA
M_{\rm Higgs}^{2, {\rm tree}} &=& \ML \mpe^2 & \mpez^2 \\ 
                           \mpez^2 & \mpz^2 \MR \non\\
&=& \ML \MA^2 \SQb + \MZ^2 \CQb & -(\MA^2 + \MZ^2) \Sb \Cb \\
    -(\MA^2 + \MZ^2) \Sb \Cb & \MA^2 \CQb + \MZ^2 \SQb \MR,
\label{higgsmassmatrixtree}
\EEA
which by diagonalization according to \refeq{hHdiag} yields the
tree-level Higgs-boson masses
\BE
M_{\rm Higgs}^{2, {\rm tree}} 
   \stackrel{\al}{\longrightarrow}
   \ML m_{H,{\rm tree}}^2 & 0 \\ 0 &  m_{h,{\rm tree}}^2 \MR~.
\EE

In order to slightly simplify the \twol\ calculation,
we have chosen to perform it in the $\Pe$-$\Pz$-basis.
In this way the angle $\al$ does not appear in the calculation of the
\twol\ self-energies, but enters at the end when the rotation into
the physical basis is performed.

In order to deal with the arising divergencies and to establish the
meaning of the
physical parameters beyond the tree level, one has to
renormalize the Higgs and the scalar top sector of the MSSM. 
For the corrections of $\oaas$ to the Higgs-boson masses, 
in the focus of this discussion here,
renormalization up to the \twol\ level is needed. 
In the following we specify the renormalization for the relevant
quantities in this calculation (explicitly listed are only those terms that
actually contribute at $\oaas$). The renormalization of the 
complete one-loop contributions to the neutral $\cp$-even Higgs-boson
masses has been performed according to \citere{mhiggsf1l}. 

We use the following notation:
$\Sie$ and $\Siz$ denote the one- and \twol\ part of an unrenormalized
self-energy, $\hSie$ and $\hSiz$ denote the one- and \twol\ part of a
renormalized self-energy, and $\Sip(k^2) = \frac{\dd}{\dd k^2}\Si(k^2)$. 
$t_1$ and $t_2$ denote the unrenormalized tadpoles; $\tade$ and
$\tadz$ represent the one- and \twol\ part of an unrenormalized
tadpole, and $\htad_1, \htad_2$ denote the renormalized tadpoles.

\bigskip
The renormalization of the masses and fields is performed as follows:
\BEA 
\MZ^2 &\to& \MZ^2 + \deMZe + \deMZz, \\
\MA^2 &\to& \MA^2 + \deMAe + \deMAz, \\
\varphi_1 &\to& \varphi_1 Z_{H_1}^{1/2}, 
   \quad \varphi_1 = \phi_1^0, \chi_1^0, \phi_1^-,\\
\varphi_2 &\to& \varphi_2 Z_{H_2}^{1/2}, 
   \quad \varphi_2 = \phi_2^0, \chi_2^0, \phi_2^+,\\
    && Z_{H_i} = 1 + \deZHie + \deZHiz\\
\Tb &\to& \Tb (1 + \deTbe + \deTbz), \\
t_i &\to& t_i + \detie + \detiz ~~(i = 1,2)~.
\EEA

This yields for the renormalized \twol\ self-energies of $\Pe$ and $\Pz$:
\BEA
\label{P1seren}
\hSiz_{\Pe}(k^2) &=& \Siz_{\Pe}(k^2) + k^2 \deZHez - \deVez, \\
\label{P2seren}
\hSiz_{\Pz}(k^2) &=& \Siz_{\Pz}(k^2) + k^2 \deZHzz - \deVzz, \\
\label{P1P2seren}
\hSiz_{\PePz}(k^2) &=& \Siz_{\PePz}(k^2) - \deVezz ,
\EEA
where it is understood that the unrenormalized self-energies at
two-loop order also 
contain the contributions arising from the subloop renormalization.
The expressions $\deVez, \deVzz$ and $\deVezz$ are the \twol\
counterterm contributions from the Higgs potential:
\BEA
\label{P1potct}
\deVez      &=& \deMZz   \CQb 
             + \deMAz   \SQb
             - \detez  \frac{e}{2 \MW \sw} \Cb (1 + \SQb) \non\\
         &&  + \detzz  \frac{e}{2 \MW \sw} \CQb \Sb 
             + \deZHez     \KL \MZ^2 \CQb + \MA^2 \SQb \KR \non\\
         &&  + \deTbz   \CQb \SQb (\MA^2 - \MZ^2), \\
\label{P2potct}
\deVzz      &=& \deMZz   \SQb 
             + \deMAz   \CQb
             - \detzz  \frac{e}{2 \MW \sw} \Sb (1 + \CQb) \non\\
         &&  + \detez  \frac{e}{2 \MW \sw} \SQb \Cb 
             + \deZHzz     \KL \MZ^2 \SQb + \MA^2 \CQb \KR \non\\
         &&  - \deTbz   \CQb \SQb (\MA^2 - \MZ^2), \\
\label{P1P2potct}
\deVezz       &=& - \deMZz   \Sb\Cb 
               - \deMAz   \Sb\Cb
               - \detez  \frac{e}{2 \MW \sw} \SDb \non \\
         &&    - \detzz  \frac{e}{2 \MW \sw} \CDb
               - \deZHez     \frac{\Sb \Cb}{2} \KL \MA^2 - \MZ^2 \KR\non\\
         &&    - \deZHzz     \frac{\Sb \Cb}{2} \KL \MA^2 - \MZ^2 \KR\non\\
         &&    - \deTbz \edz \SZb\CZb (\MA^2 + \MZ^2), 
\EEA
with the electroweak mixing angle
$\sw^2 = 1 - \cw^2$, $\cw^2 = \MW^2/\MZ^2$.

\bigskip
The counterterms are fixed by imposing on-shell renormalization conditions for 
the renormalized self-energies. For the $A$ boson this reads:
\BE
\label{higgsren1}
\re\hSi_A(\MA^2) = 0 ~.
\EE
The tadpole conditions are:
\BE
\label{higgsrentad}
\htad_1 = 0, \quad \htad_2 = 0~.
\EE
The conditions for the tadpoles have the consequence that the $v_i$
remain the minima of the Higgs potential also at the \twol\ level.

The resulting expressions for the renormalization constants
contributing to the leading 
\twol\ corrections to the neutral $\cp$-even Higgs-boson masses,
expressed in terms of unrenormalized self-energies and tadpoles,
are given in Sec.~\ref{subsec:mhleadingtl}.


\subsection{The scalar quark sector of the MSSM}
\label{sec:squarkren}

Renormalization in the squark sector is needed in the present
calculation at the \onel\ level, i.e.\ at $\oas$. As above, we
work in the on-shell scheme. In the following the formulas are written
for one flavor. 

The squark mass term of the MSSM Lagrangian is given by
\BE
{\cal L}_{m_{\tilde{f}}} = -\frac{1}{2} 
   \Big( \tilde{f}_L^{\dag},\tilde{f}_R^{\dag} \Big)\; {\bf Z} \; 
   \VL \tilde{f}_L \\[0.5ex] \tilde{f}_R \VR~,
\EE
where
\BE
\renewcommand{\arraystretch}{1.5}
{\bf Z} = \ML M_{\tilde{Q}}^2 + M_Z^2\CZb(I_3^f-Q_f\sw^2) + m_f^2 
    & m_f (A_f - \mu \{\CTb;\Tb\}) \\
    m_f (A_f - \mu \{\cot\be;\tan\be\}) 
    & M_{\tilde{Q}'}^2 + M_Z^2 \CZb Q_f\sw^2 + m_f^2 \VR ,
\label{squarkmassenmatrix}
\EE
and $\{\CTb;\Tb\}$ corresponds to $\{u;d\}$-type squarks.
The soft SUSY breaking term $M_{\tilde{Q}'}$ is given by:
\BEA
M_{\tilde{Q}'} &=& \left\{ \begin{array}{cl} 
M_{\tilde{U}} & \quad\mbox{for right handed $u$-type squarks} \\
M_{\tilde{D}} & \quad\mbox{for right handed $d$-type squarks}
\end{array} \right. .
\EEA

In order to diagonalize the mass matrix and to determine the physical mass
eigenstates the following rotation has to be performed:
\BE
\VL \sfe \\ \sfz \VR = \ML \costf & \sintf \\ -\sintf & \costf \MR 
                       \VL \sfl \\ \sfr \VR .
\label{squarkrotation}
\EE
The mixing angle $\tsf$ is given for $\Tb > 1$ by:
\BEA
\costf &=& \sqrt{
    \frac{(m_{\tilde{f}_R}^2 - m_{\tilde{f}_1}^2)^2}
          {m_f^2 \, (A_f - \mu \{\CTb ; \Tb \})^2 
    + (m_{\tilde{f}_R}^2-m_{\tilde{f}_1}^2)^2} } \\
\sintf &=& \mp\; sgn \Big[ 
    A_f - \mu \{\CTb ; \Tb \} \Big] \non \\ 
& & \times \; \sqrt{ 
    \frac{m_f^2 \, (A_f - \mu \{\CTb ; \Tb \})^2}
    {m_f^2 \, (A_f - \mu \{\CTb ; \Tb \})^2 
    + (m_{\tilde{f}_R}^2 - m_{\tilde{f}_1}^2)^2} }~. 
\label{stt}
\EEA
The negative sign in (\ref{stt}) corresponds to $u$-type squarks, the
positive sign 
to $d$-type ones. 
$m_{\tilde{f}_R}^2 =M_{\tilde{Q}'}^2  + M_Z^2 \CZb Q_f\sw^2 + m_f^2 $
denotes the lower right entry in the 
squark mass matrix~(\ref{squarkmassenmatrix}).
The masses are given by the eigenvalues of the mass matrix:
\BEA
m_{\tilde{f}_{1,2}}^2 &=& \edz \KKL M_{\tilde{Q}}^2 + M_{\tilde{Q}'}^2 \KKR
  + \frac{1}{2} M_Z^2 \CZb I^f_3 + m_f^2 \non \\
& & \left\{ \begin{array}{l} \displaystyle \pm\; \frac{c_f}{2}\, 
    \sqrt{ \Big[ M_{\tilde{Q}}^2 - M_{\tilde{Q}'}^2 + 
                 M_Z^2 \CZb (I^f_3-2Q_f\sw^2)\Big]^2 
    + 4 m_f^2 \Big( A_u - \mu \CTb \Big)^2} \\[2ex]
    \displaystyle \pm\; \frac{c_f}{2}\, 
    \sqrt{ \Big[ M_{\tilde{Q}}^2 - M_{\tilde{Q}'}^2 + 
                 M_Z^2 \CZb (I^f_3-2Q_f\sw^2)\Big]^2 
    + 4 m_f^2 \Big( A_d - \mu \Tb \Big)^2} 
    \end{array} \right. \\
c_f &=& sgn \KKL M_{\tilde{Q}}^2 - M_{\tilde{Q}'}^2 + 
              M_Z^2 \CZb (I^f_3-2Q_f\sw^2) \KKR
\label{Squarkmasse} 
\EEA
for $u$-type and $d$-type squarks, respectively.
For most of our discussions (see Sec.~\ref{sec:numresults}) we make
the choice 
\BE
M_{\tilde{Q}} = M_{\tilde{Q}'} =: \msq~.
\EE
Since the non-diagonal entry of the mass matrix
\refeq{squarkmassenmatrix} is proportional to the fermion mass, 
mixing becomes particularly important for $\Sferm = \Stop$, 
in the case of $\Tb \gg 1$ also for $\Sferm = \Sbot$.

\bigskip
For an on-shell renormalization it is convenient to express the
squark mass matrix
in terms of the physical masses $\msfe, \msfz$ and the mixing angle $\tsf$:
\BE
{\bf Z} = \ML \cosQtf \msfe^2 + \sinQtf \msfz^2 & 
              \sintf \costf (\msfe^2 - \msfz^2) \\
              \sintf \costf (\msfe^2 - \msfz^2) &
              \sinQtf \msfe^2 + \cosQtf \msfz^2 
          \MR~.
\label{smm:physparam}
\EE
$A_f$ can be written as follows:
\BE
A_f = \frac{\sintf \costf (\msfe^2 - \msfz^2)}{\mf} + \mu \{\CTb; \Tb\}.
\label{eq:af}
\EE

The renormalization of the fields, the masses, and the mixing angle is then 
performed via
\BEA 
\sfl &\to& \sfl (1 + \edz \de Z_{\sfl} ) \\
\sfr &\to& \sfr (1 + \edz \de Z_{\sfr} ) \\
\msfi^2 &\to& \msfi^2 + \de \msfi^2 \\
\tsf &\to& \tsf + \de \tsf .
\EEA
In the mass eigenstate basis, the field renormalization reads:
\BE
\VL \sfe \\ \sfz \VR 
\to
\ML 1 + \edz \de Z_{\sfe} & \edz \de Z_{\sfez} \\
    \edz \de Z_{\sfze} & 1 + \edz \de Z_{\sfz} \MR
\VL \sfe \\ \sfz \VR,
\EE
with
\BEA
\VL \de Z_{\sfe} \\ \de Z_{\sfz} \VR &=& 
    \ML \cosQtf & \sinQtf \\ \sinQtf & \cosQtf \MR 
    \VL \de Z_{\sfl} \\ \de Z_{\sfr} \VR \\
\de Z_{\sfez} &=& \sintf \costf (\de Z_{\sfr} - \de Z_{\sfl}) 
               = \de Z_{\sfze} \\
 &=& \frac{\sintf \costf}{\cosQtf - \sinQtf} 
     (\de Z_{\sfz} - \de Z_{\sfe}). \non 
\EEA

The renormalized diagonal and non-diagonal self-energies in this basis
have the following structure:
\BEA
\hSi_{\sfe}(k^2) &=& \Si_{\sfe}(k^2) - \de\msfe^2 
                     + (k^2 - \msfe^2) \de Z_{\sfe} \\
\hSi_{\sfz}(k^2) &=& \Si_{\sfz}(k^2) - \de\msfz^2
                     + (k^2 - \msfz^2) \de Z_{\sfz} \\
\hSi_{\sfe\sfz}(k^2) &=& \Si_{\sfe\sfz}(k^2) 
                         - (\msfe^2 - \msfz^2) \de\tsf
                         + (k^2 - \edz (\msfe^2 + \msfz^2)) \de Z_{\sfez} .
\EEA

We impose the following on-shell renormalization conditions:
\BEA
\label{sfren1}
\re\hSi_{\sfe}(\msfe^2) &=& 0 \\
\label{sfren2}
\re\hSip_{\sfe}(\msfe^2) &=& 0 \\
\label{sfren3}
\re\hSi_{\sfz}(\msfz^2) &=& 0 \\
\label{sfren4}
\re\hSip_{\sfz}(\msfe^2) &=& -\re\Sip_{\sfe}(\msfe^2) +
                               \re\Sip_{\sfz}(\msfz^2)\\ 
\label{sfren5}
\re\hSi_{\sfe\sfz}(\msfe^2) &=& 0 ,
\EEA

which determines the renormalization constants to be
\BEA
\label{sfrenkonst1}
\de\msfe^2 &=& \re\Si_{\sfe}(\msfe^2) \\
\label{sfrenkonst2}
\de\msfz^2 &=& \re\Si_{\sfe}(\msfz^2) \\
\label{sfrenkonst3}
\de Z_{\sfe} &=& - \Sip_{\sfe}(\msfe^2) \\
\label{sfrenkonst4}
\de Z_{\sfz} &=& \de Z_{\sfe} \quad \Rightarrow \quad \de Z_{\sfez} = 0\\
\label{sfrenkonst5}
\de\tsf &=& \frac{1}{\msfe^2 - \msfz^2} \Si_{\sfe\sfz}(\msfe^2) .
\EEA
The unsymmetric renormalization condition~(\ref{sfren4}) is 
chosen for convenience since it leads to $\de Z_{\sfz} = \de Z_{\sfe}$ and 
accordingly to $\de Z_{\sfez} = 0$, which simplifies the expression for
the counterterm of the mixing angle. 
In \refeq{sfren5} we have imposed the condition that the non-diagonal
self-energy vanishes at $q^2 = \msfe^2$. Alternatively one could choose
$q^2 = \msfz^2$, instead; 
the numerical difference 
arising from these different choices is irrelevant for the results of the
Higgs-boson masses, as we have checked explicitly.

Taking into account that neither $\de\mu$ nor $\de\Tb$ are of $\oas$,
one obtains from \refeq{eq:af}:
\BE
\de A_f = \frac{\sintf \costf (\msfe^2 - \msfz^2)}{\mf}
          \KKL \frac{1 - 2 \sinQtf}{\sintf \costf} \de\tsf
              + \frac{\de\msfe^2 - \de\msfz^2}{\msfe^2 - \msfz^2}
              - \frac{\de\mf}{\mf} \KKR .
\label{deltaAf}
\EE

\bigskip
For completeness we also list the expression for the quark mass
counterterm in the on-shell scheme,
\BE
\de m_f = m_f \left( \Si_f^V(m_f^2) +  \Si_f^S(m_f^2) \right),
\EE
where the scalar functions in the decomposition of the fermion
self-energy $\Si_f(p)$ are defined according to 
\BE
\Si_f(p) = \pslash \Si_f^V(p^2) + \pslash \ga_5 \Si_f^A(p^2) + 
m_f \Si_f^S(p^2) .
\EE


\section{Calculation of the neutral $\cp$-even Higgs-boson\\ masses}
\label{sec:calc}

\subsection{Leading \twol\ contributions to the Higgs-boson\\
self-energies}
\label{subsec:mhleadingtl}

The dominant \onel\ contributions to the Higgs-boson
mass matrix in \refeq{higgsmassmatrixtree} are given by terms of the form 
$G_F \mt^4 \ln (\mste \mstz/\mt^2)$, which arise from $t$- and
$\Stop$-loops.
They can be obtained by evaluating the contribution of the
$t$--$\Stop$-sector to the $\Pez$ self-energies at zero external
momentum from the Yukawa part of the theory (neglecting the gauge
couplings). Accordingly, the leading contributions to the \onel\
corrected Higgs-boson masses are derived by diagonalizing the matrix
\BE
M^{2, \rm{1-loop}}_{\rm Higgs}
= \VL \mpe^2 - \hSie_{\Pe}(0)\;\;\;\;\;\; \mpez^2 - \hSie_{\PePz}(0) \\
     \mpez^2 - \hSie_{\PePz}(0)\;\;\;\;\;\; \mpz^2 - \hSie_{\Pz}(0) \VR ,
\label{higgsmassmatrixnondiag}
\EE
where the $\hSie$ denote the \onel\ Yukawa contributions of the
$t$--$\Stop$-sector to the renormalized \onel\ $\Pez$ self-energies. 
For completeness, we list here the explicit form of these dominant
\onel\ corrections (in the numerical results given in
Sec.~\ref{sec:numresults}  we use the
complete \onel\ on-shell result as given in \citere{mhiggsf1l}):
\BEA
\hSie_{\Pe}(0) &=& \frac{3 G_F \mt^4}{\wz \pi^2 \SQb}
          \frac{\mu^2 (A_t - \mu \CTb)^2}{(\mste^2 - \mstz^2)^2}
               \KL 1 - \frac{\mste^2 + \mstz^2}{\mste^2 - \mstz^2}
                   \ln \frac{\mste}{\mstz} \KR, \non \\
\hSie_{\PePz}(0) &=& \frac{3 G_F \mt^4}{2 \wz \pi^2 \SQb}
          \KKL - \frac{\mu (A_t - \mu \CTb)}{\mste^2 - \mstz^2}
               \ln \frac{\mste^2}{\mstz^2} \right. \non\\
 &&       \left. - \frac{2 \mu A_t (A_t - \mu \CTb)^2}{(\mste^2 - \mstz^2)^2}
               \KL 1 - \frac{\mste^2 + \mstz^2}{\mste^2 - \mstz^2}
                   \ln \frac{\mste}{\mstz} \KR \KKR, \non\\
\hSie_{\Pz}(0) &=& \frac{3 G_F \mt^4}{\wz \pi^2 \SQb}
          \KKL \ln \KL \frac{\mste \mstz}{\mt^2} \KR
               + \frac{A_t (A_t - \mu \CTb)}{\mste^2 - \mstz^2}
               \ln \frac{\mste^2}{\mstz^2} \right. \non\\
 &&       \left. + \frac{A_t^2 (A_t - \mu \CTb)^2}{(\mste^2 - \mstz^2)^2}
               \KL 1 - \frac{\mste^2 + \mstz^2}{\mste^2 - \mstz^2}
                   \ln \frac{\mste}{\mstz} \KR \KKR .
\EEA

By comparison with the full \onel\
result~\cite{mhoneloop,mhiggsf1l,pierce} it has been 
shown that these contributions indeed contain the bulk of the \onel\
corrections. 
They typically approximate the full \onel\ result within $5 \gev$.

\bigskip
In order to derive the leading two-loop contributions to the masses of
the neutral $\cp$-even Higgs bosons we have evaluated the QCD
corrections to
\refeq{higgsmassmatrixnondiag}~\cite{mhiggsletter,mhiggsletter2}. 
Accordingly, we have calculated the $\oaas$ contribution of the
$t$--$\Stop$-sector to the $\Pez$ self-energies at zero momentum transfer,
neglecting the gauge couplings.
Because of the large value of the strong coupling constant these are
expected to be the most sizable two-loop corrections (see also
\citere{hoanghempfling}). 

The leading two-loop contributions to the $\Pez$ self-energies are 
given, according to\\ Eqs.~(\ref{P1seren})-(\ref{P1P2seren}), by
\BEA
\label{P1serenb}
\hSiz_{\Pe}(0) &=& \Siz_{\Pe}(0) - \deVez, \\
\label{P2serenb}
\hSiz_{\Pz}(0) &=& \Siz_{\Pz}(0) - \deVzz, \\
\label{P1P2serenb}
\hSiz_{\PePz}(0) &=& \Siz_{\PePz}(0) - \deVezz ,
\EEA
and for the leading contributions the potential counterterms 
eqs.~(\ref{P1potct})--(\ref{P1P2potct}) simplify to
\BEA
\label{P1potctMW0Yuk2l}
\deVez  &=& 
             + \deMAz   \SQb
             - \detez  \frac{e}{2 \MW \sw} \Cb (1 + \SQb) \\
         &&  + \detzz  \frac{e}{2 \MW \sw} \CQb \Sb, \non\\
\label{P2potctMW0Yuk2l}
\deVzz  &=& 
             + \deMAz   \CQb
             - \detzz  \frac{e}{2 \MW \sw} \Sb (1 + \CQb) \\
         &&  + \detez  \frac{e}{2 \MW \sw} \SQb \Cb, \non\\
\label{P1P2potctMW0Yuk2l}
\deVezz  &=& 
               - \deMAz   \Sb\Cb
               - \detez  \frac{e}{2 \MW \sw} \SDb \\
         &&    - \detzz  \frac{e}{2 \MW \sw} \CDb . \non
\EEA

{}From the on-shell renormalization conditions
\refeqs{higgsren1}--(\ref{higgsrentad}) we obtain for the counterterms
in \refeqs{P1potctMW0Yuk2l}--(\ref{P1P2potctMW0Yuk2l})
\BE
\deMAz = \Siz_A(0)
\EE
and
\BE
\detez = -\tadz_1, \quad \detzz = -\tadz_2 .
\EE


\subsection{Evaluation of the relevant Feynman diagrams}

The calculations have been performed using Dimensional Reduction
(DRED)~\cite{dred}, which is necessary in order to preserve the
relevant SUSY relations. Naive application (without an appropriate 
shift in the couplings) of Dimensional Regularization
(DREG)~\cite{dreg}, on the other hand, does not lead to a finite result.
The same observation has also been made in~\citere{hoanghempfling}.

\begin{figure}[ht]
\begin{center}
\mbox{
\psfig{figure=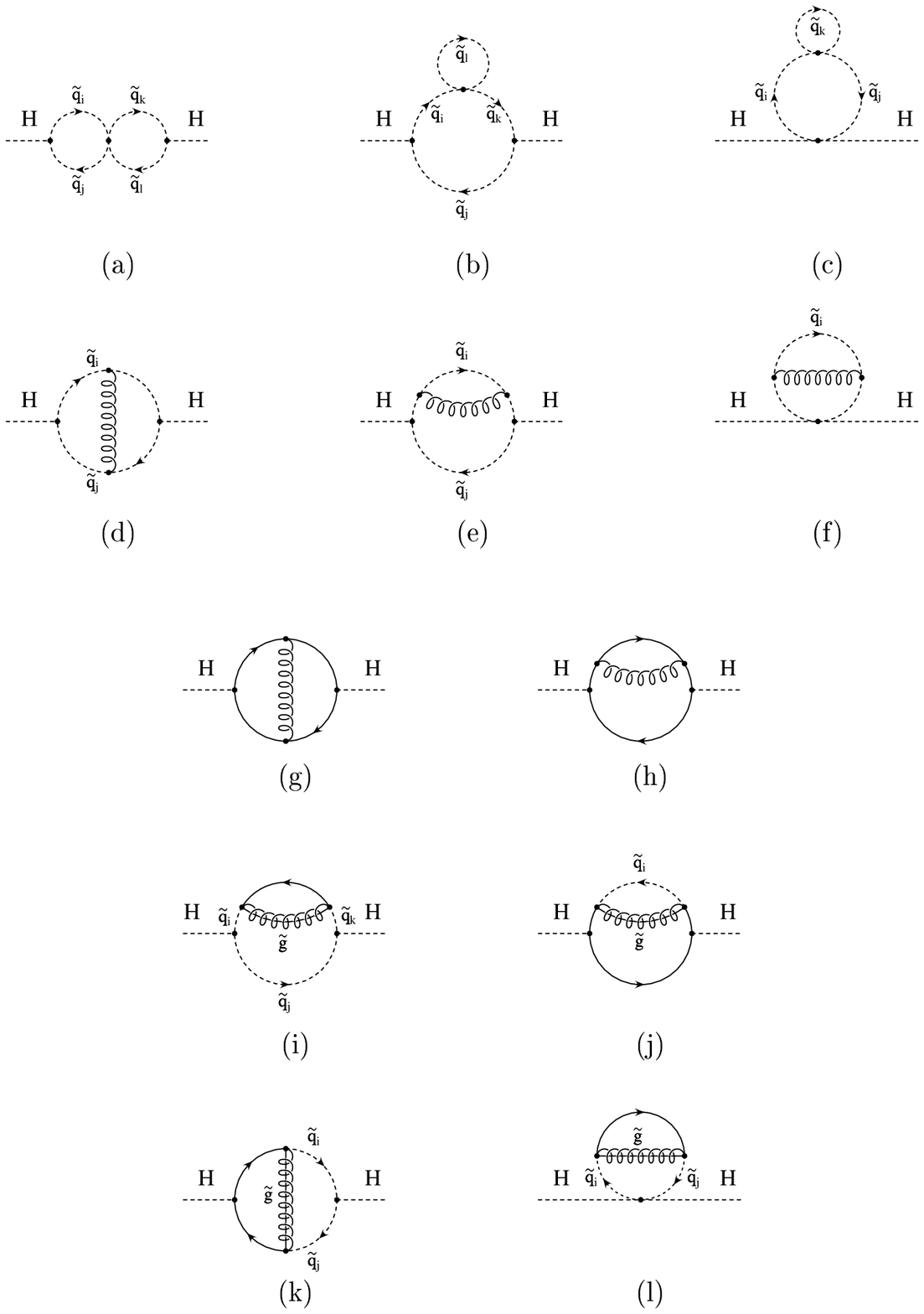,width=12cm,bbllx=90pt,bblly=185pt,%
                                        bburx=510pt,bbury=750pt}}
\end{center}
\caption{
Feynman diagrams for the contribution of squark and quark loops to the
Higgs-boson self-energies at the \twol\ level ($H = \Pe, \Pz, A$).}
\label{fig:fdhb2l}
\end{figure}

\begin{figure}[ht]
\begin{center}
\mbox{
\psfig{figure=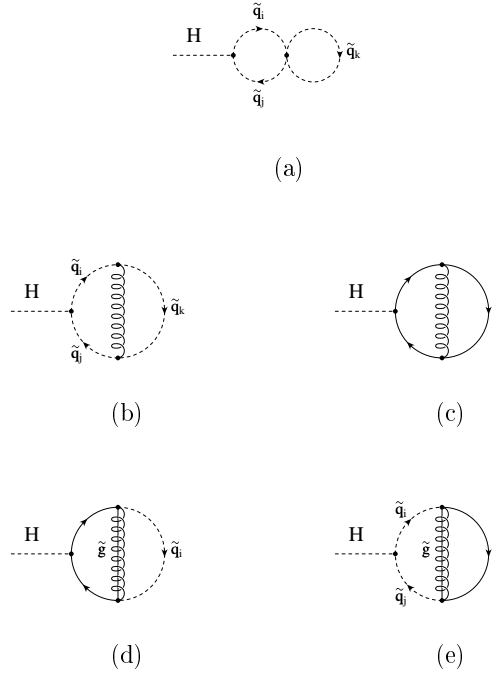,width=12cm,bbllx=90pt,bblly=400pt,
                                        bburx=510pt,bbury=740pt}}
\end{center}
\caption{
Feynman diagrams for the contributions of squark and quark loops to the
Higgs-boson tadpoles at the \twol\ level ($H = \Pe, \Pz$).}
\label{fig:fdhbtp2l}
\end{figure}

The Feynman diagrams contributing to the $\Pe, \Pz$ and $A$
self-energies are depicted in \reffi{fig:fdhb2l}.%
\footnote{The diagrams with a closed gluon line give zero contribution in
DREG and DRED, they are omitted here.}
The Feynman diagrams for the tadpole diagrams are shown in
\reffi{fig:fdhbtp2l}. 

\begin{figure}[ht]
\begin{center}
\mbox{
\psfig{figure=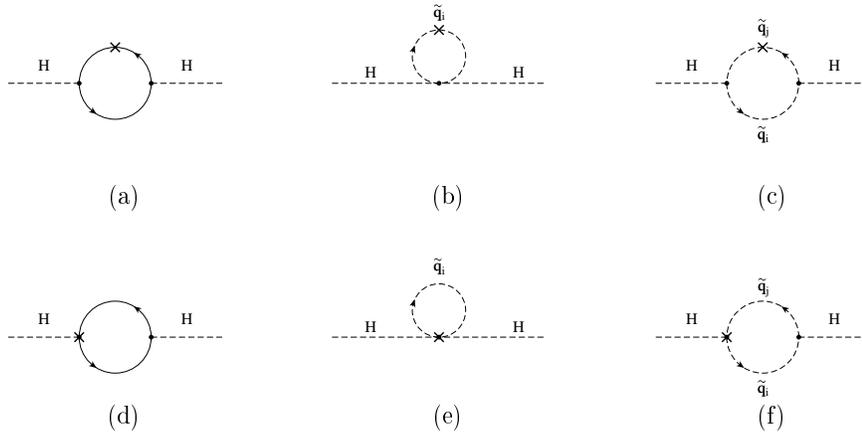,width=12cm,bbllx=90pt,bblly=510pt,%
                                   bburx=510pt,bbury=740pt}}
\end{center}
\caption{
One-loop counterterm contributions to the Higgs boson self-energies at
the \twol\ level ($H = \Pe, \Pz, A$).}
\label{fig:fdhbct2l}
\end{figure}

\begin{figure}[ht]
\begin{center}
\mbox{
\psfig{figure=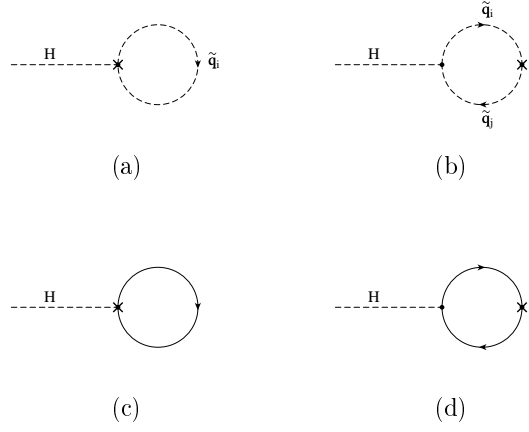,width=12cm,bbllx=90pt,bblly=480pt,
                                        bburx=510pt,bbury=740pt}}
\end{center}
\caption{
One-loop counterterm contributions to the Higgs-boson tadpoles at the
\twol\ level ($H = \Pe, \Pz$).}
\label{fig:fdhbtpct2l}
\end{figure}

\begin{figure}[ht]
\begin{center}
\mbox{
\psfig{figure=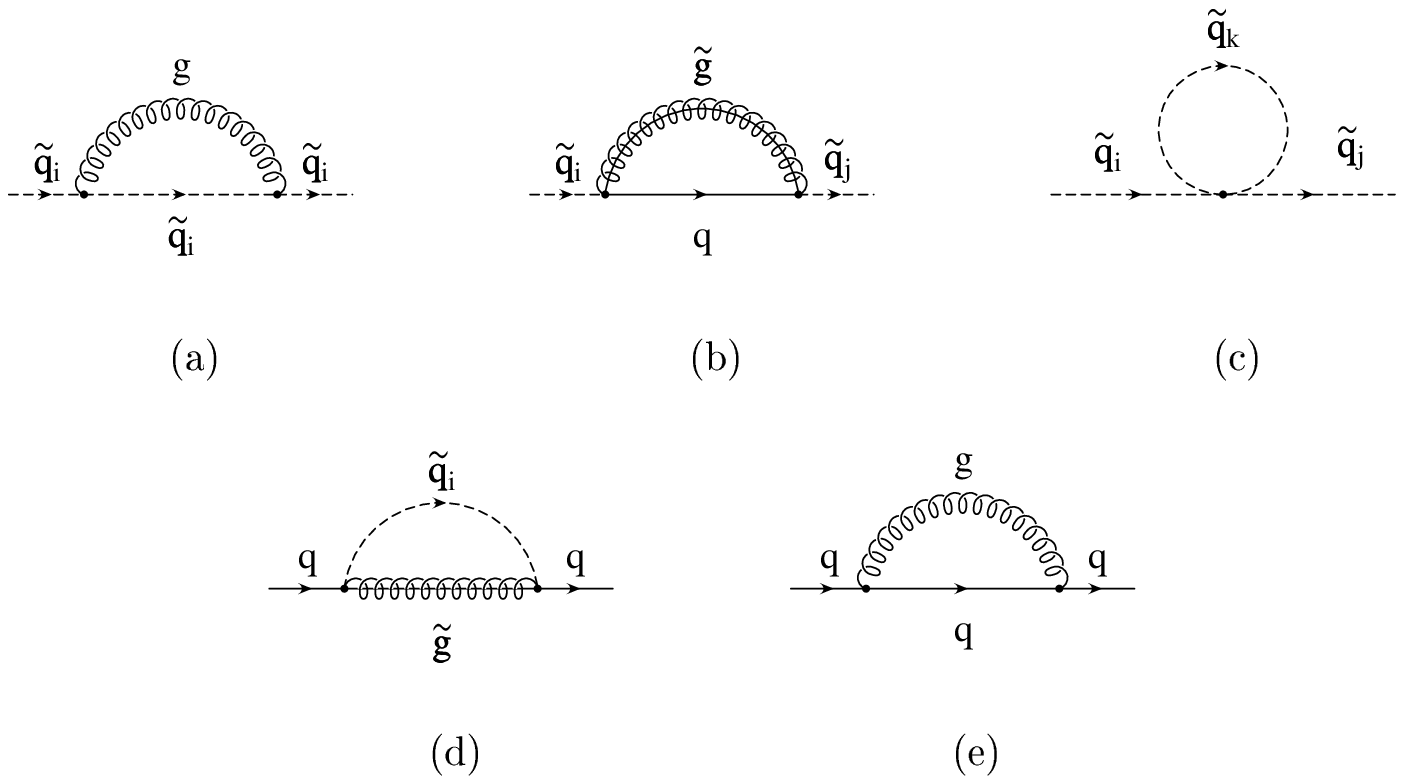,width=12cm,bbllx=90pt,bblly=530pt,%
                                   bburx=520pt,bbury=750pt}}
\end{center}
\caption{
One-loop diagrams for the squark and quark mass counterterms and for the
squark mixing-angle counterterm.}
\label{fig:fdmassct2}
\end{figure}

There are three classes of diagrams: pure scalar diagrams
(Fig.~\ref{fig:fdhb2l}a--c, Fig~\ref{fig:fdhbtp2l}a), 
diagrams with gluon exchange (Fig.~\ref{fig:fdhb2l}d--h,
Fig~\ref{fig:fdhbtp2l}b--c), 
and diagrams with gluino exchange (Fig.~\ref{fig:fdhb2l}i--l,
Fig~\ref{fig:fdhbtp2l}d--e). 
These diagrams have to be supplemented by the corresponding \onel\
diagrams with counterterm insertions, which are depicted in
\reffi{fig:fdhbct2l} and in \reffi{fig:fdhbtpct2l}. The counterterm
insertions are generated by the renormalization in the top and scalar
top sector (see \refse{sec:squarkren}). They are calculated from
the Feynman diagrams in \reffi{fig:fdmassct2}.

The gluon-exchange contribution of $\oas$ to the quark mass counterterm
reads in DRED:
\BE
\de^g\mq = \frac{\als}{\pi} \mq \left(-\frac{1}{\de} + \ga_{\mathrm{E}}
+ \ln\left(\frac{\mq^2}{4 \pi \mu^2}\right) - \frac{5}{3} \right) +
{\cal O}(\de) ,
\EE
where $2 \de = 4 - n$ with $n$ the space--time dimension, $\ga_{\mathrm{E}}$
is Euler's constant, and $\mu$ is the 't~Hooft scale. The explicit form
of the other counterterms of the quark and scalar quark sector can be
found in~\citere{drhosuqcdb}.

Some of the diagrams shown in \reffis{fig:fdhb2l}, \ref{fig:fdhbtp2l}
vanish when they are
combined with the corresponding counterterm contributions of
\reffis{fig:fdhbct2l}, \ref{fig:fdhbtpct2l}.
{}From the pure scalar diagrams only \reffi{fig:fdhb2l}a yields a
non-vanishing contribution.  
The diagrams \reffi{fig:fdhb2l}b--c are canceled exactly with their
corresponding counterterm diagrams. Here the mass renormalization for
the diagonal terms (with two identical squarks) and the mixing-angle
renormalization for the non-diagonal terms (with two different
squarks) are needed.
The same applies for the tadpole diagram \reffi{fig:fdhbtp2l}a
together with the counterterm diagram \reffi{fig:fdhbtpct2l}b.
The diagrams \reffi{fig:fdhb2l}f are exactly canceled with the
corresponding diagram with counterterm insertion
\reffi{fig:fdhbct2l}b.
The same applies for the tadpole diagrams \reffi{fig:fdhbtp2l}b
together with the counterterm diagram \reffi{fig:fdhbtpct2l}b.

\bigskip
We now briefly describe the evaluation of the \twol\ diagrams.
As explained above, the calculation involves irreducible \twol\ diagrams 
at zero momentum-transfer and counterterm diagrams. 
In deriving our results we have made strong use of computer algebra
tools: 
the diagrams were generated with the Mathematica
package \fa~\cite{FA}. For this purpose we have implemented a model file
which contains the relevant part of the MSSM Lagrangian,
i.e.\ all SUSY propagators 
($\tilde{t}_1, \tilde{t}_2, \tilde{b}_1, \tilde{b}_2, \tilde{g}$) 
needed for the QCD-corrections and the appropriate
vertices (Higgs boson-squark vertices, squark-gluon and squark-gluino
vertices). The program inserts propagators and vertices into the 
graphs in all possible ways and creates the amplitudes including all
symmetry factors. The evaluation of the \twol\ diagrams and counterterms
was performed  with the Mathematica package \tc~\cite{TC}.
By means of \twol\ tensor integral decompositions it reduces
the amplitudes to a minimal set of standard scalar integrals, consisting
in this case of products of the basic one-loop integrals $A_0, 
B_0$~\cite{HoVe} (the $B_0$ functions originate from the counterterm 
contributions only) and the \twol\ function $T_{134}$, which is
the genuine \twol\ scalar integral at zero momentum-transfer (vacuum
integral). This integral is known for arbitrary internal masses and
admits a compact representation for $\de \to 0$  in terms of logarithms
and dilogarithms (see for instance Ref.~\cite{Davydychev}).
It should be noted that from the expansion of
the one-loop two-point function~$B_0$,
\BE
B_0(q^2,m_a,m_b) = \frac{1}{\de} + B_{0}^{\rm fin}(q^2,m_a,m_b)+
\de\,B_{0}^{\de}(q^2,m_a,m_b) ,
\EE
only the term $B_{0}^{\rm fin}$ contributes, while $B_{0}^{\de}$ drops
out in our final result.
{}From the output generated with \tc\ a {\tt FORTRAN} code was
created which allows a fast calculation for a given set of
parameters. This code has been implemented into the {\tt FORTRAN} program
\fh~\cite{feynhiggs}, see below. 

Our results for the \twol\
$\phi_{1,2}$ self-energies are given in terms of the
SUSY parameters $\Tb$, $\MA$, $\mu$, $\mste$, $\mstz$, $\tst$, and 
$\mgl$. In the general case the results
are by far too lengthy to be given here
explicitly. In the special case of vanishing mixing in the 
$\Stop$-sector, $\mu = 0$, and $\mste = \mstz = \mst$, a relatively 
compact expression can be derived which is given in
\citere{mhiggsletter}. 
We have performed an expansion of this result for
large values of $\mgl$. It yields for the leading terms
\BE
\hSi_{\Pz}^{(2)} = C_F \frac{\al}{\pi} \frac{\als}{\pi}
                   \frac{3\, \mt^2}{2\,\sw^2\SQb}
                   \KKL -\ln^2(\mgl^2) + \ln(\mgl^2) \KKR
                   + {\cal O}(\mgl^0) .
\EE
This shows that the gluino does not decouple from the \twol\ result,
contrary to the case of the \twol\ QCD contributions to the
$\rho$-parameter in the MSSM~\cite{drhosuqcdb,drhosuqcda}.

\smallskip
In \citere{hoanghempfling} a result for the limiting case 
\BE
\mst = \mste = \mstz = \mgl \gg \mt, \qquad \Tb \to \infty
\EE
has been given. In this limit we obtain
\BEA
&& \hSiz_{\Pe}(0) = 0 ,\;\;\;\; \hSiz_{\PePz}(0) = 0 , \non \\
\hSiz_{\Pz}(0) &=& 
-C_F \frac{\al}{\pi} \frac{\als}{\pi} \frac{9\,\mt^4}{4 \sw^2 \MW^2}
\ln\KL \frac{\mst^2}{\mt^2} \KR 
\KKL \ln\KL \frac{\mst^2}{\mt^2} \KR + 2 \KKR ,
\label{LLexpansion}
\EEA 
which agrees with the corresponding result given in
\citere{hoanghempfling}.

\clearpage


\subsection{Determination of the Higgs-boson masses}

In the Feynman-diagrammatic approach the
Higgs-boson masses are derived beyond tree level 
by determining the poles of the $h-H$-propagator
matrix whose inverse 
is given by
\BE
\left(\Delta_{\rm Higgs}\right)^{-1}
= - i \ML q^2 -  m_{H,{\rm tree}}^2 + \hSi_{H}(q^2) &  \hSi_{hH}(q^2) \\
     \hSi_{hH}(q^2) & q^2 -  m_{h,{\rm tree}}^2 + \hSi_{h}(q^2) \MR,
\label{higgspropagatormatrixnondiag}
\EE
where again the $\hSi$ denote the renormalized Higgs-boson
self-energies, now in the $h-H$-basis.

Determining the
poles of the matrix $\Delta_{\rm Higgs}$ in
\refeq{higgspropagatormatrixnondiag} is equivalent to solving
the equation
\BE
\left[q^2 - \mh^2 + \hat\Sigma_{hh}(q^2) \right]
\left[q^2 - \mH^2 + \hat\Sigma_{HH}(q^2) \right] - 
\left[\hat\Sigma_{hH}(q^2)\right]^2 = 0 . 
\EE

In our calculation the complete one-loop result for the
Higgs-boson self-energies in the on-shell scheme~\cite{mhiggsf1l} 
is combined with 
the leading \twol\ contributions, which have been outlined in
the previous section.
The matrix \refeq{higgspropagatormatrixnondiag} therefore
contains the renormalized Higgs-boson self-energies
\BE
\hSi_s(q^2) = \hSie_s(q^2) + \hSiz_s(0), \quad s = h, H, hH,
\label{besttwoloopse}
\EE
where the momentum dependence is neglected only in the \twol\
contribution.

Since the \twol\ contribution has been calculated in the $\Pe$-$\Pz$-basis,
a rotation into the $h$-$H$-basis, according to \refeq{hHdiag}, has to
be performed:
\BEA
\hSiz_{H} &=& \CQa \hSiz_{\Pe} + \SQa \hSiz_{\Pz} + 
              2 \Sa \Ca \hSiz_{\PePz} \non \\
\hSiz_{h} &=& \SQa \hSiz_{\Pe} + \CQa \hSiz_{\Pz} - 
              2 \Sa \Ca \hSiz_{\PePz} \non \\
\hSiz_{hH} &=& - \Sa \Ca \KL \hSiz_{\Pe} - \hSiz_{\Pz} \KR + 
              (\CQa - \SQa) \hSiz_{\PePz} .
\label{higgsserotation}
\EEA

\bigskip
We have implemented two further corrections beyond $\oaas$
into the prediction
for $\mh$, which are illustrated  in
\reffis{fig:mh_MtLRdivmq}, \ref{fig:mH_MtLRdivmq}, 
\ref{fig:mh_mq} and \ref{fig:mh_MA}. 
The leading \twol\ Yukawa correction of ${\cal O}(G_F^2 \mt^6)$ 
is taken over from the result obtained by renormalization
group methods. It reads~\cite{mhiggsRG1a,ccpw}
\BEA
\label{yukawaterm}
\Delta\mh^2 &=& \frac{9}{16\pi^4} G_F^2 \mt^6
               \KKL \tilde{X} t + t^2 \KKR \\
\mbox{with} && \tilde{X} = \Bigg[
                \KL \frac{\mstz^2 - \mste^2}{4 \mt^2} \sinQZtt \KR^2
                \KL 2 - \frac{\mstz^2 + \mste^2}{\mstz^2 - \mste^2}
                      \log\KL \frac{\mstz^2}{\mste^2} \KR \KR \non\\
            && \mbox{}\hspace{1cm}  
               + \frac{\mstz^2 - \mste^2}{2 \mt^2} \sinQZtt
                      \log\KL \frac{\mstz^2}{\mste^2} \KR \Bigg], \\
 && t = \frac{1}{2} \log \KL \frac{\mste^2 \mstz^2}{\mt^4} \KR .
\EEA

\bigskip
The second higher-order contribution which has been implemented
concerns leading QCD corrections beyond
\twol\ order, taken into account by using the $\overline{\rm{MS}}$
top mass%
\footnote{
The functional dependence of $\mtms(\mt)$ is known up to 
${\cal O}(\als^2)$~\cite{mtmsbar2loop}. Since $\mtms(\mt)$ enters only
at the \twol\ level, we have incorporated only the \onel\ correction
to $\mtms(\mt)$,
thus neglecting only contributions of ${\cal O}(\al\als^3)$ in $\mh$.}
\BE
\mtms = \mtms(\mt) \approx \frac{\mt}{1 + \frac{4}{3\,\pi} \als(\mt)}
\label{mtrun}
\EE
for the \twol\ contributions instead of the pole mass, $\mt = 175 \gev$.
In the $\Stop$ mass matrix, however, we continue to use the pole mass
as an input parameter.
Only when performing the comparison with the RG results we use
$\mtms$ in the $\Stop$ mass matrix for the \twol\ result, since in the
RG results the running masses appear everywhere.
This three-loop effect gives rise to a shift up to $1.5 \gev$ in the
prediction for $\mh$.

\bigskip
The complete \onel\ calculation together with the leading \twol\
corrections and the other corrections beyond $\oaas$
have been implemented into
the {\tt FORTRAN} code \fh~\cite{feynhiggs}.
This code can be linked to existing programs as a subroutine, thus
providing an accurate calculation of $\mh$ and $\mH$ which can be used
for further phenomenological analyses. 
\fh\ is available via its WWW page\\
{\tt http://www-itp.physik.uni-karlsruhe.de/feynhiggs}.


\section{Numerical results for $\mh$ and $\mH$}
\label{sec:numresults}

\subsection{Dependence of the results on the MSSM parameters}
\label{subsec:mhparameterdependence}

In this subsection we give a detailed discussion of the dependence of
$\mh$ on the parameters of the MSSM. For $\Tb$ we restrict ourselves
to two typical values which
are favored by SUSY-GUT scenarios~\cite{su5so10}: $\Tb = 1.6$ for
the $SU(5)$ scenario and $\Tb = 40$ for the $SO(10)$ scenario. 
Other parameters are $\MZ = 91.187 \gev, \MW = 80.375 \gev, 
G_F = 1.16639 \, 10^{-5} \gev^{-2}, \als(\mt) = 0.1095$, 
$\mt = 175 \gev$, and $\mb = 4.5 \gev$ (if not indicated differently).
The parameter $M$ appearing in the plots is the $SU(2)$ gaugino mass
parameter. The other gaugino mass parameter, $M_1$, is fixed via the
GUT relation
\BE
M_1 = \frac{5}{3} \frac{\sw^2}{\cw^2} M~.
\label{m1}
\EE
The scalar top masses and the mixing angle are related to the
parameters $M_{{\tilde t}_L}$, $M_{{\tilde t}_R}$ and $\Mtlr$ of the 
$\Stop$ mass matrix, which reads
\BE
\label{stopmassenmatrix}
{\cal M}^2_{\Stop} = 
  \ML \MstL^2 + \mt^2 + \CZb (\edz - \frac{2}{3} \sw^2) \MZ^2 &
      \mt \Mtlr \\
      \mt \Mtlr &
      \MstR^2 + \mt^2 + \frac{2}{3} \CZb \sw^2 \MZ^2 
  \MR~,
\EE
with
\BE
\Mtlr = A_t - \mu \CTb~.
\label{mtlr}
\EE
In the figures below we have chosen 
$\msq \equiv \MstL = \MstR$ (if not indicated differently).

\begin{figure}[ht!]
\begin{center}
\mbox{
\psfig{figure=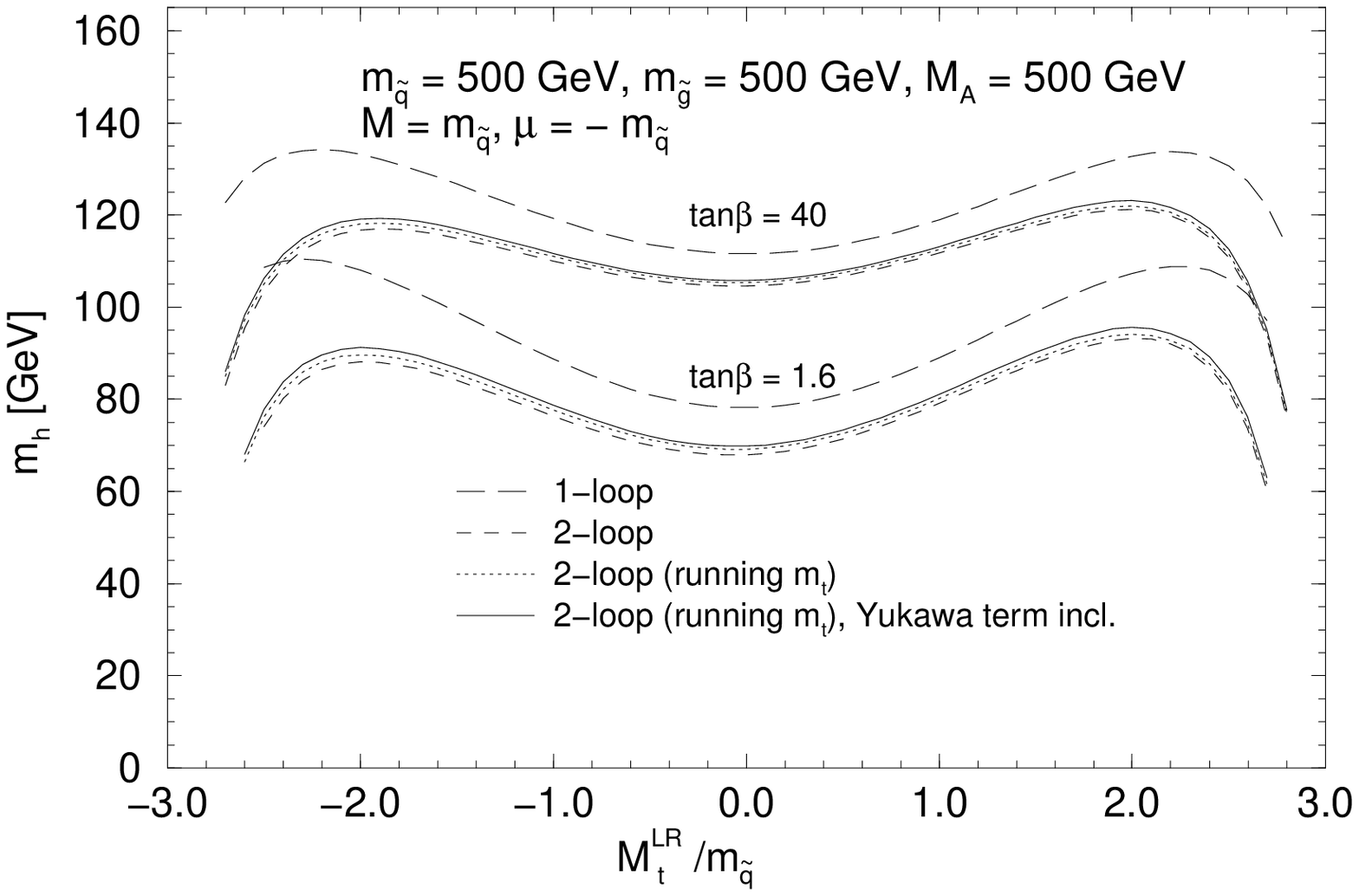,width=7.3cm,height=8cm,
                      bbllx=150pt,bblly=100pt,bburx=450pt,bbury=420pt}}
\end{center}
\caption[]{One- and \twol\ results for $\mh$ as a function of
$\Mtlr/\msq$ for two values of $\Tb$. The corrections beyond $\oaas$
discussed in the text are shown separately.}
\label{fig:mh_MtLRdivmq}
\end{figure}

\bigskip
Fig.~\ref{fig:mh_MtLRdivmq} shows the result for $\mh$ obtained
from the diagrammatic calculation of the full \onel\ and leading
\twol\ contributions. The two contributions beyond $\oaas$
discussed above are
shown in separate curves. For comparison the pure \onel\ result is also
given. The results are plotted as a function of $\Mtlr/\msq$, where
$\msq$ is fixed to $500 \gev$. 
The \twol\ contributions give rise to a large reduction of the 
\onel\ result of 10--20 GeV. The two corrections beyond $\oaas$
both increase $\mh$ by up to $2 \gev$. 
A minimum occurs around  $\Mtlr = 0 \gev$ which we refer to as `no
mixing'.  
A maximum in the \twol\ result for $\mh$ is reached for about 
$\Mtlr/\msq \approx \pm 2$ in the $\Tb = 1.6$ scenario as well as in the 
$\Tb = 40$ scenario. This case we refer to as `maximal mixing'. 
The position of the maximum is shifted compared to its \onel\ value of about
$\Mtlr/\msq \approx \pm 2.4$. The Yukawa correction and the insertion
of the running top mass
have only a negligible effect on the location of the maximum.

\begin{figure}[ht!]
\begin{center}
\mbox{
\psfig{figure=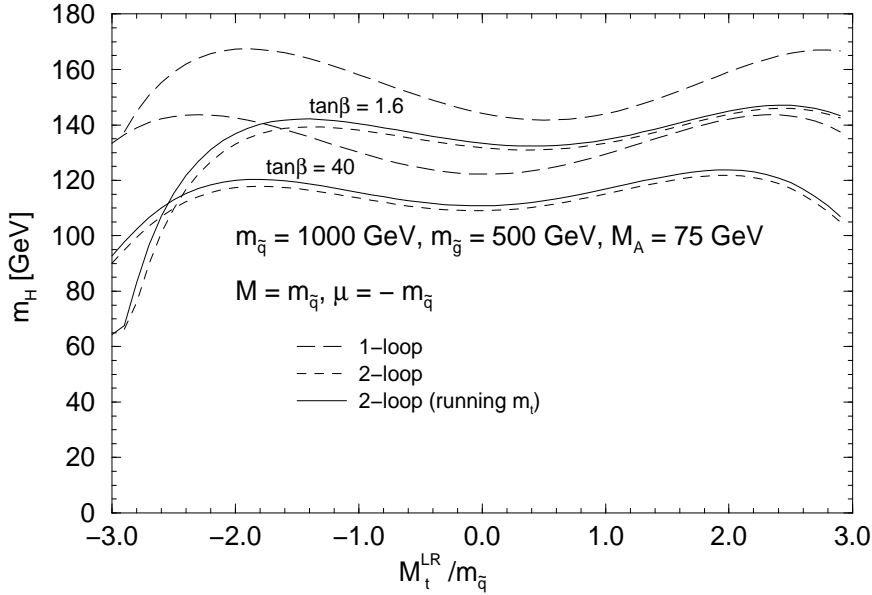,width=7.3cm,height=8cm,
                      bbllx=150pt,bblly=100pt,bburx=450pt,bbury=420pt}}
\end{center}
\caption[]{One- and \twol\ results for $\mH$ as a function of
$\Mtlr/\msq$ for two values of $\Tb$. The running top mass correction
discussed in the text is shown separately.}
\label{fig:mH_MtLRdivmq}
\end{figure}

\begin{figure}[ht!]
\begin{center}
\mbox{
\psfig{figure=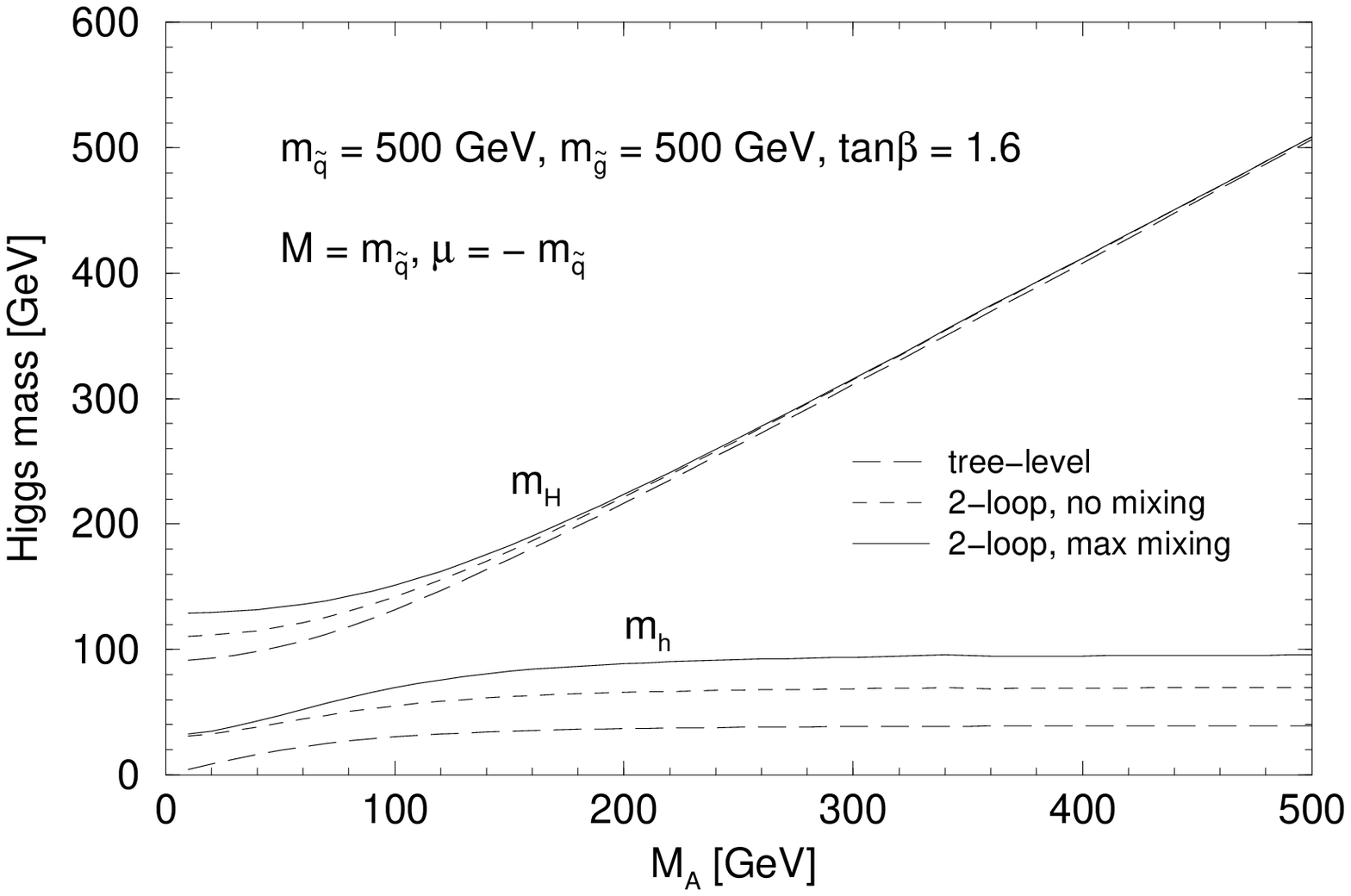,width=5.3cm,height=8cm,
                      bbllx=150pt,bblly=100pt,bburx=450pt,bbury=420pt}}
\hspace{7.5em}
\mbox{
\psfig{figure=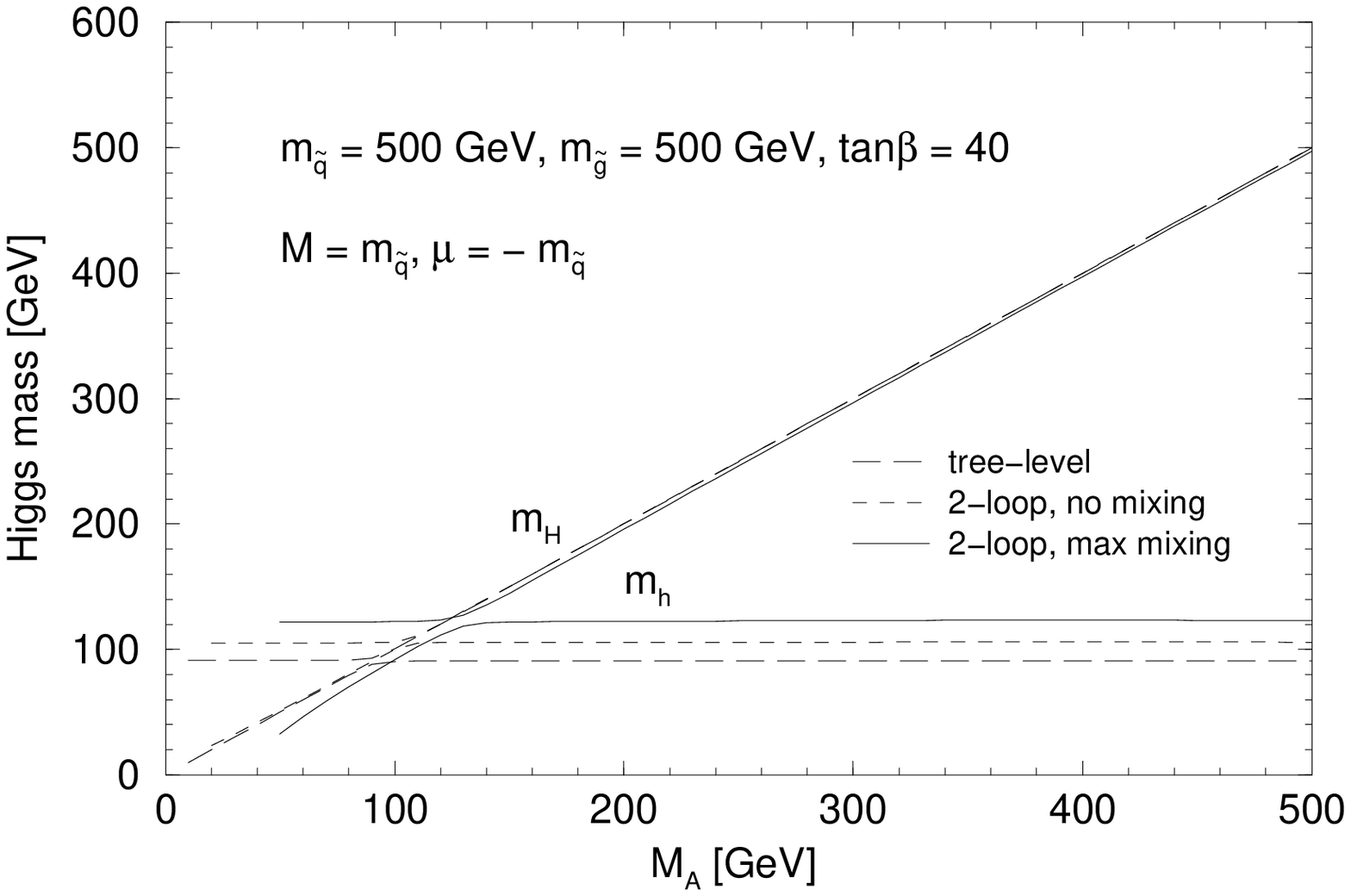,width=5.3cm,height=8cm,
                      bbllx=150pt,bblly=100pt,bburx=450pt,bbury=420pt}}
\end{center}
\caption[]{
$\mh$ and $\mH$ are shown as a function of $\MA$ for $\Tb = 1.6$ and
$\Tb = 40$. 
The \twol\ value for $\mh$ contains both corrections beyond $\oaas$, 
whereas for $\mH$ only the running top mass effect is included. 
The maximal-mixing scenario corresponds to the definition 
according to the discussion of \reffi{fig:mh_MtLRdivmq}. 
}
\label{fig:mhH_MA}
\end{figure}

\reffi{fig:mH_MtLRdivmq} depicts the result for the heavy
Higgs-boson mass, $\mH$, obtained in the same way as $\mh$ above. The only
difference is that no Yukawa term has been included. In the plot
we have chosen the small value $\MA = 75 \gev$, close to the lower
experimental bound,  since only for a light
$A$ boson the higher-order corrections give a sizable
contribution (see also \reffi{fig:mhH_MA}). Here the values for $\mH$
obtained for small $\Tb$ are larger than for $\Tb = 40$. 
The values of $\Mtlr/\msq$ for which $\mH$ is maximal depend in this
case on $\Tb$ and the sign of $\Mtlr$.

Both Higgs-boson masses are shown in \reffi{fig:mhH_MA} for low and high
$\Tb$ and the no-mixing and the maximal-mixing case, where the latter case
corresponds to the definition according to \reffi{fig:mh_MtLRdivmq}
for the light Higgs boson. For $\mH$ sizable corrections at the one- and
\twol\ level are obtained only for $\MA \lsim 200 \gev$ for $\Tb = 1.6$
and for $\MA \lsim 120 \gev$ for $\Tb = 40$.

\begin{figure}[ht!]
\begin{center}
\hspace{1em}
\mbox{
\psfig{figure=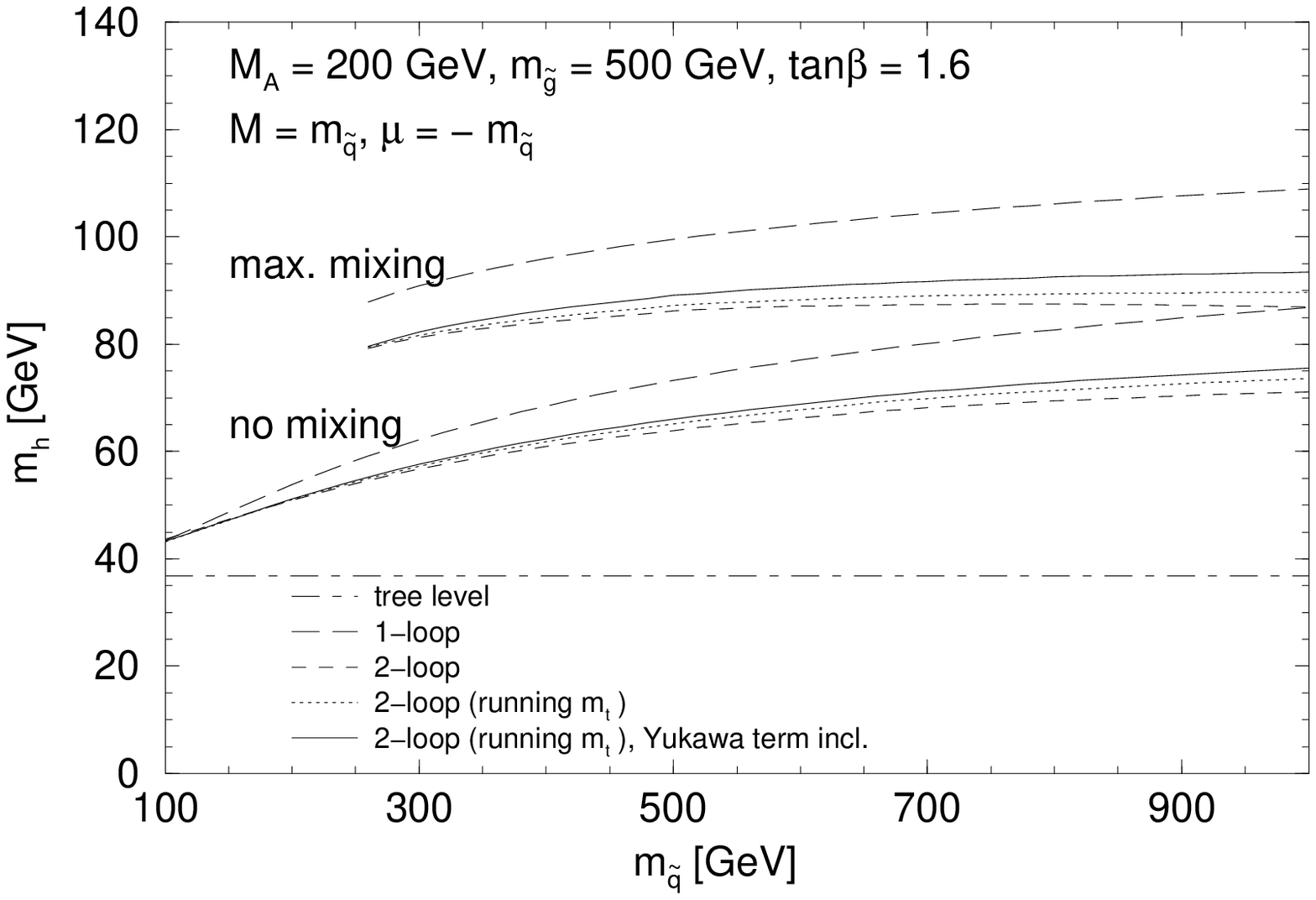,width=5.3cm,height=8cm,
                      bbllx=150pt,bblly=100pt,bburx=450pt,bbury=420pt}}
\hspace{7.5em}
\mbox{
\psfig{figure=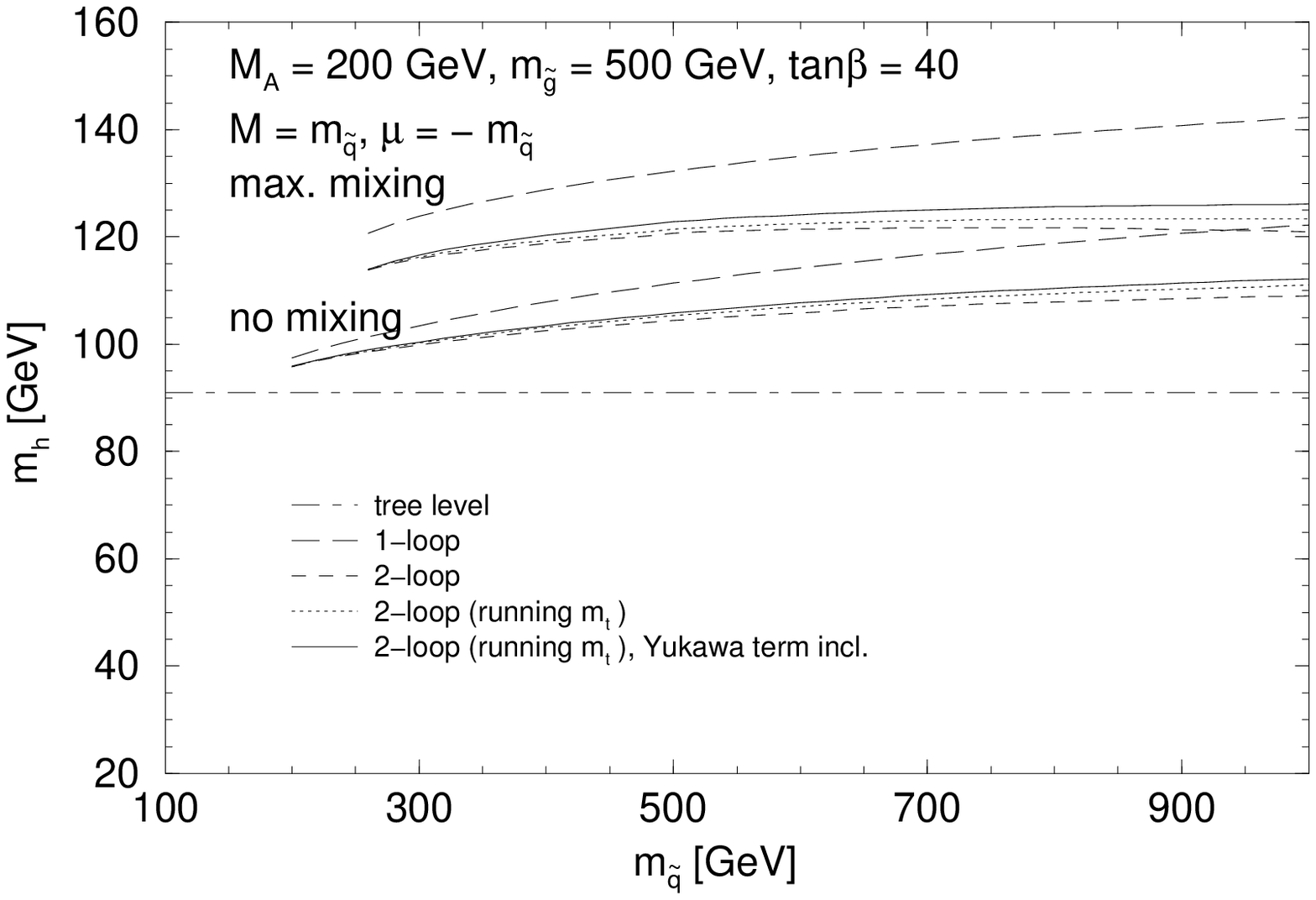,width=5.3cm,height=8cm,
                      bbllx=150pt,bblly=100pt,bburx=450pt,bbury=420pt}}
\end{center}

\vspace{1cm}

\begin{center}
\hspace{1em}
\mbox{
\psfig{figure=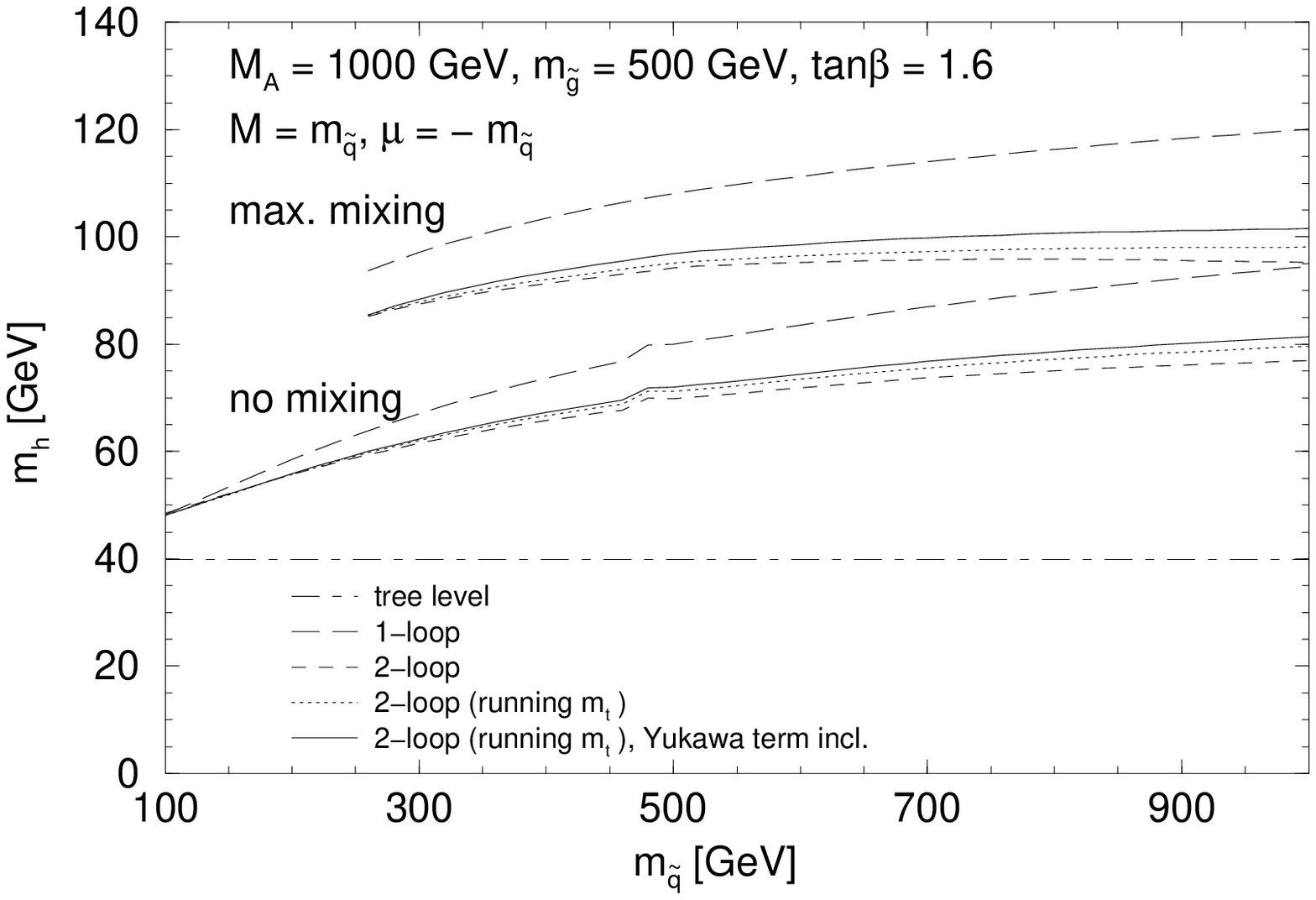,width=5.3cm,height=8cm,
                      bbllx=150pt,bblly=100pt,bburx=450pt,bbury=420pt}}
\hspace{7.5em}
\mbox{
\psfig{figure=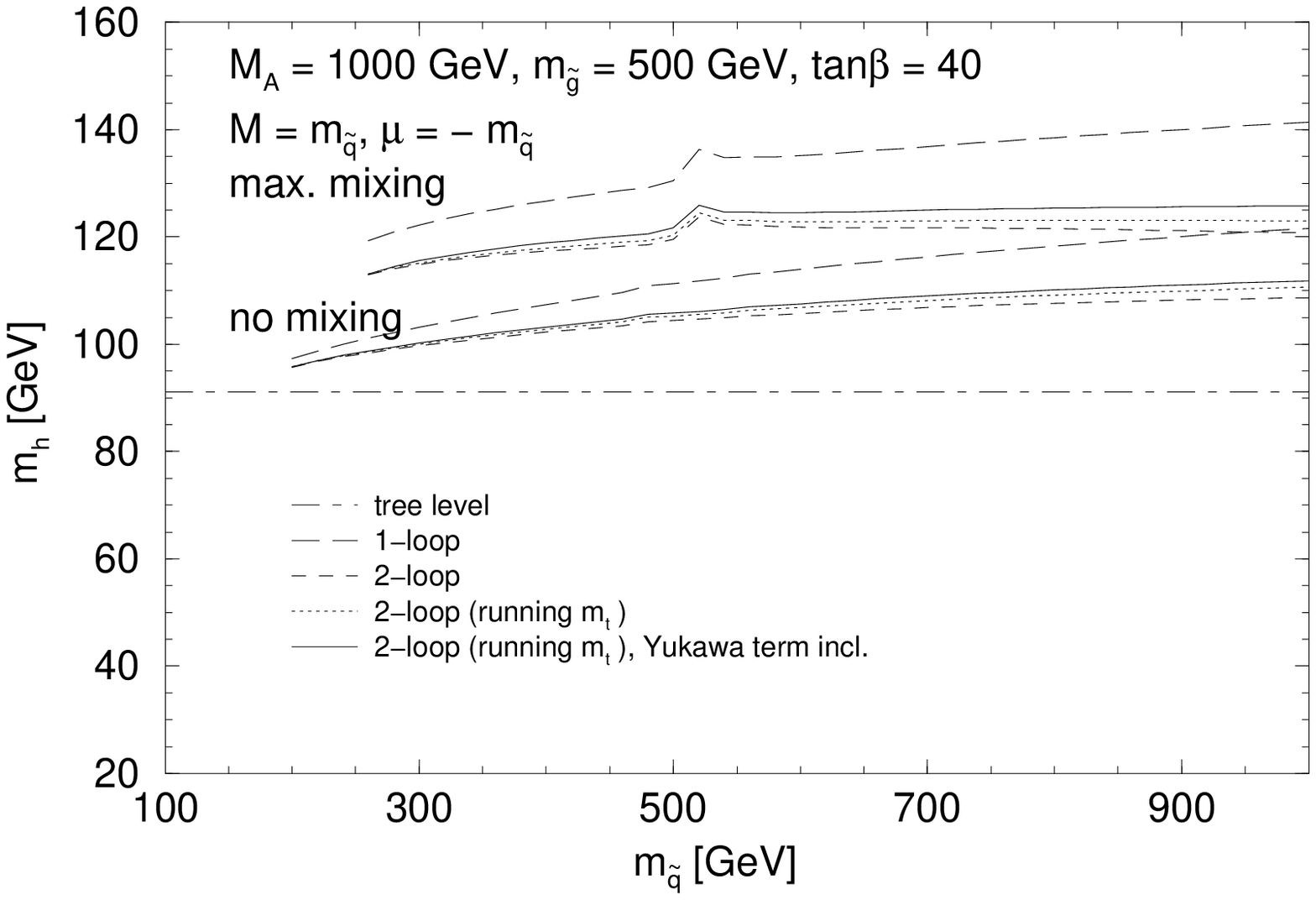,width=5.3cm,height=8cm,
                      bbllx=150pt,bblly=100pt,bburx=450pt,bbury=420pt}}
\end{center}
\caption[]{
The mass of the lightest Higgs boson for $\Tb = 1.6$ and $\Tb = 40$. 
The tree-level, the \onel\ and the \twol\ results for $\mh$ are shown 
as a function of $\msq$ for the no-mixing and the maximal-mixing case,
and for $\MA = 200,1000 \gev$.
} 
\label{fig:mh_mq}
\end{figure}

\bigskip
More relevant for todays' colliders is the mass of the lighter Higgs
boson, $\mh$, on which we will focus in the following discussion.
In \reffi{fig:mh_mq} $\mh$ is shown in the two scenarios with 
$\Tb = 1.6$ and $\Tb = 40$ as a function of $\msq$ for no mixing and 
maximal mixing and for $\MA = 200,1000 \gev$. The tree-level, the \onel\ and
the \twol\ results with the two corrections beyond $\oaas$
are shown (the values of $\msq$ are such
that the corresponding $\Stop$-masses lie within the experimentally
allowed region). 
In all scenarios of \reffi{fig:mh_mq} the \twol\
corrections give rise to a large reduction 
of the \onel\ value of $\mh$. The effect is generally larger in the 
$\Tb = 1.6$ scenario, and for maximal mixing and large $\MA$. 
The inclusion of the Yukawa correction and the running top mass
leads to a slight shift in $\mh$ towards higher values.
This effect amounts up to $20\%$ of the \twol\ correction. In the
$\Tb = 1.6$ scenario with $\msq = 1 \tev$, $\mh$ reaches about 
$75~(81) \gev$ for 
$\MA = 200~(1000) \gev$ in the no-mixing case,
and $94~(101) \gev$ in the 
maximal-mixing case. For $\Tb = 40$ the respective values of $\mh$ are
$112 \gev$ in the no-mixing case, and $126 \gev$ in the
maximal-mixing case for both values of $\MA$.
The peaks in the plots for $\MA = 1 \tev$ are due to the threshold
$\MA = \msbe + \msbz$ in the
\onel\ contribution, originating from the sbottom-loop diagram in
the $A$~self-energy.

\begin{figure}[ht!]
\begin{center}
\hspace{1em}
\mbox{
\psfig{figure=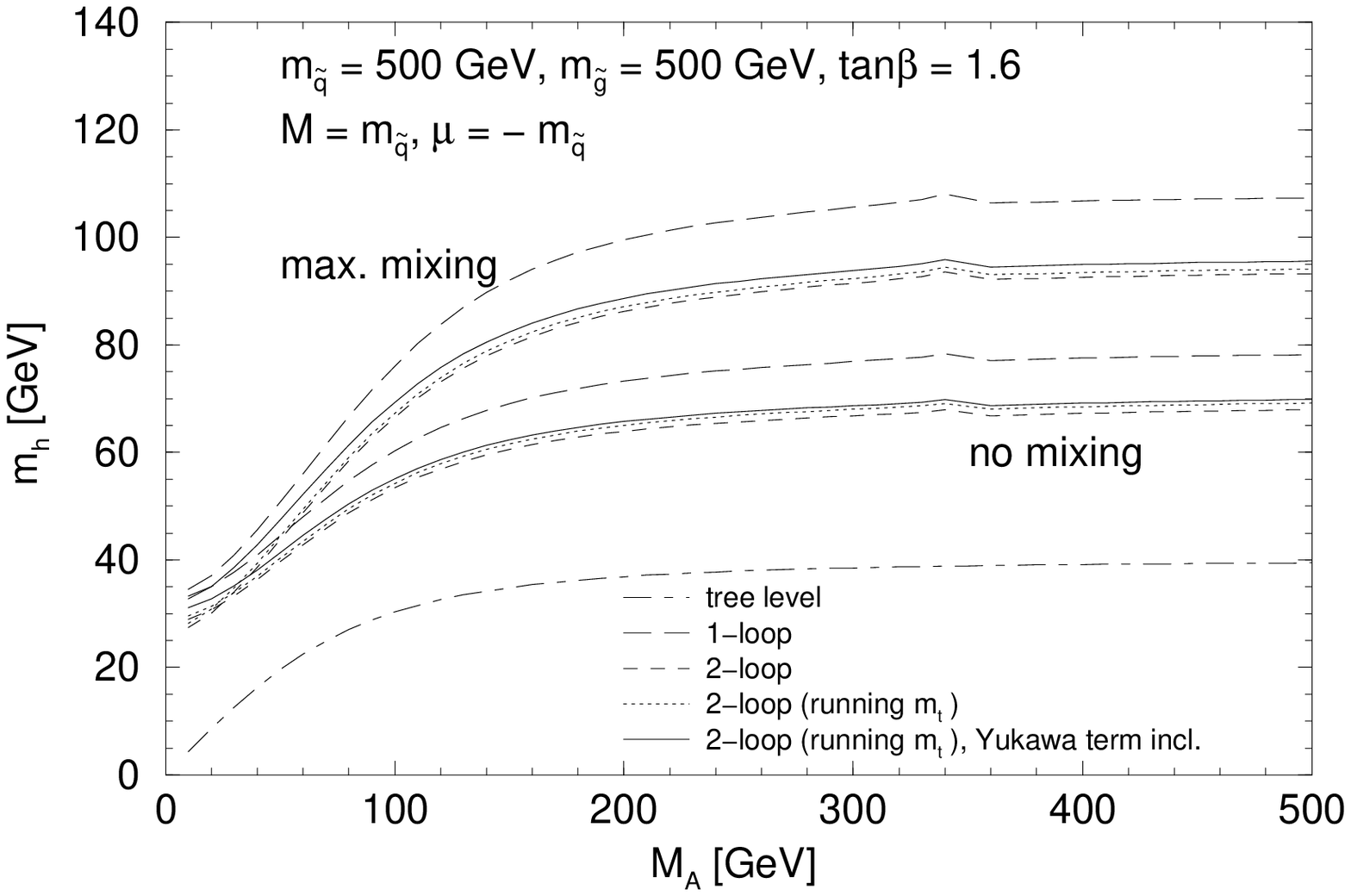,width=5.3cm,height=8cm,
                      bbllx=150pt,bblly=100pt,bburx=450pt,bbury=420pt}}
\hspace{7.5em}
\mbox{
\psfig{figure=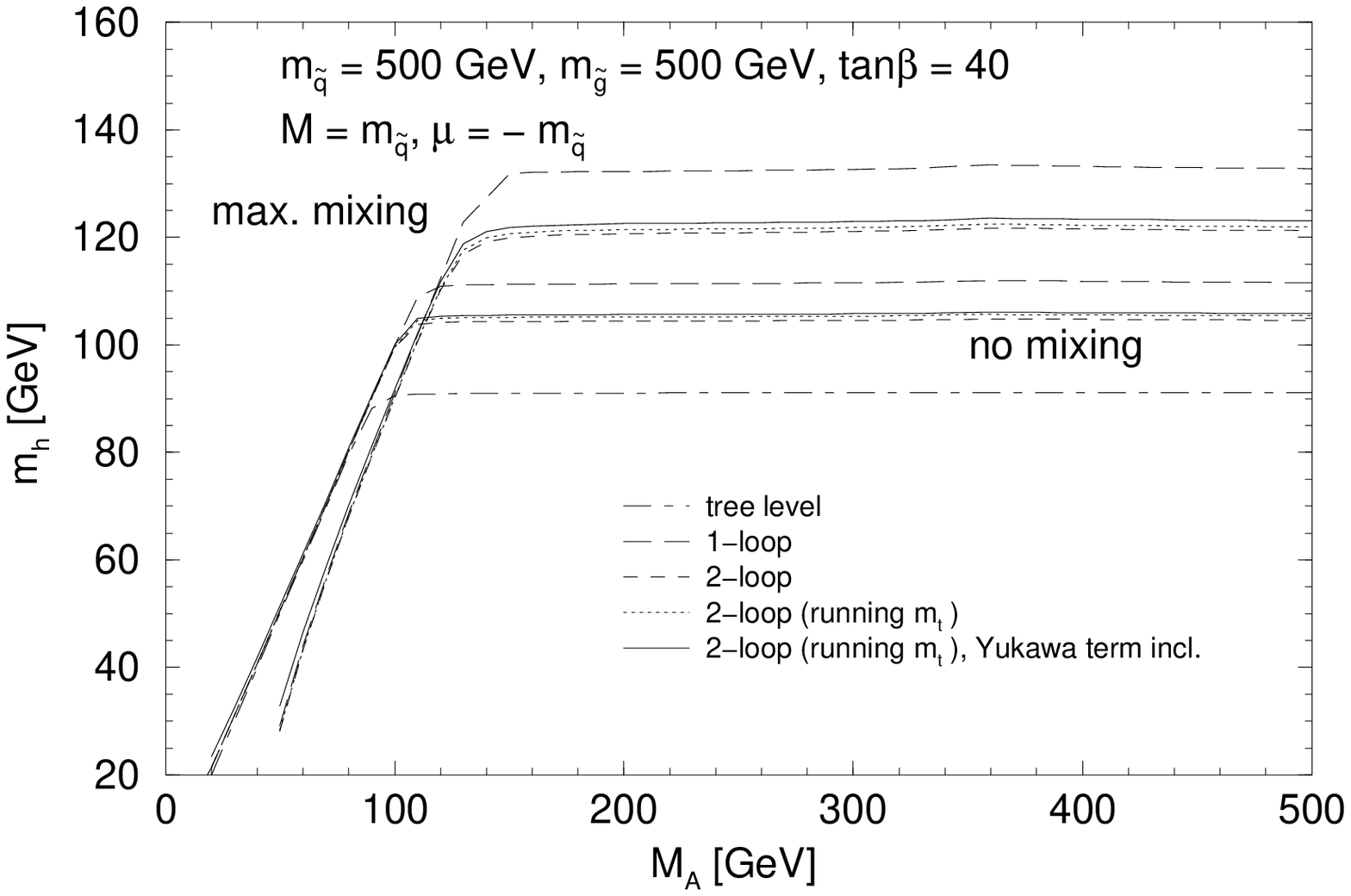,width=5.3cm,height=8cm,
                      bbllx=150pt,bblly=100pt,bburx=450pt,bbury=420pt}}
\end{center}

\vspace{1cm}

\begin{center}
\hspace{1em}
\mbox{
\psfig{figure=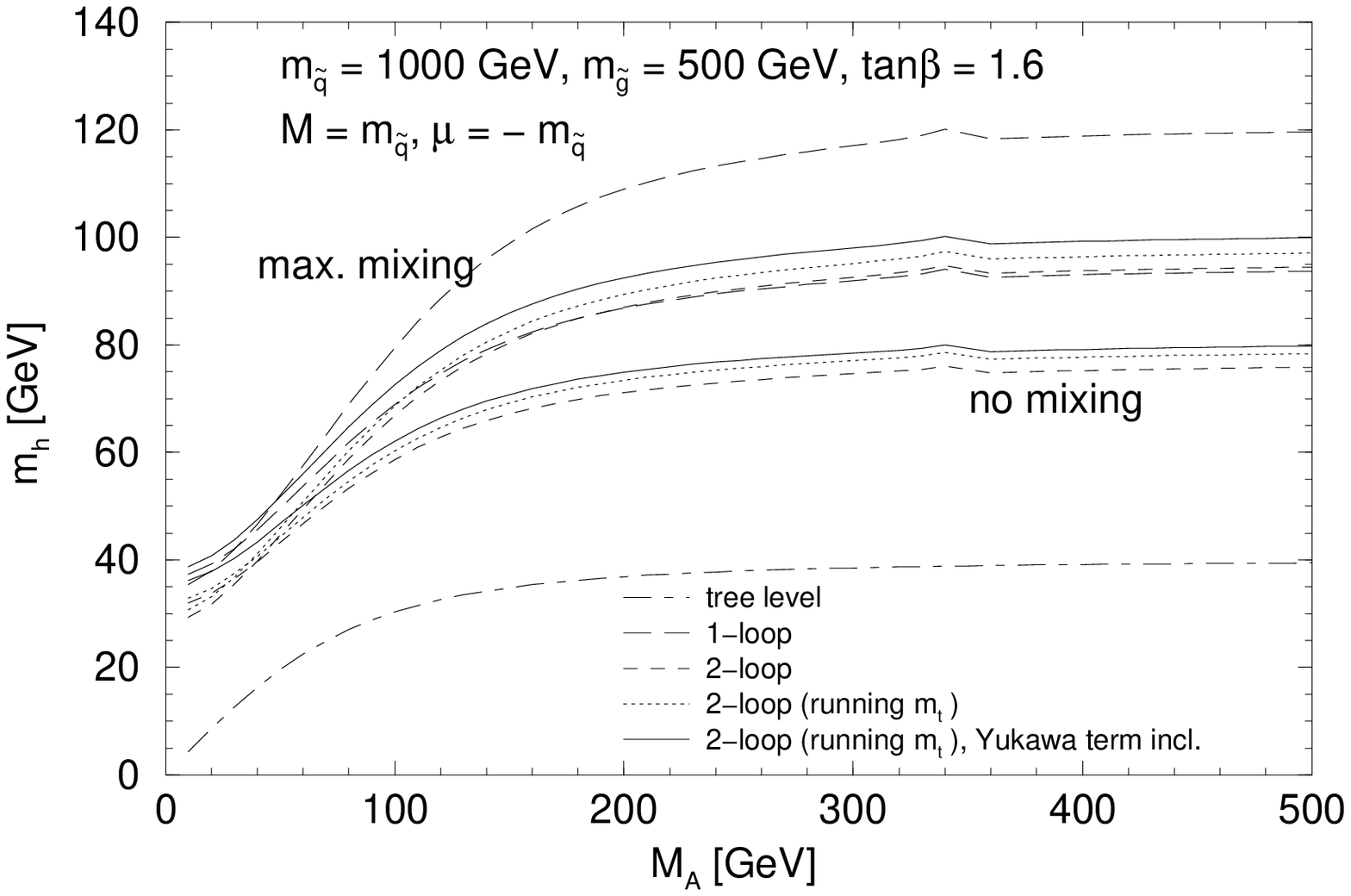,width=5.3cm,height=8cm,
                      bbllx=150pt,bblly=100pt,bburx=450pt,bbury=420pt}}
\hspace{7.5em}
\mbox{
\psfig{figure=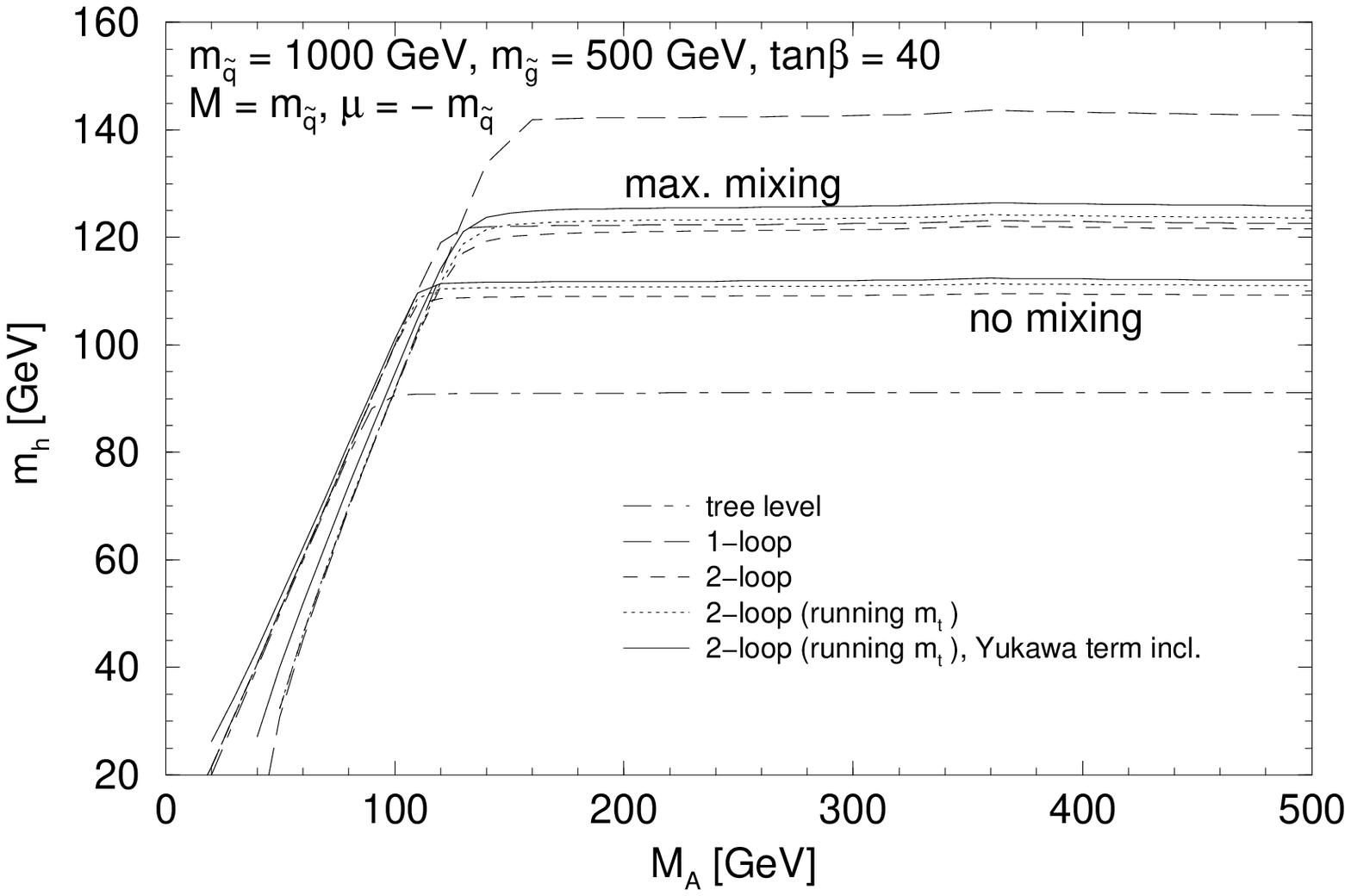,width=5.3cm,height=8cm,
                      bbllx=150pt,bblly=100pt,bburx=450pt,bbury=420pt}}
\end{center}
\caption[]{
The mass of the lightest Higgs boson as a function of $\MA$ for 
$\Tb = 1.6$ and $\Tb = 40$.  
The tree-level, the \onel\ and the \twol\ results for $\mh$ are shown 
in the no-mixing and the maximal-mixing case,
and for $\msq = 500,1000 \gev$.
} 
\label{fig:mh_MA}
\end{figure}

\smallskip
The dependence of $\mh$ on $\MA$ is depicted in \reffi{fig:mh_MA}
in the two scenarios with 
$\Tb = 1.6$ and $\Tb = 40$ for no mixing and 
maximal mixing for $\msq = 500,1000 \gev$.
In all scenarios of \reffi{fig:mh_MA} the \twol\
corrections give rise to a large reduction 
of the \onel\ value of $\mh$.
A saturation effect can be observed for
$\MA \gsim 300~(150) \gev$ in the $\Tb = 1.6~(40)$ scenario.
The peaks in the plots for $\MA = 350 \gev$ are due to the threshold
$\MA = 2\,\mt$ in the
\onel\ contribution, originating from the top-loop diagram in
the $A$~self-energy.

\begin{figure}[ht!]
\begin{center}
\hspace{1em}
\mbox{
\psfig{figure=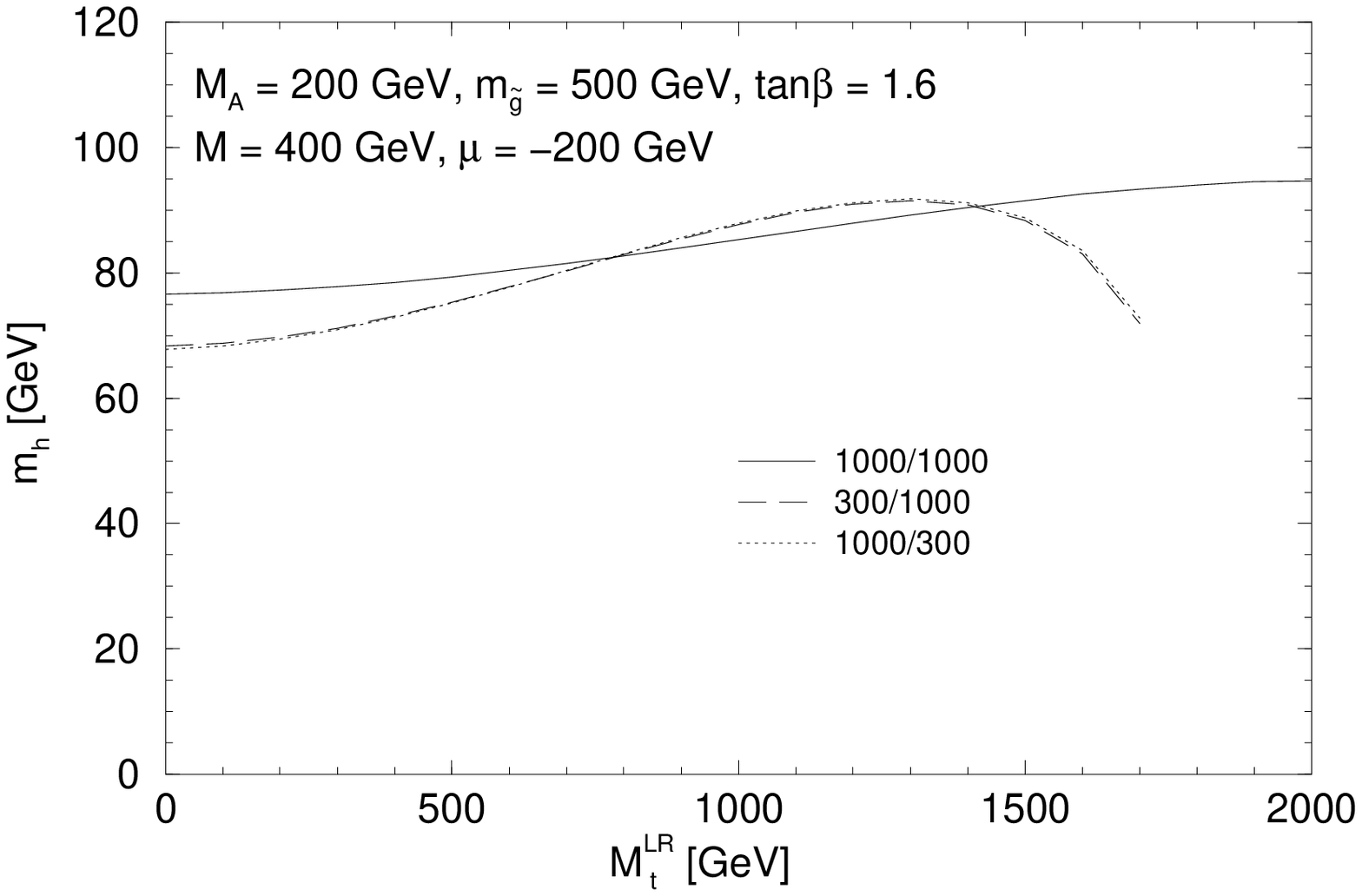,width=5.3cm,height=8cm,
                      bbllx=150pt,bblly=100pt,bburx=450pt,bbury=420pt}}
\hspace{7.5em}
\mbox{
\psfig{figure=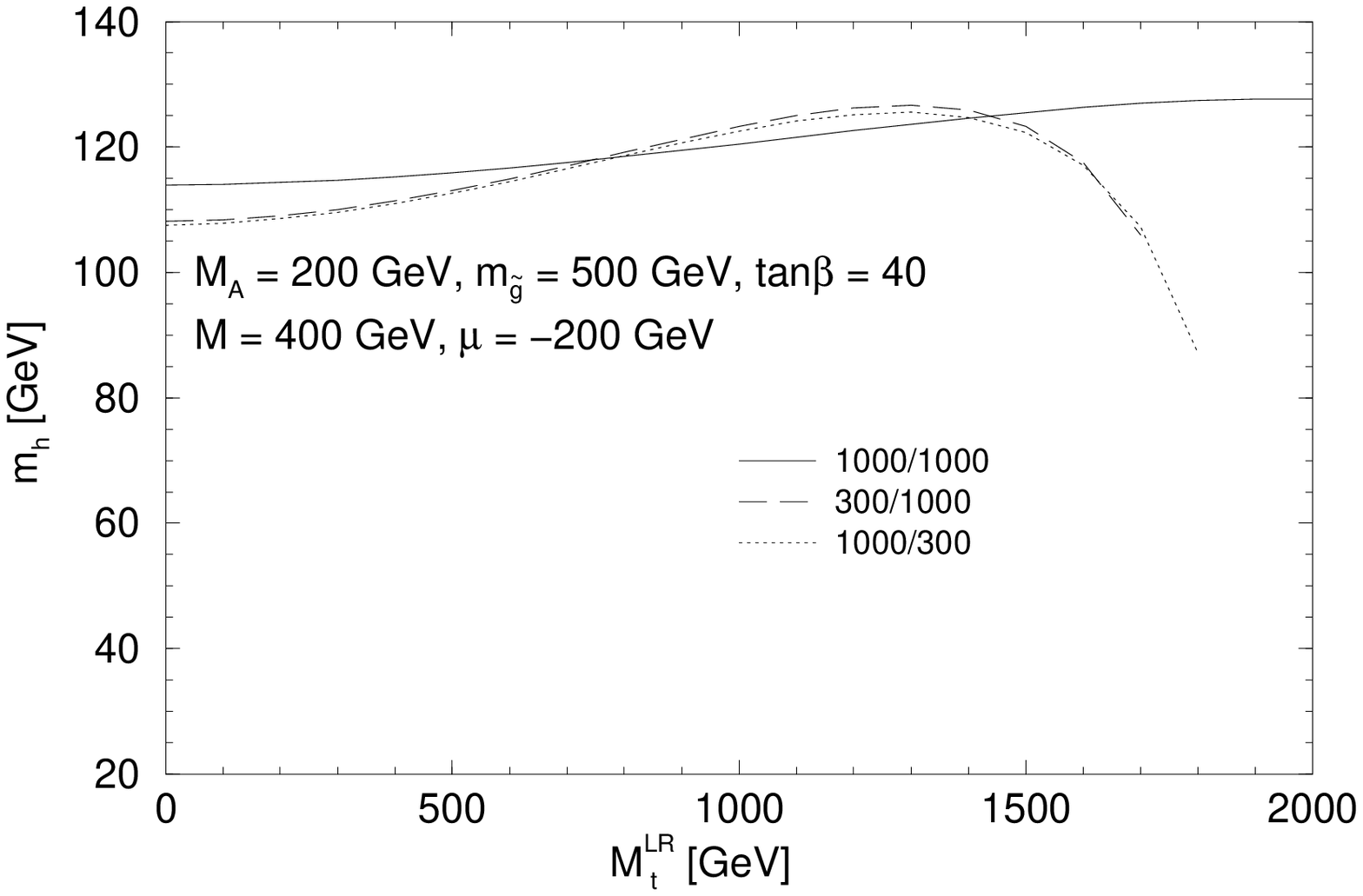,width=5.3cm,height=8cm,
                      bbllx=150pt,bblly=100pt,bburx=450pt,bbury=420pt}}
\end{center}

\vspace{1cm}

\begin{center}
\hspace{1em}
\mbox{
\psfig{figure=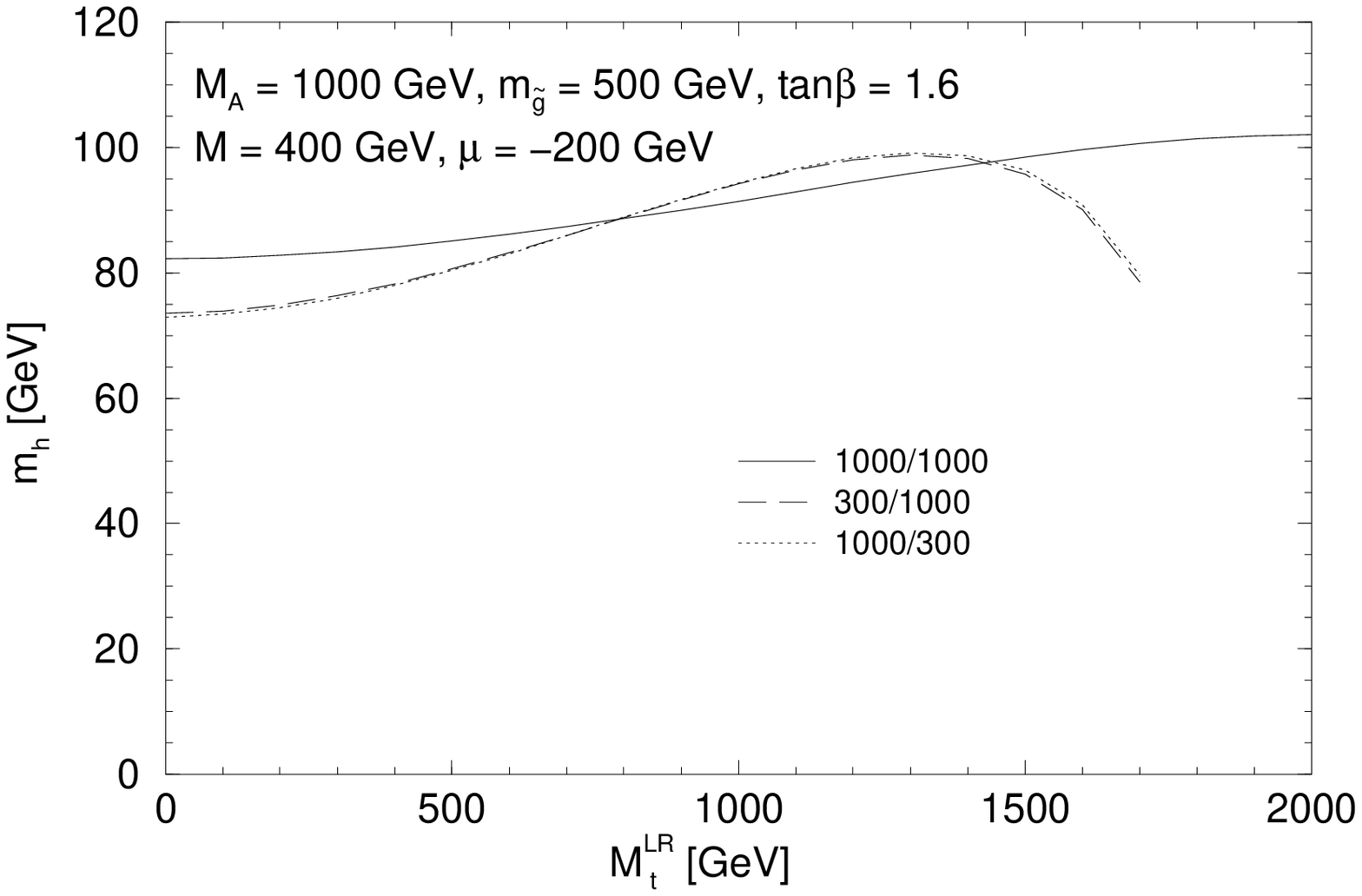,width=5.3cm,height=8cm,
                      bbllx=150pt,bblly=100pt,bburx=450pt,bbury=420pt}}
\hspace{7.5em}
\mbox{
\psfig{figure=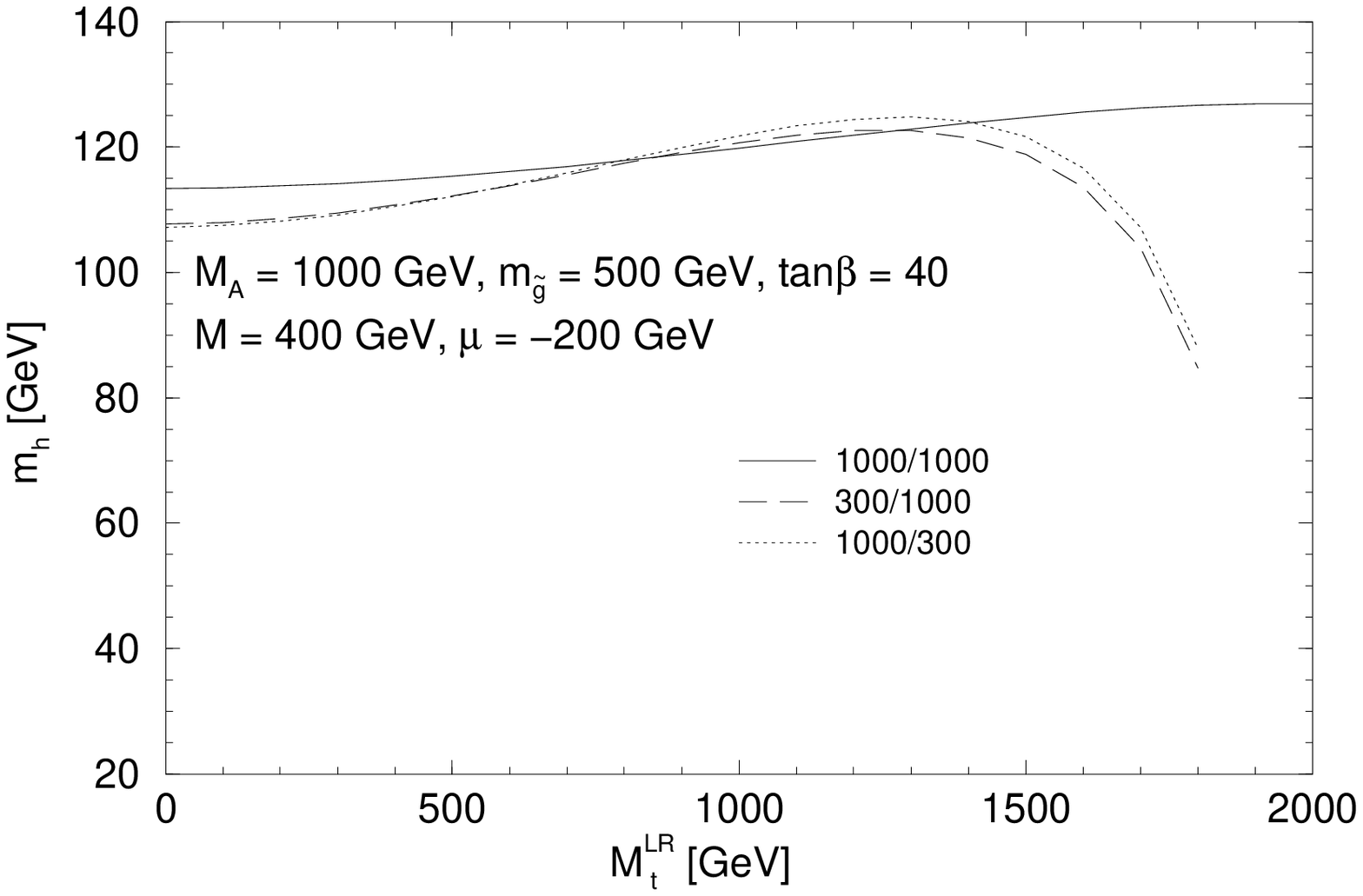,width=5.3cm,height=8cm,
                      bbllx=150pt,bblly=100pt,bburx=450pt,bbury=420pt}}
\end{center}
\caption[]{
The mass of the lightest Higgs boson as a function of $\Mtlr$ for
differently split values of the soft 
SUSY breaking terms. The curves in the plots correspond to the values
$\MstL/\MstR = 1000/1000,~300/1000$ and $1000/300$.
} 
\label{fig:mh_MtLR}
\end{figure}

\bigskip
Allowing for a splitting between the parameters 
$\MstL$, $\MstR$ in the $\Stop$ mass matrix yields maximal values of
$\mh$ which are approximately the same as for the case $\msq = \MstL =
\MstR$, provided that one sets 
\BE
m_{\tilde{q}_{\bigr|\MstL = \MstR}} = 
 {\rm max}\{\MstL, \MstR\}_{\Bigr|\MstL \neq \MstR}~,
\EE
see \reffi{fig:mh_MtLR}. 
However, the location of the maximal Higgs-boson mass, depending on
$\Mtlr$, is shifted towards smaller values, typically by about 40\%.
The numerical difference in $\mh$ in the two splitting scenarios
$\MstL/\MstR = 300/1000$ and $\MstL/\MstR = 1000/300$ is
small. They differ by up to $2 \gev$ only in the large $\Tb$ scenario when
$\Mtlr > 1000 \gev$.

\begin{figure}[ht!]
\begin{center}
\hspace{1em}
\mbox{
\psfig{figure=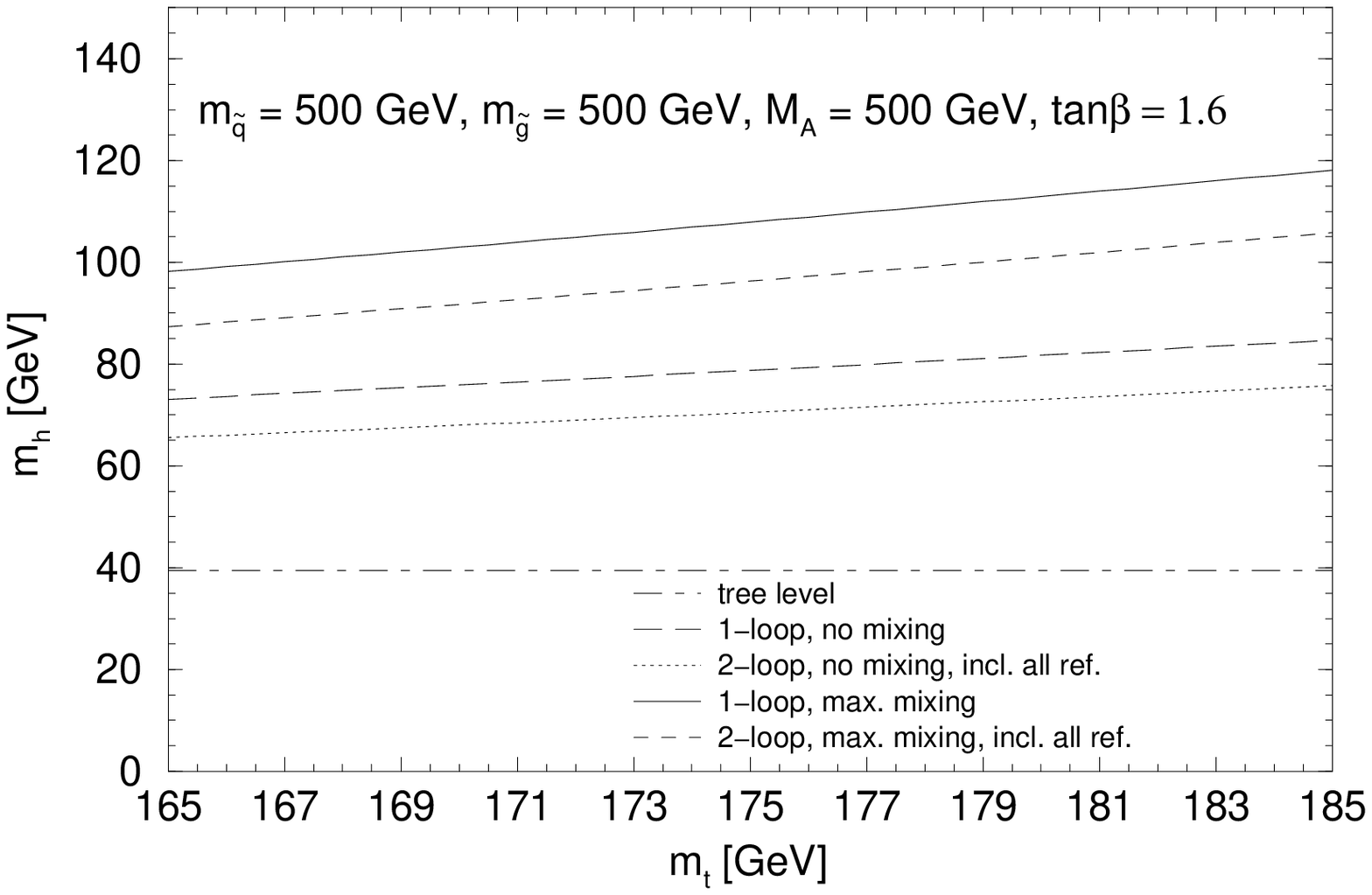,width=5.3cm,height=8cm,
                      bbllx=150pt,bblly=100pt,bburx=450pt,bbury=420pt}}
\hspace{7.5em}
\mbox{
\psfig{figure=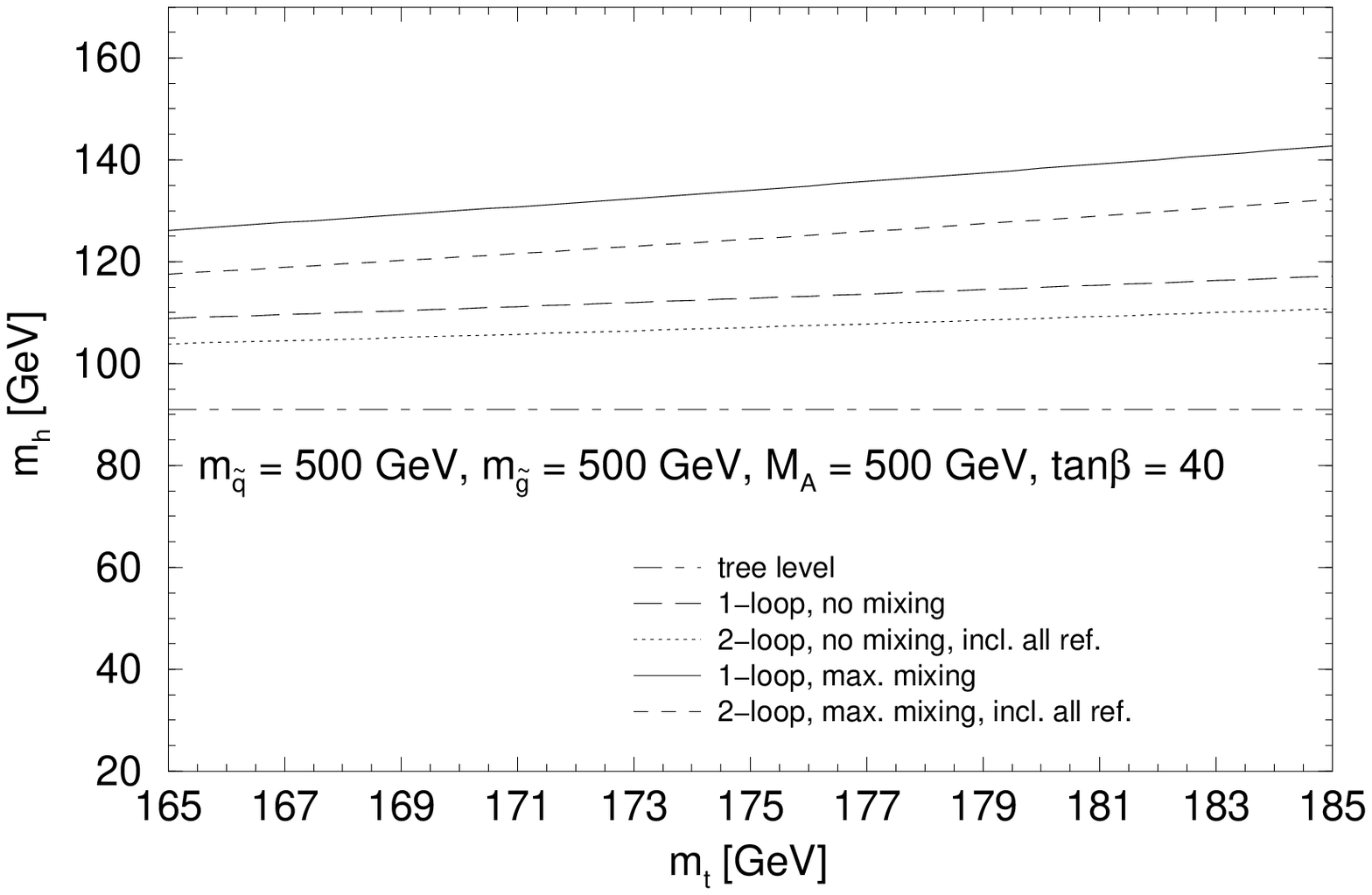,width=5.3cm,height=8cm,
                      bbllx=150pt,bblly=100pt,bburx=450pt,bbury=420pt}}
\end{center}

\vspace{1cm}

\begin{center}
\hspace{1em}
\mbox{
\psfig{figure=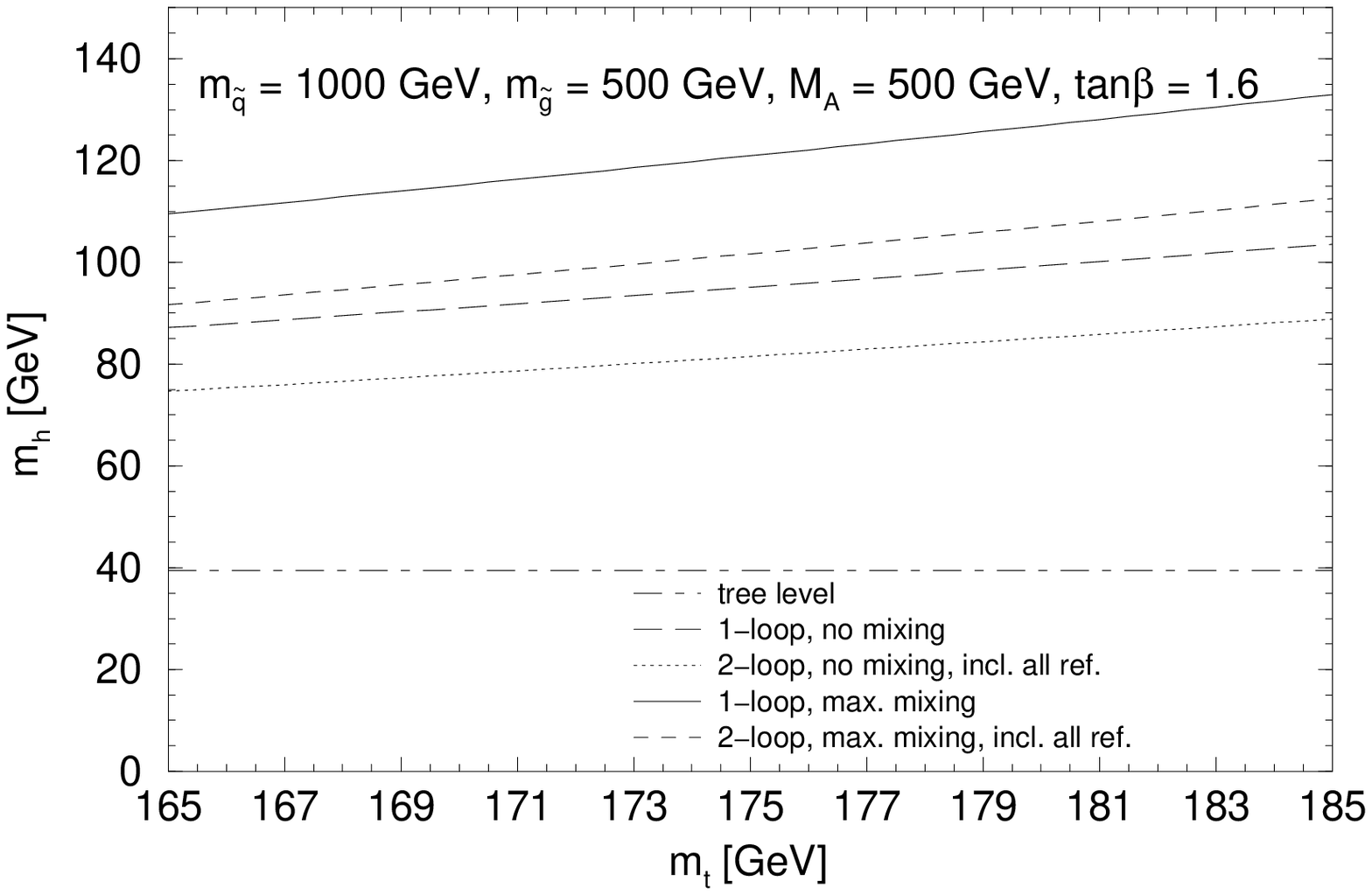,width=5.3cm,height=8cm,
                      bbllx=150pt,bblly=100pt,bburx=450pt,bbury=420pt}}
\hspace{7.5em}
\mbox{
\psfig{figure=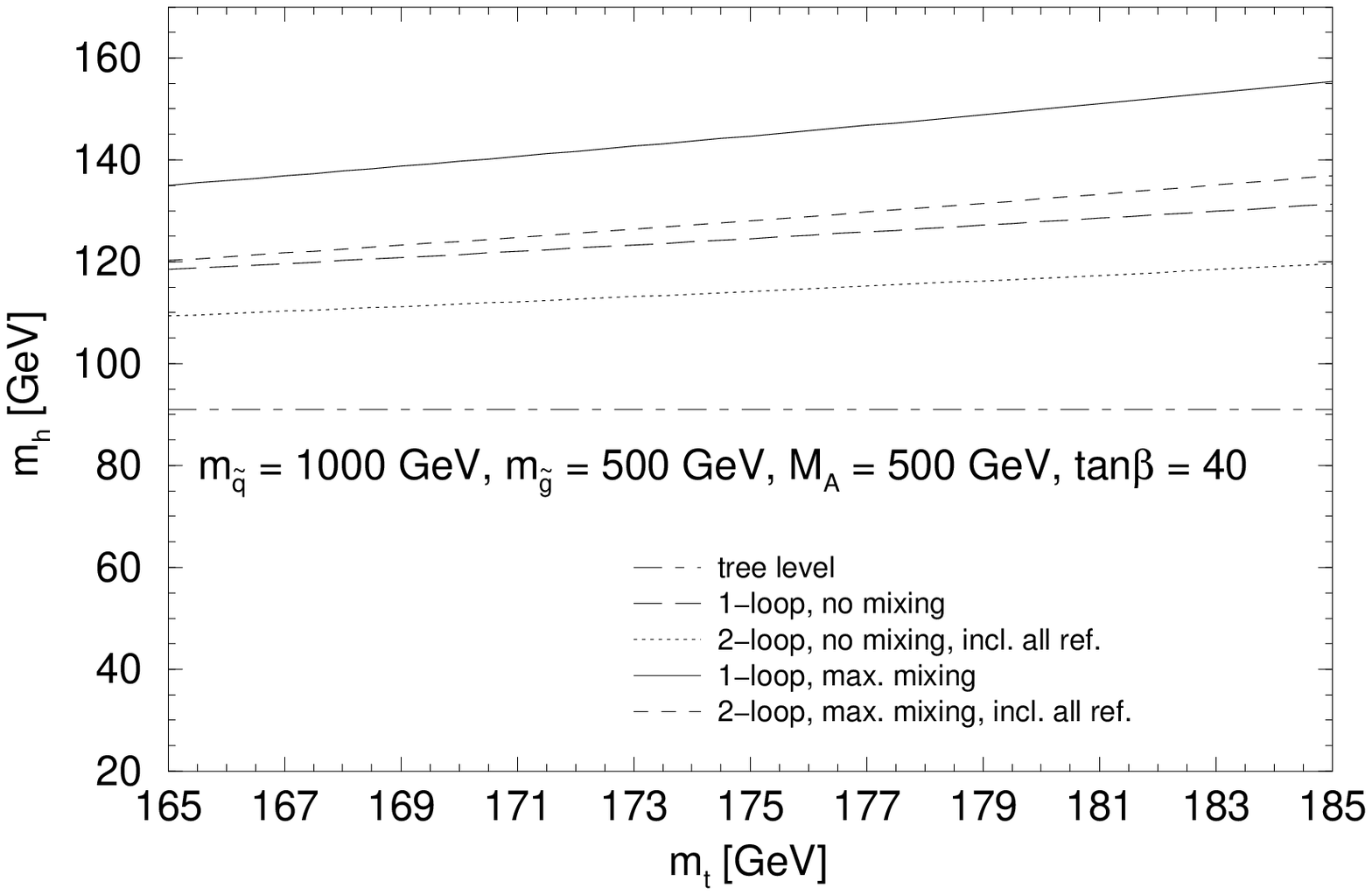,width=5.3cm,height=8cm,
                      bbllx=150pt,bblly=100pt,bburx=450pt,bbury=420pt}}
\end{center}
\caption[]{
The mass of the lightest Higgs boson is shown as a function of $\mt$:
the scenarios of no mixing and maximal mixing are depicted for low and
high $\Tb$, and  $\mu = -200 \gev, M = 400 \gev$.
} 
\label{fig:mh_mt}
\end{figure}

\bigskip
The variation of $\mh$ with $\mt$ is rather strong. The scenarios for
no mixing and maximal mixing and for $\Tb = 1.6$ and $\Tb = 40$ are
shown in \reffi{fig:mh_mt}, where $\mt$ is varied around the central value of 
$\mt = 175 \gev$ by $\pm 10 \gev$. The variation of $\mh$ is stronger for low
$\Tb$ and larger $\msq$: in the $\Tb = 1.6$ scenario $\mh$ varies by
more than $10 \gev$ and about $20 \gev$ for no-mixing and
maximal-mixing, respectively. 
In the $\Tb = 40$ scenario the respective values are less
than $10 \gev$ and about $15 \gev$.

\begin{figure}[ht!]
\begin{center}
\hspace{1em}
\mbox{
\psfig{figure=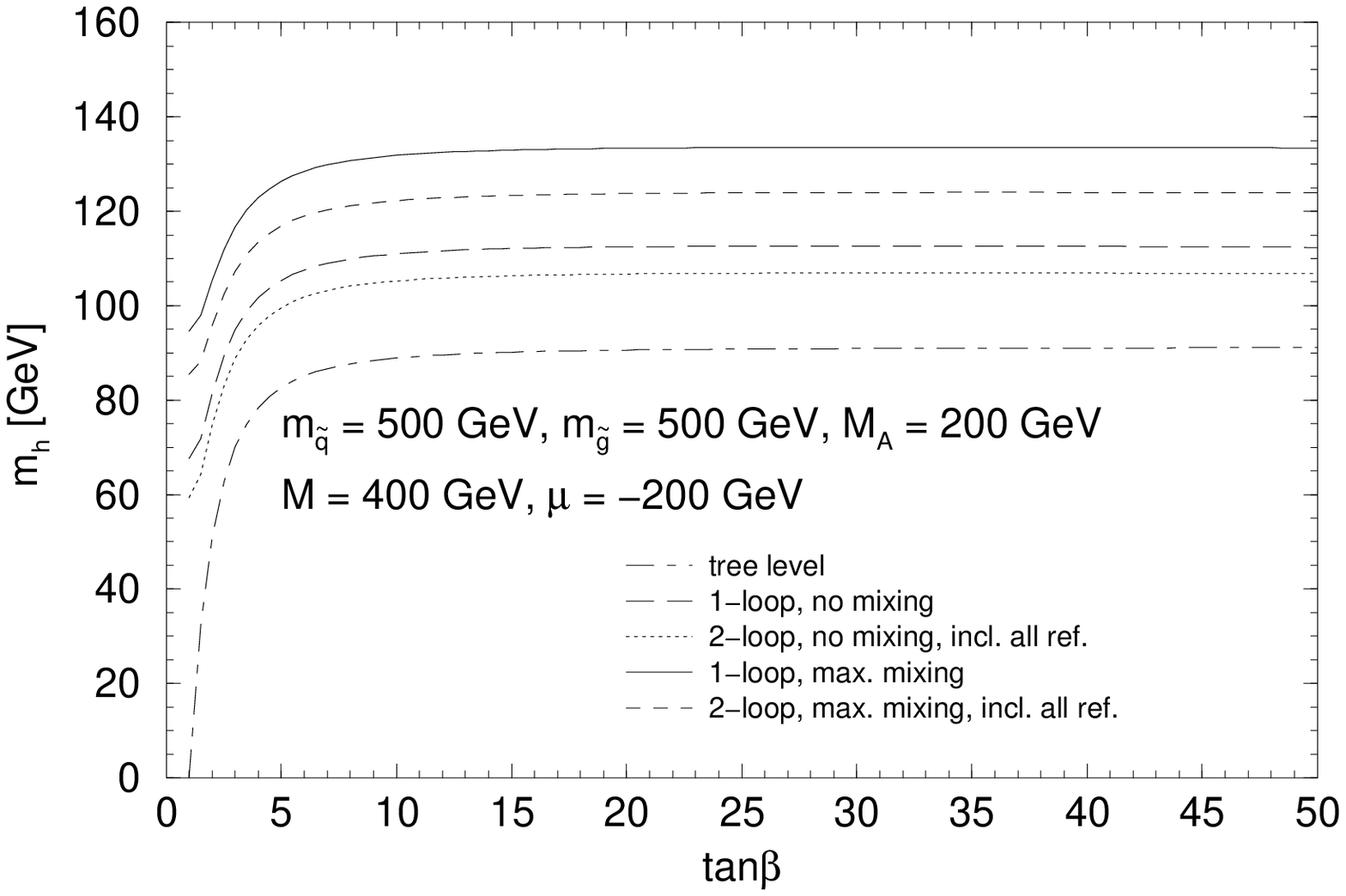,width=5.3cm,height=8cm,
                      bbllx=150pt,bblly=100pt,bburx=450pt,bbury=420pt}}
\hspace{7.5em}
\mbox{
\psfig{figure=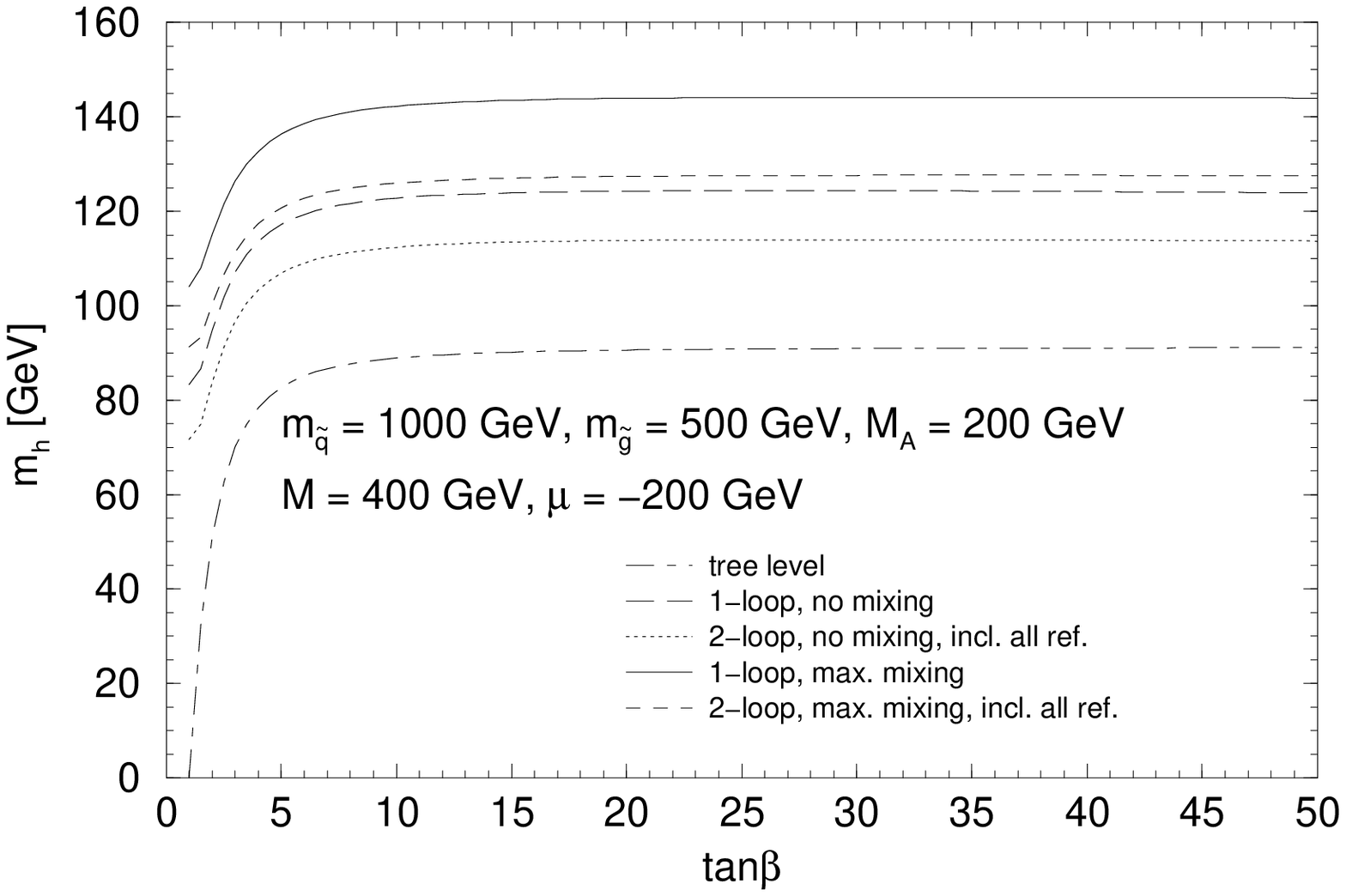,width=5.3cm,height=8cm,
                      bbllx=150pt,bblly=100pt,bburx=450pt,bbury=420pt}}
\end{center}

\vspace{1cm}

\begin{center}
\hspace{1em}
\mbox{
\psfig{figure=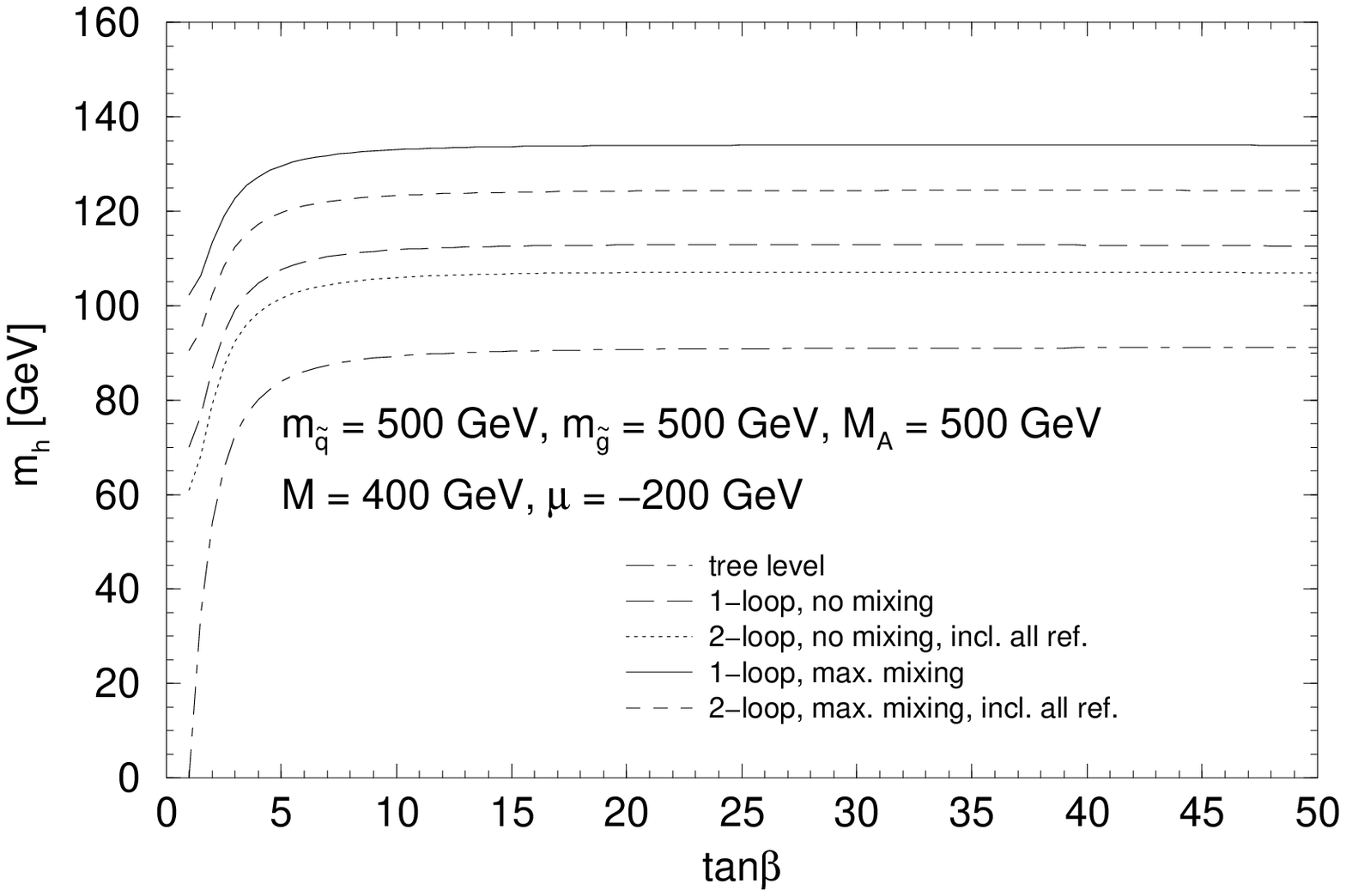,width=5.3cm,height=8cm,
                      bbllx=150pt,bblly=100pt,bburx=450pt,bbury=420pt}}
\hspace{7.5em}
\mbox{
\psfig{figure=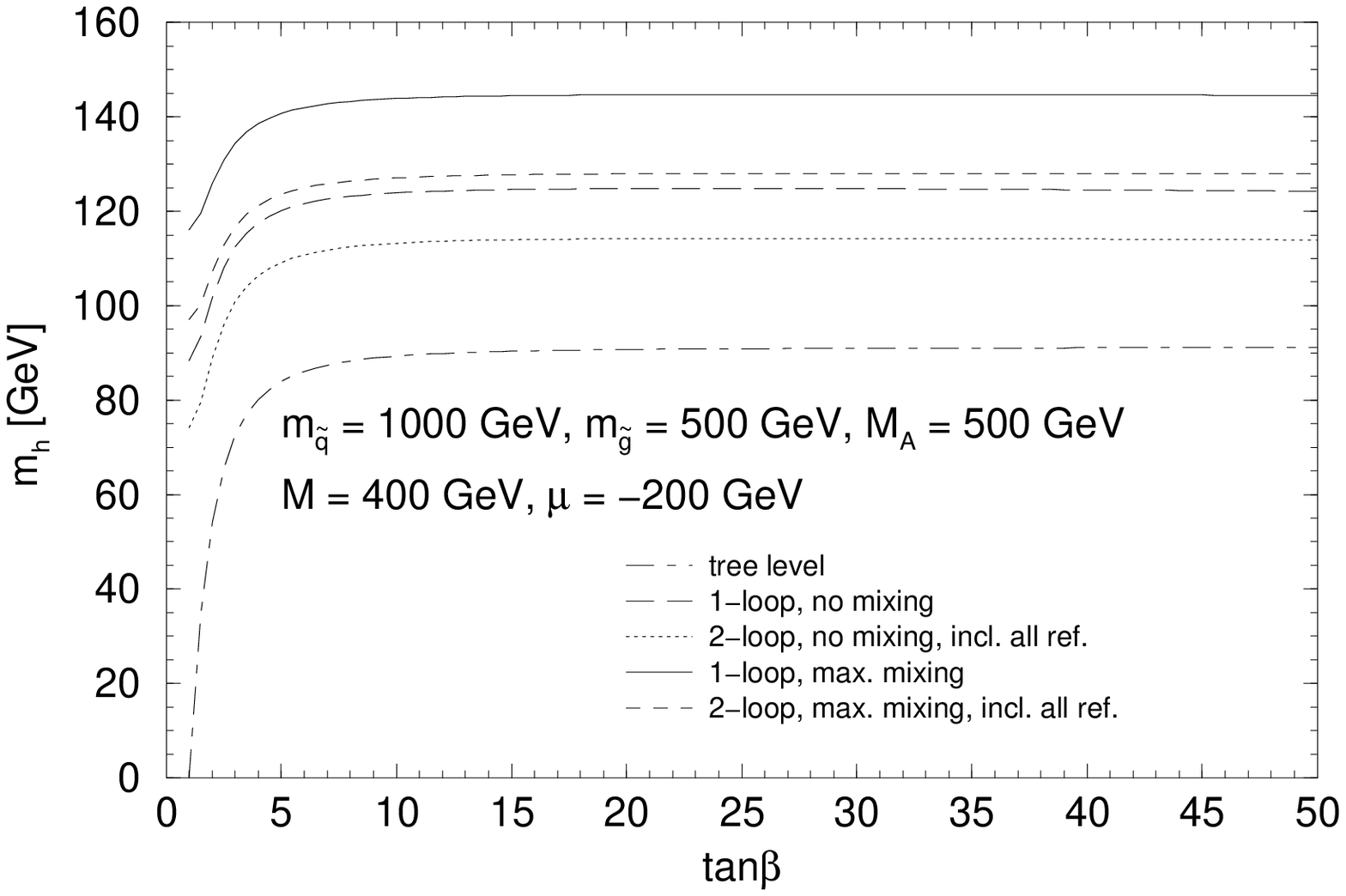,width=5.3cm,height=8cm,
                      bbllx=150pt,bblly=100pt,bburx=450pt,bbury=420pt}}
\end{center}
\caption[]{
The mass of the lightest Higgs boson is shown as a function of
$\Tb$. $\mh$ is 
plotted for the scenarios with $\MA = 200,500 \gev$ and 
$\msq = 500,1000 \gev$ in the no-mixing and the maximal-mixing case.
} 
\label{fig:mh_tb}
\end{figure}

\bigskip
Varying $\Tb$ around the value $\Tb = 1.6$ has a relatively large
impact on $\mh$ (higher values for $\mh$ are obtained
for larger $\Tb$), while the effect of varying $\Tb$ around $\Tb = 40$
is marginal. This is shown in \reffi{fig:mh_tb} for 
$\MA = 200, 1000 \gev$, $\msq = 200, 1000 \gev$ for the no-mixing
and the maximal-mixing scenario.
For $\Tb > 15$ the variation is less than $1 \gev$%
\footnote{
A non-negligible effect can arise for large $\Tb$ if $\mu$ is also
large. This is briefly discussed below in the context of the
$\mu$-dependence of $\mh$.
}.

\begin{figure}[ht!]
\begin{center}
\mbox{
\psfig{figure=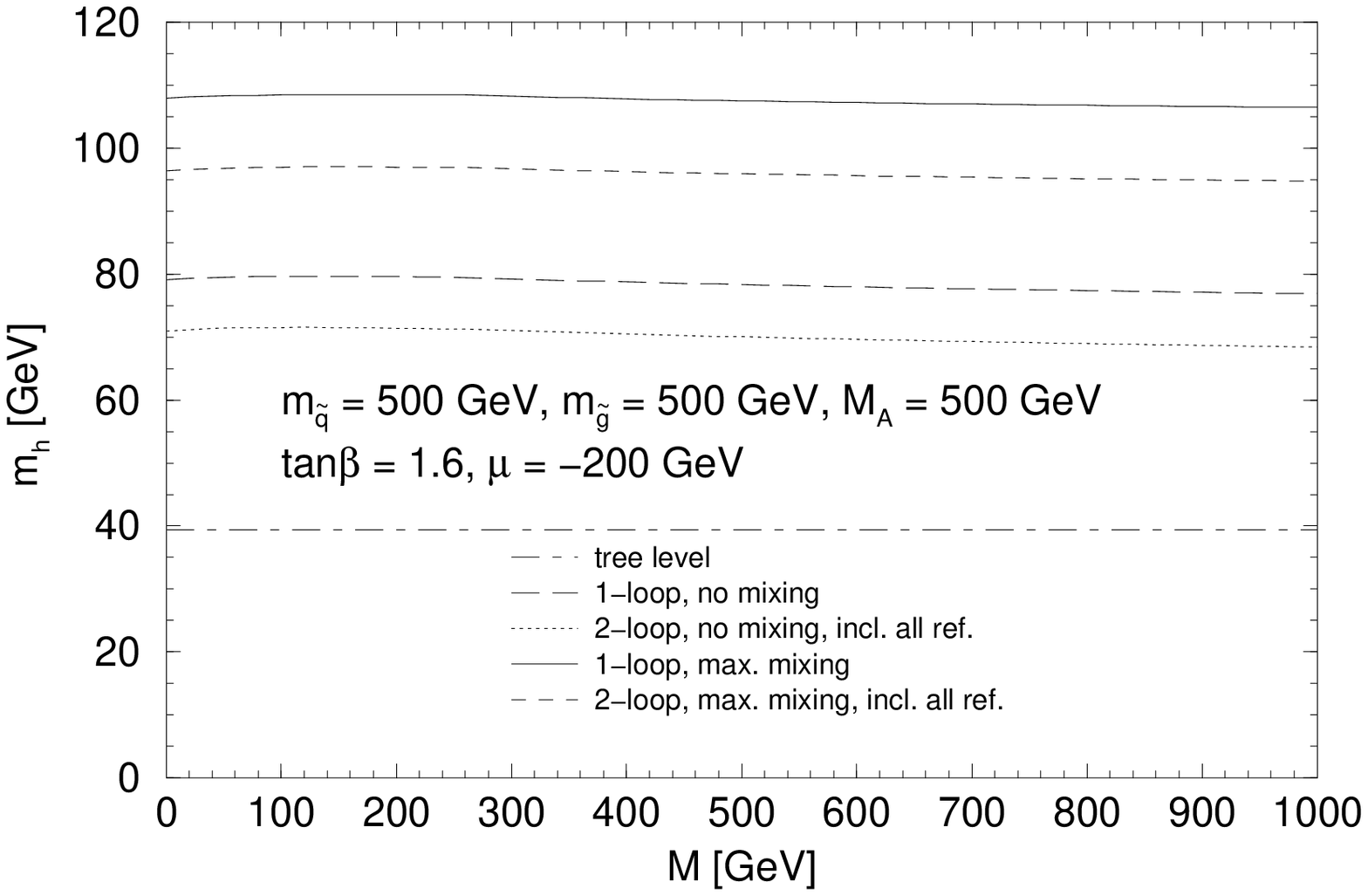,width=5.3cm,height=8cm,
                      bbllx=150pt,bblly=100pt,bburx=450pt,bbury=420pt}}
\hspace{7.5em}
\mbox{
\psfig{figure=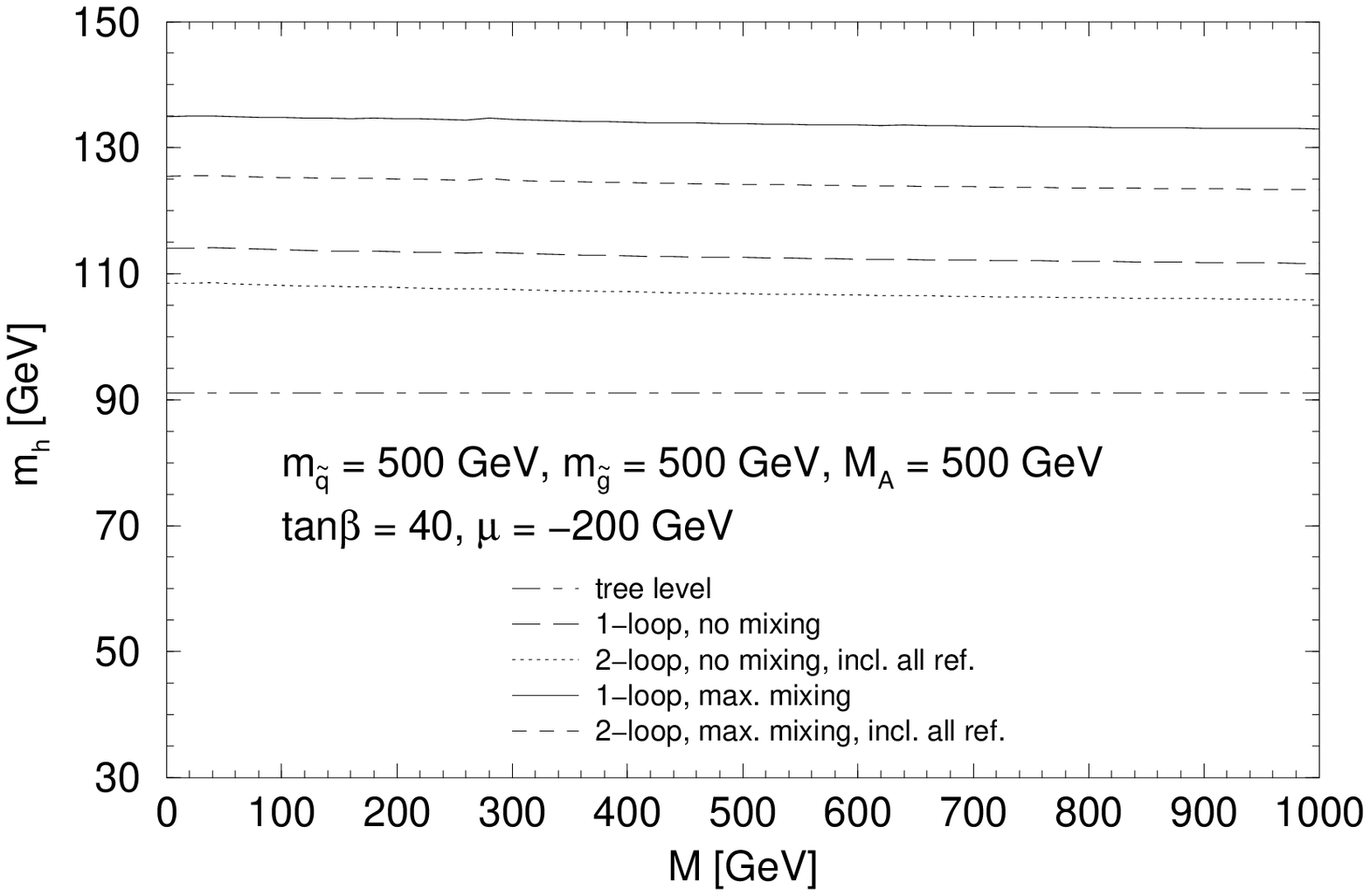,width=5.3cm,height=8cm,
                      bbllx=150pt,bblly=100pt,bburx=450pt,bbury=420pt}}
\end{center}
\caption[]{
The mass of the lightest Higgs boson is shown as a function of $M$,
the soft SUSY breaking parameter of the chargino and neutralino sector.
}
\label{fig:mh_M}
\end{figure}

\bigskip
In \reffi{fig:mh_M} $\mh$ is shown as a function of $M$, the soft SUSY
breaking parameter in the chargino and neutralino sector (see
Sec.~\ref{m1}). 
In our calculation $M$ enters only in the \onel\ self-energies.
The variation is less than $4 \gev$ for the whole
$M$~parameter space. 
For increasing $M$ the result for $\mh$ decreases in general.

\begin{figure}[ht!]
\begin{center}
\mbox{
\psfig{figure=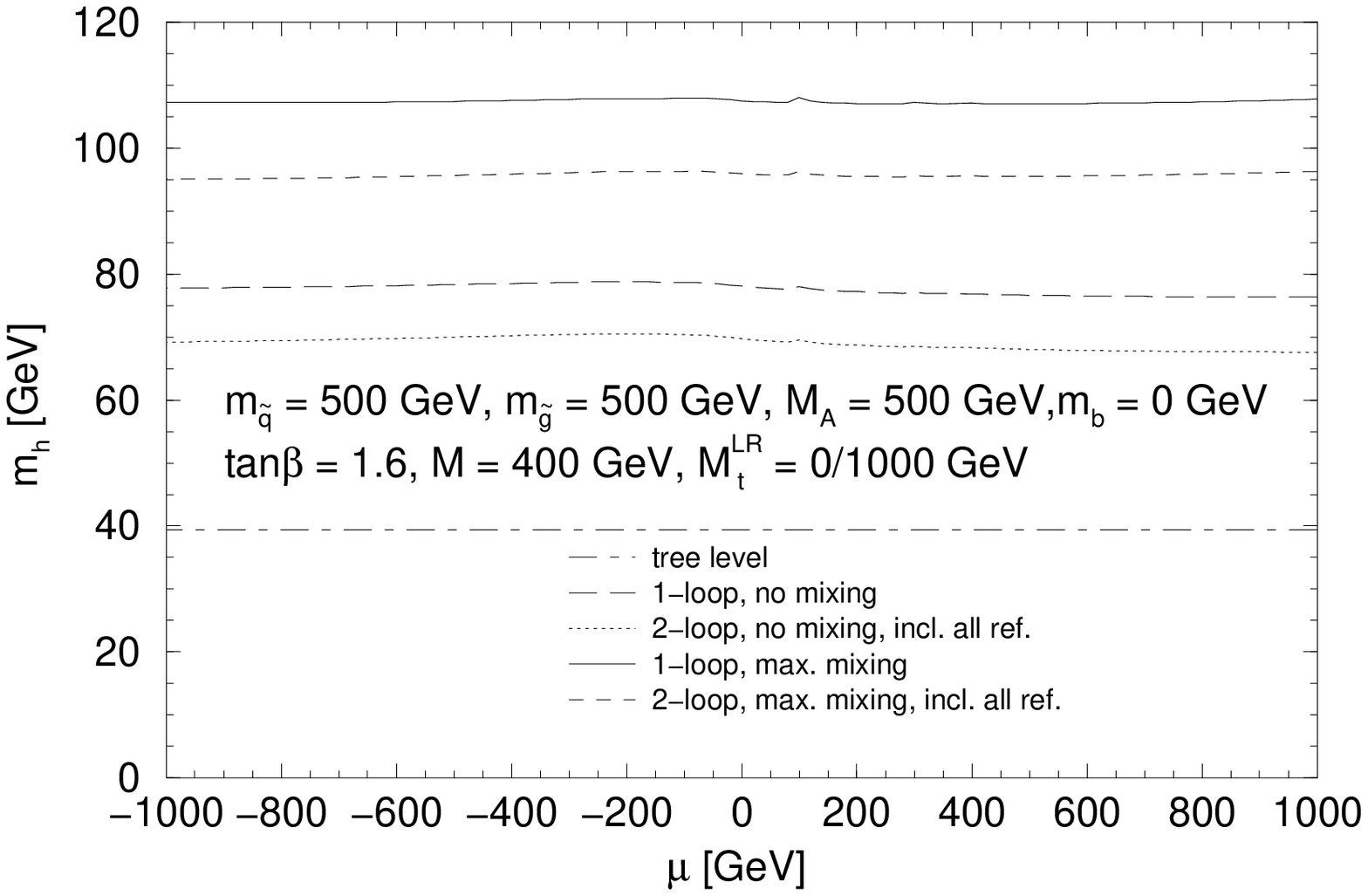,width=5.3cm,height=8cm,
                      bbllx=150pt,bblly=100pt,bburx=450pt,bbury=420pt}}
\hspace{7.5em}
\mbox{
\psfig{figure=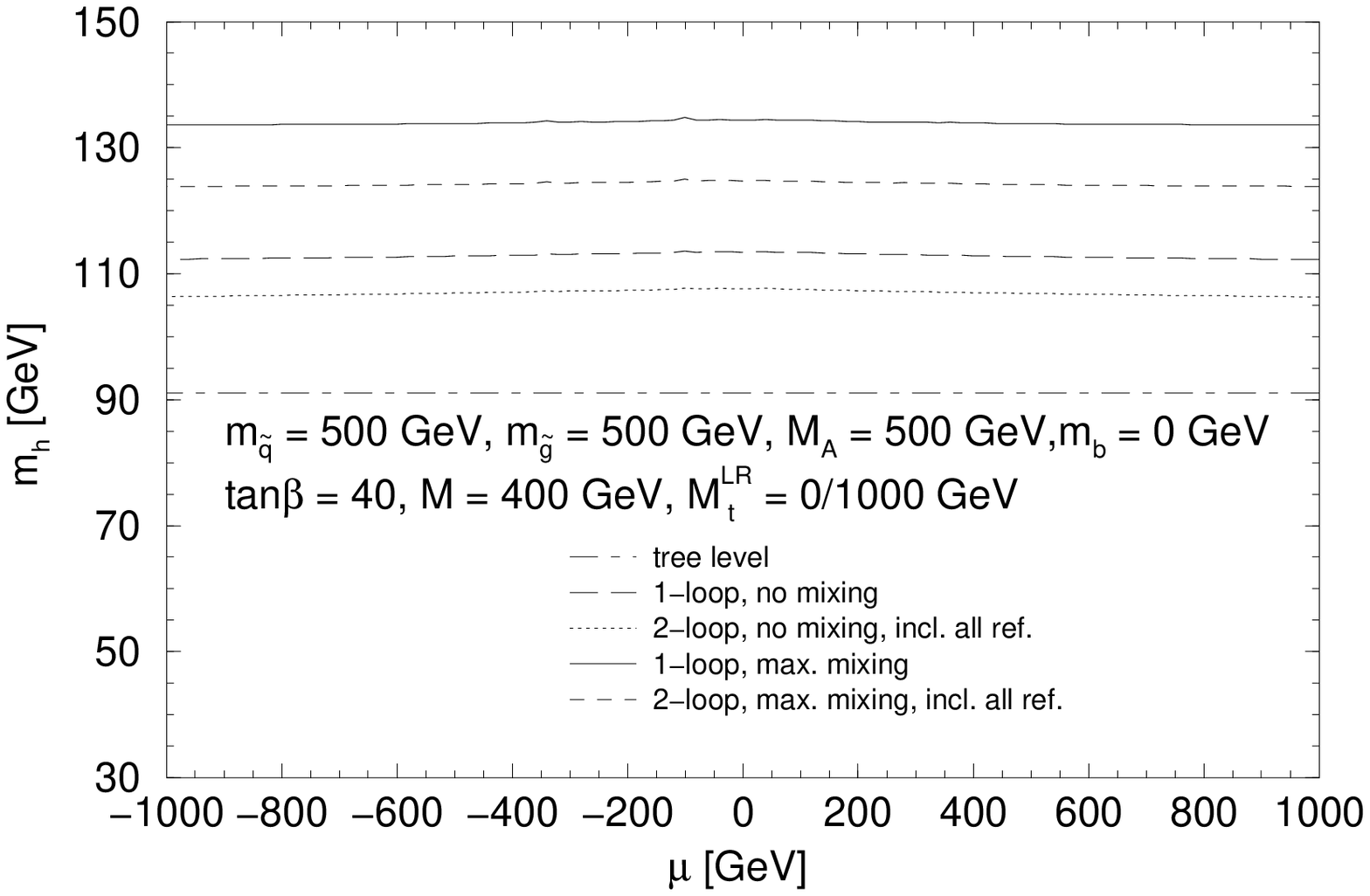,width=5.3cm,height=8cm,
                      bbllx=150pt,bblly=100pt,bburx=450pt,bbury=420pt}}
\end{center}
\caption[]{
The mass of the lightest Higgs boson is shown as a function of $\mu$,
the Higgs mixing parameter. The mass of the bottom quark, $\mb$, is set 
to zero in order to avoid unnaturally large \onel\ effects.
}
\label{fig:mh_mu}
\end{figure}

\bigskip
The dependence of $\mh$ on $\mu$, the Higgs mixing parameter, is
depicted in 
\reffi{fig:mh_mu}. The parameter $\mu$ enters via the non-diagonal
Higgs-squark coupling at 
one- and \twol\ order and via the chargino and neutralino sector in the
\onel\ self-energies.
It should be noted that for the plots in \reffi{fig:mh_mu} we have set
$\mb = 0 \gev$, thus 
suppressing the contribution of the $b-\Sbot$-sector. The reason is
that for large $\mu$ and for large $\Tb$ some Higgs-sbottom couplings
can become rather large, 
which makes the perturbative calculation questionable in this case.
The variation of $\mh$ with $\mu$ in \reffi{fig:mh_mu} is relatively
weak, not exceeding $3 \gev$. A maximum (for the choice 
$M = 400 \gev$) for $\mh$ lies between $\mu = -200 \gev$ and 
$\mu = -100 \gev$.
For decreasing $M$ the maximum is reached for slightly smaller values
of $\mu$, see also Sec.~\ref{subsec:upperboundformh}.

\begin{figure}[ht!]
\begin{center}
\hspace{1em}
\mbox{
\psfig{figure=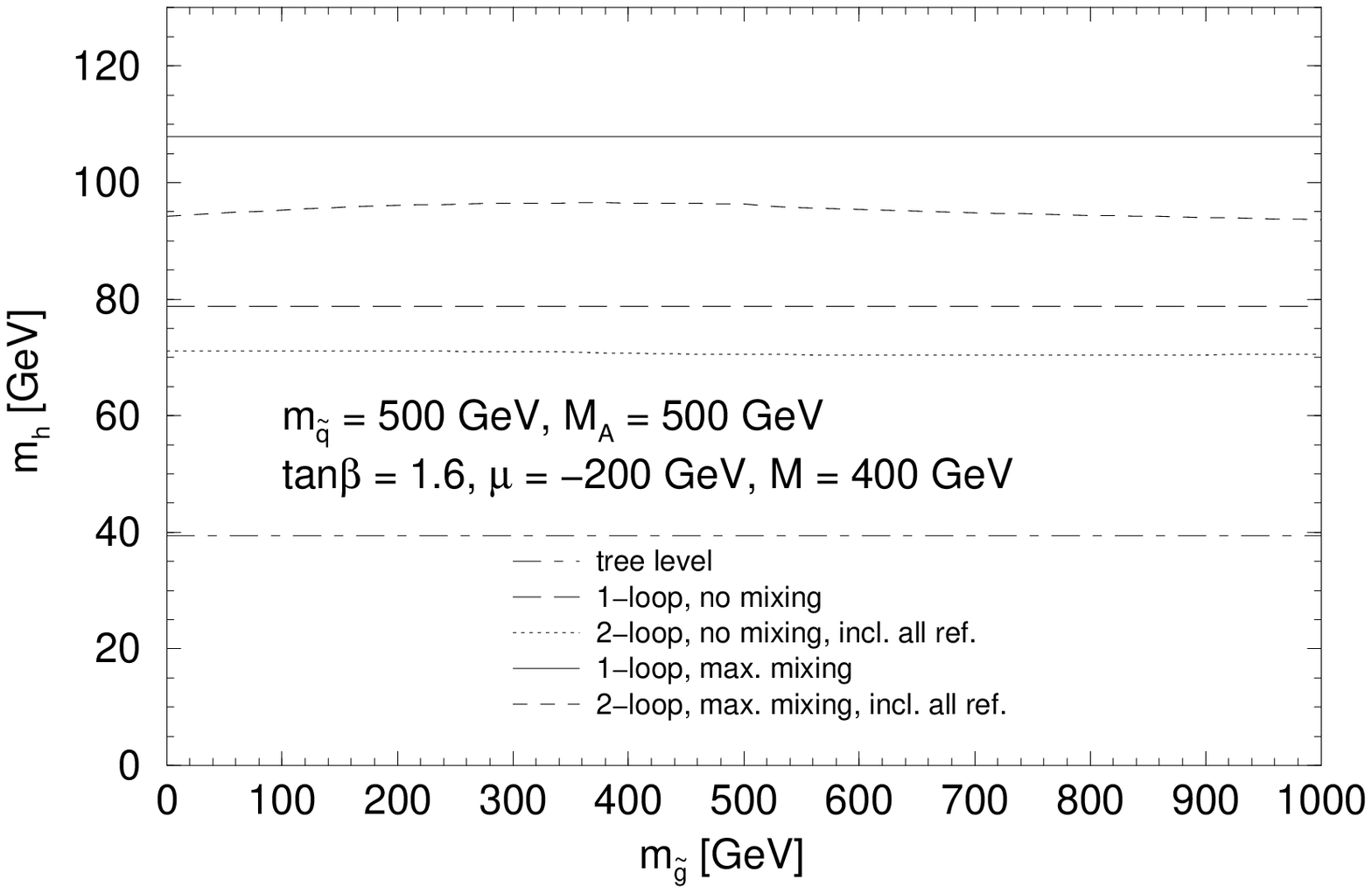,width=5.3cm,height=8cm,
                      bbllx=150pt,bblly=100pt,bburx=450pt,bbury=420pt}}
\hspace{7.5em}
\mbox{
\psfig{figure=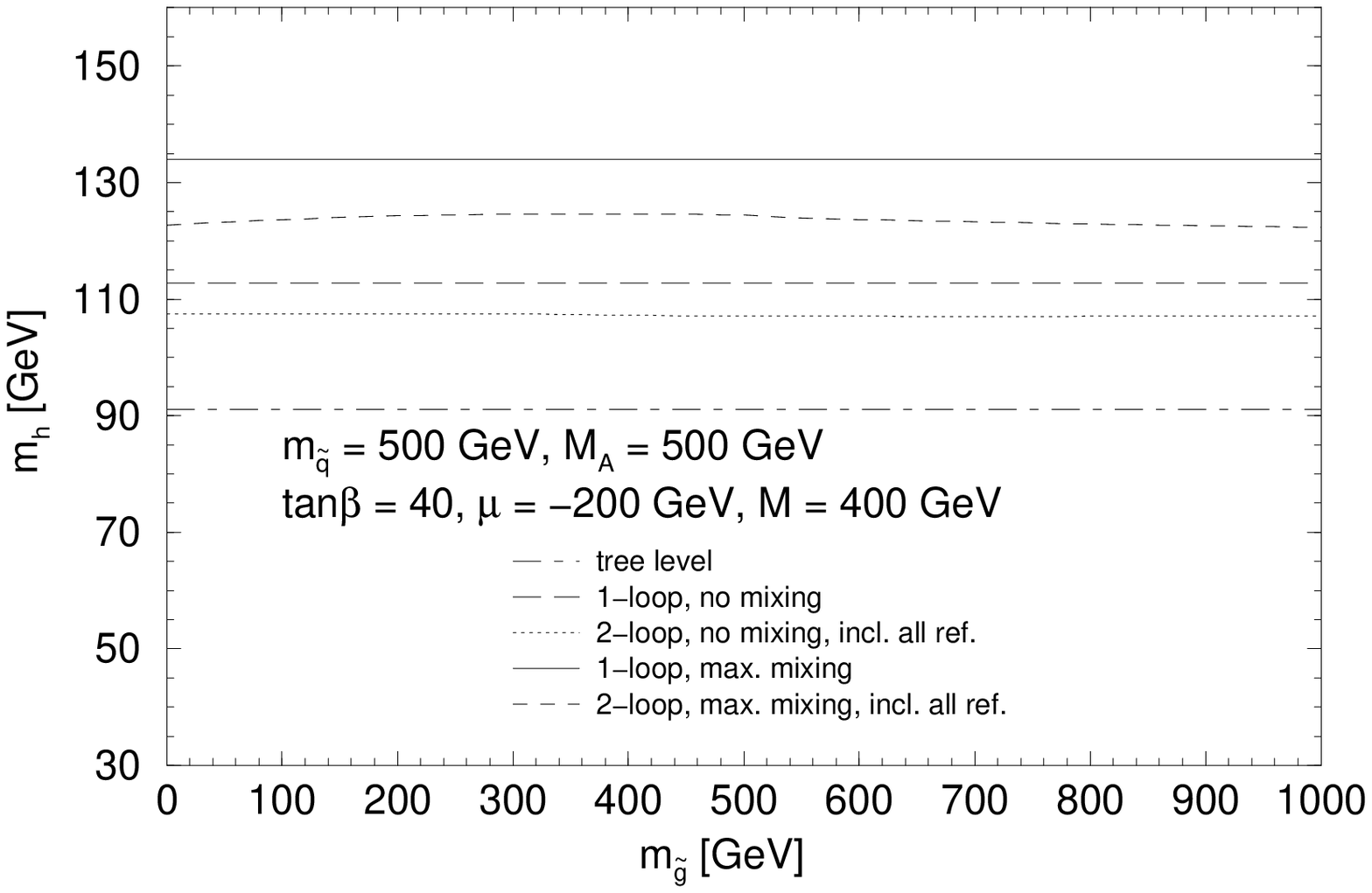,width=5.3cm,height=8cm,
                      bbllx=150pt,bblly=100pt,bburx=450pt,bbury=420pt}}
\end{center}

\vspace{1cm}

\begin{center}
\hspace{1em}
\mbox{
\psfig{figure=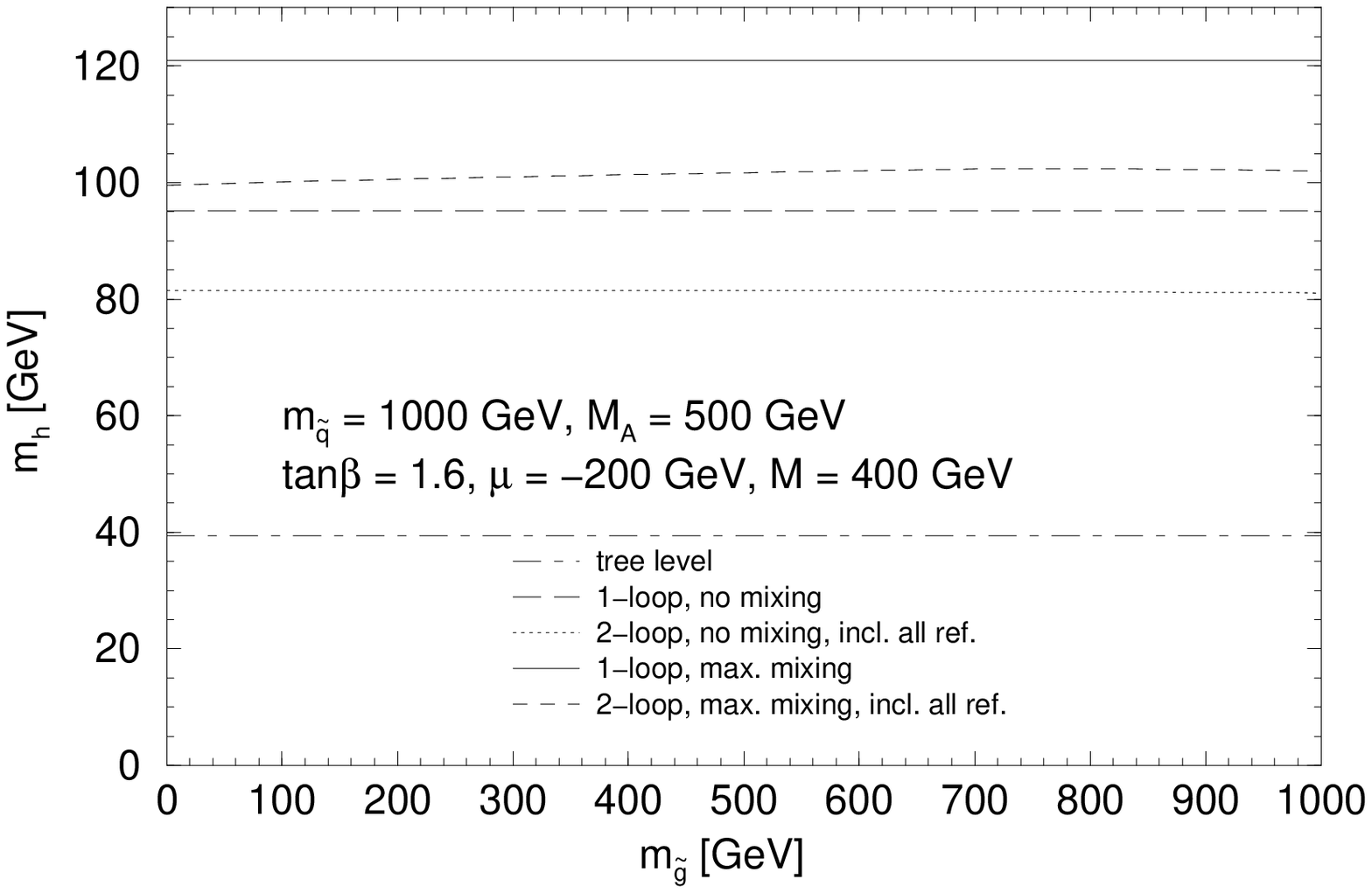,width=5.3cm,height=8cm,
                      bbllx=150pt,bblly=100pt,bburx=450pt,bbury=420pt}}
\hspace{7.5em}
\mbox{
\psfig{figure=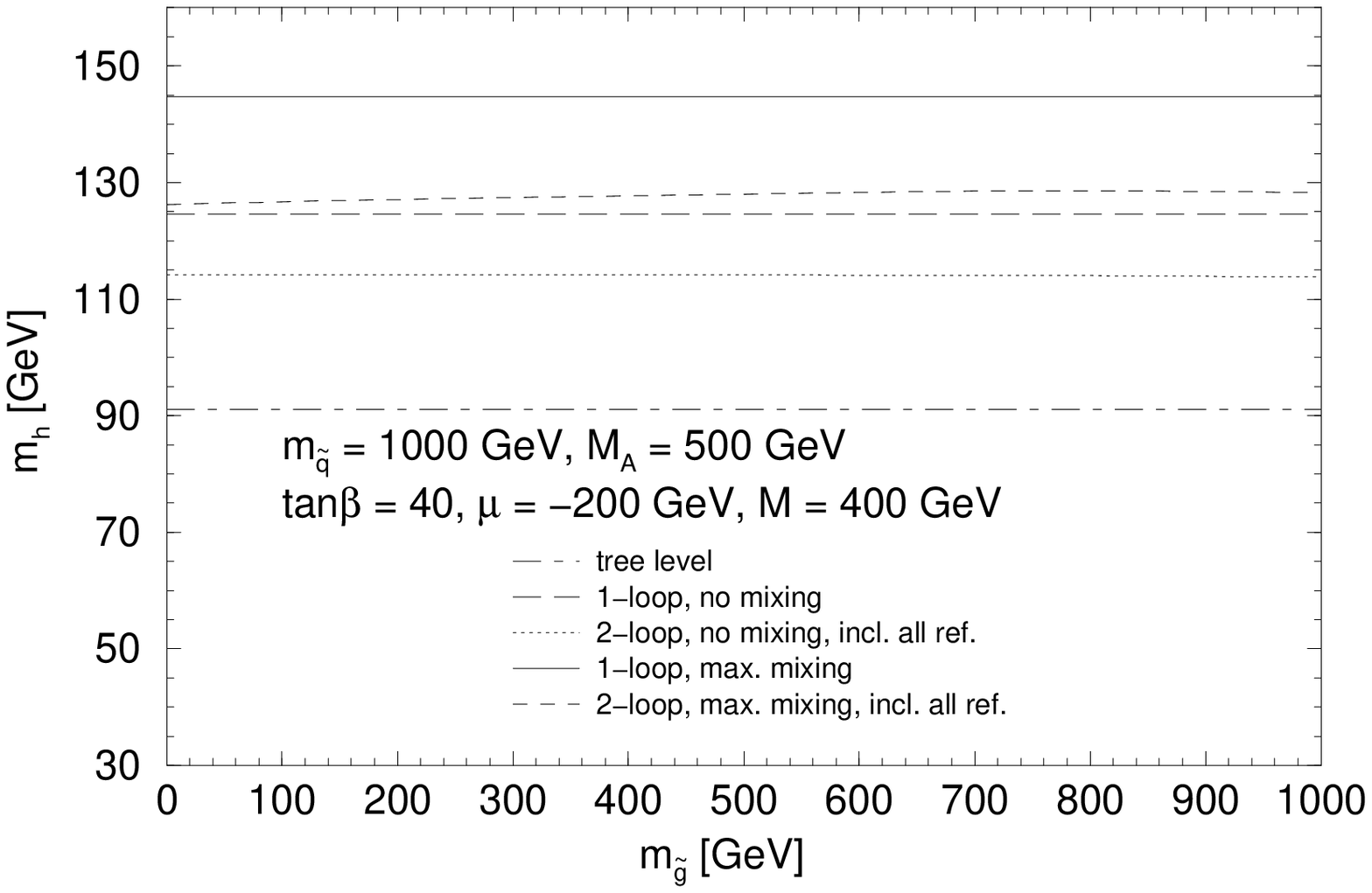,width=5.3cm,height=8cm,
                      bbllx=150pt,bblly=100pt,bburx=450pt,bbury=420pt}}
\end{center}
\caption[]{
The mass of the lightest Higgs boson as a function of $\mgl$ for a
common value of $\MA = 500 \gev, \msq = 500,1000 \gev$ for the
no-mixing and the maximal-mixing case and for low and high $\Tb$.
} 
\label{fig:mh_mgl}
\end{figure}

\bigskip
Finally we show the dependence on the gluino mass, $\mgl$, which
enters only at the \twol\ level. 
\reffi{fig:mh_mgl} depicts $\mh$ as a function of $\mgl$ in the
scenarios with $\msq = 500,1000 \gev$ for $\Tb = 1.6$ and $\Tb = 40$
in the no-mixing and the maximal-mixing case. 
Small variations below $1 \gev$ occur in the no-mixing scenario, while
change in $\mh$ up to $4 \gev$ arises in the maximal-mixing
scenario. $\mh$ reaches a maximum at about $\mgl \approx 0.8\,\msq$. 
Since the parameter $\mgl$ is absent in the RG approach, a variation
of $\mh$ with $\mgl$ can directly be seen as a deviation of
the diagrammatic result from the RG result, see Sec.~\ref{subsec:rgcomp}.

\begin{figure}[ht!]
\begin{center}
\hspace{1em}
\mbox{
\psfig{figure=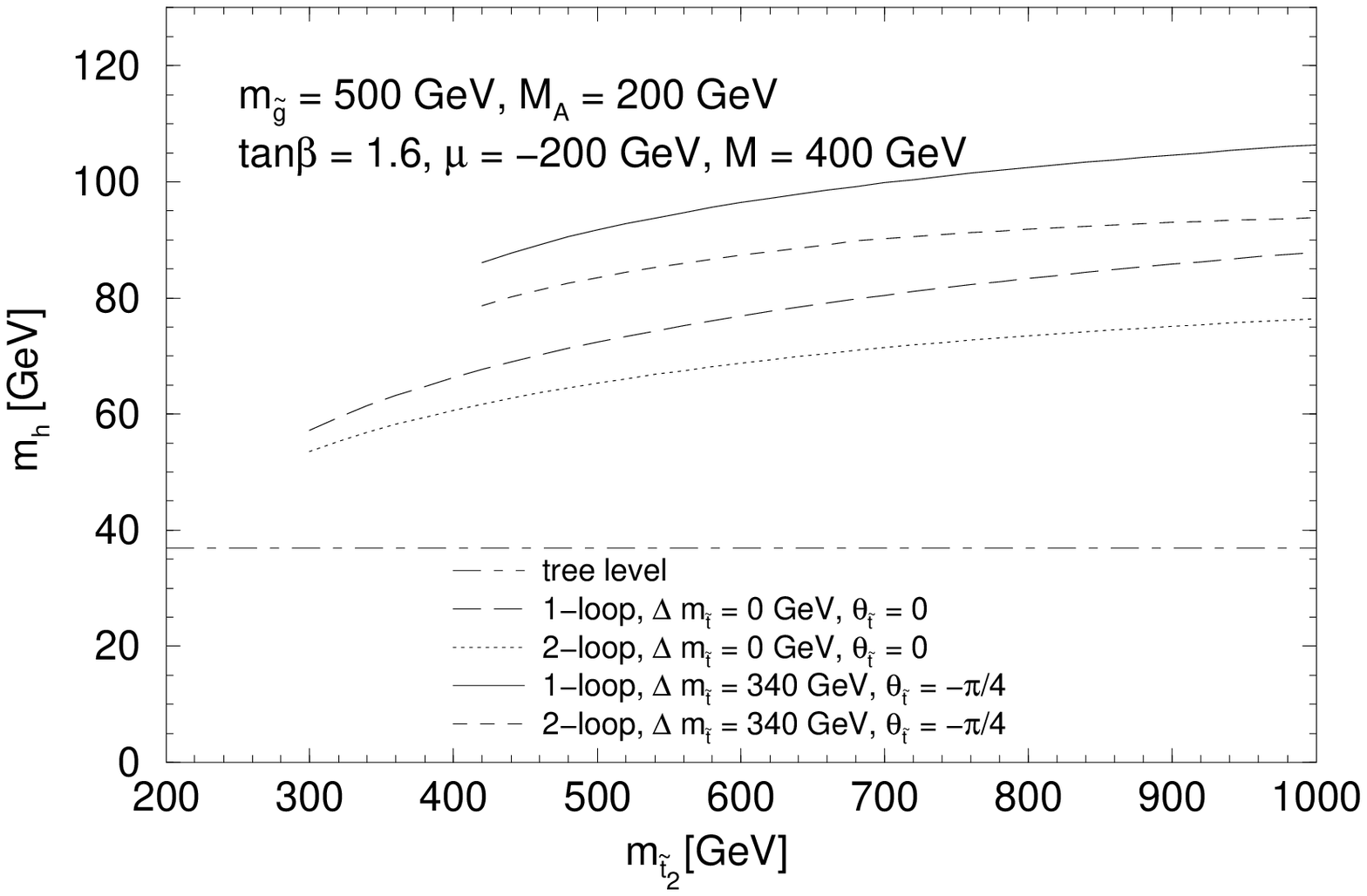,width=5.3cm,height=8cm,
                      bbllx=150pt,bblly=100pt,bburx=450pt,bbury=420pt}}
\hspace{7.5em}
\mbox{
\psfig{figure=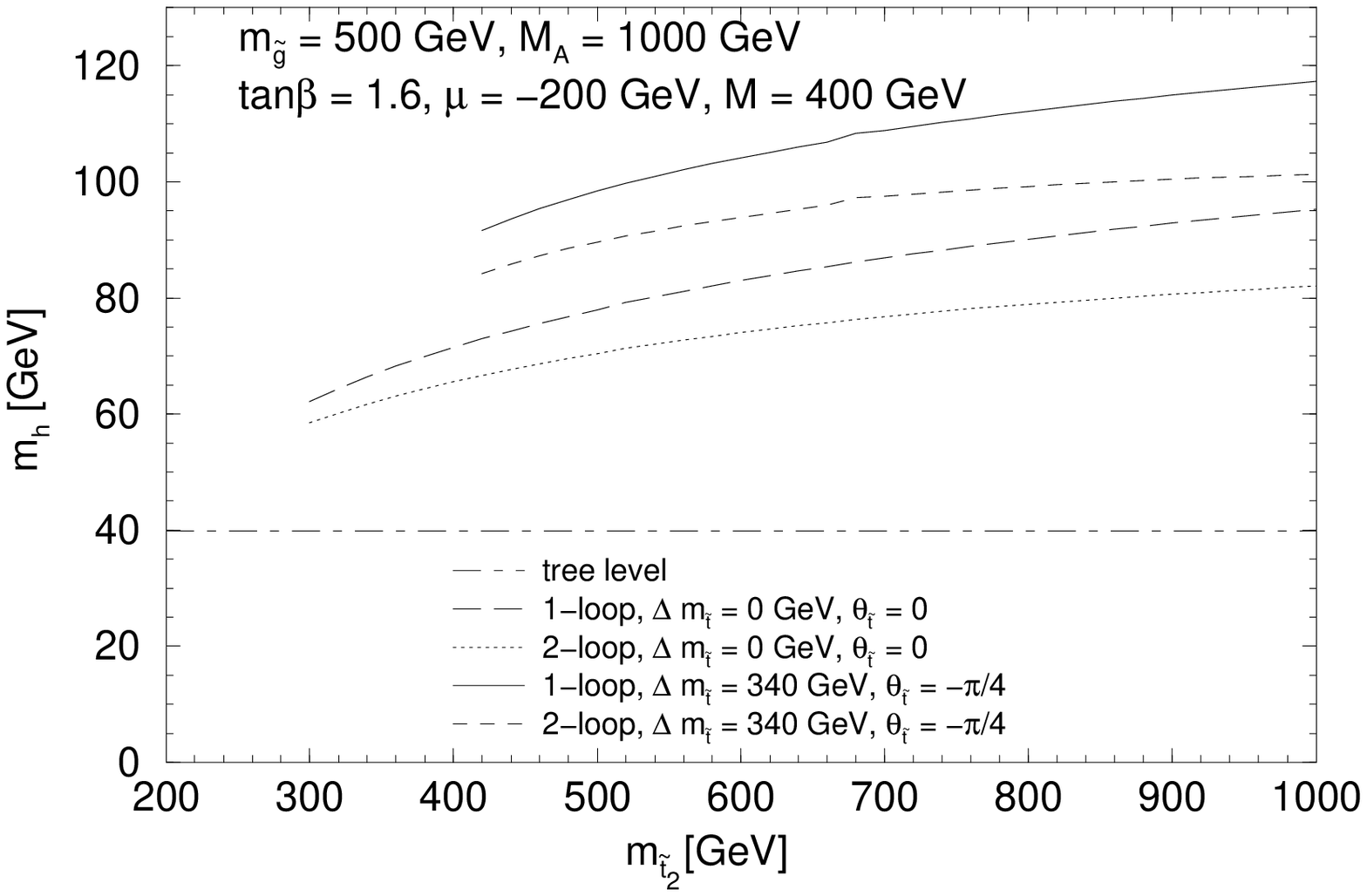,width=5.3cm,height=8cm,
                      bbllx=150pt,bblly=100pt,bburx=450pt,bbury=420pt}}
\end{center}

\vspace{1cm}

\begin{center}
\hspace{1em}
\mbox{
\psfig{figure=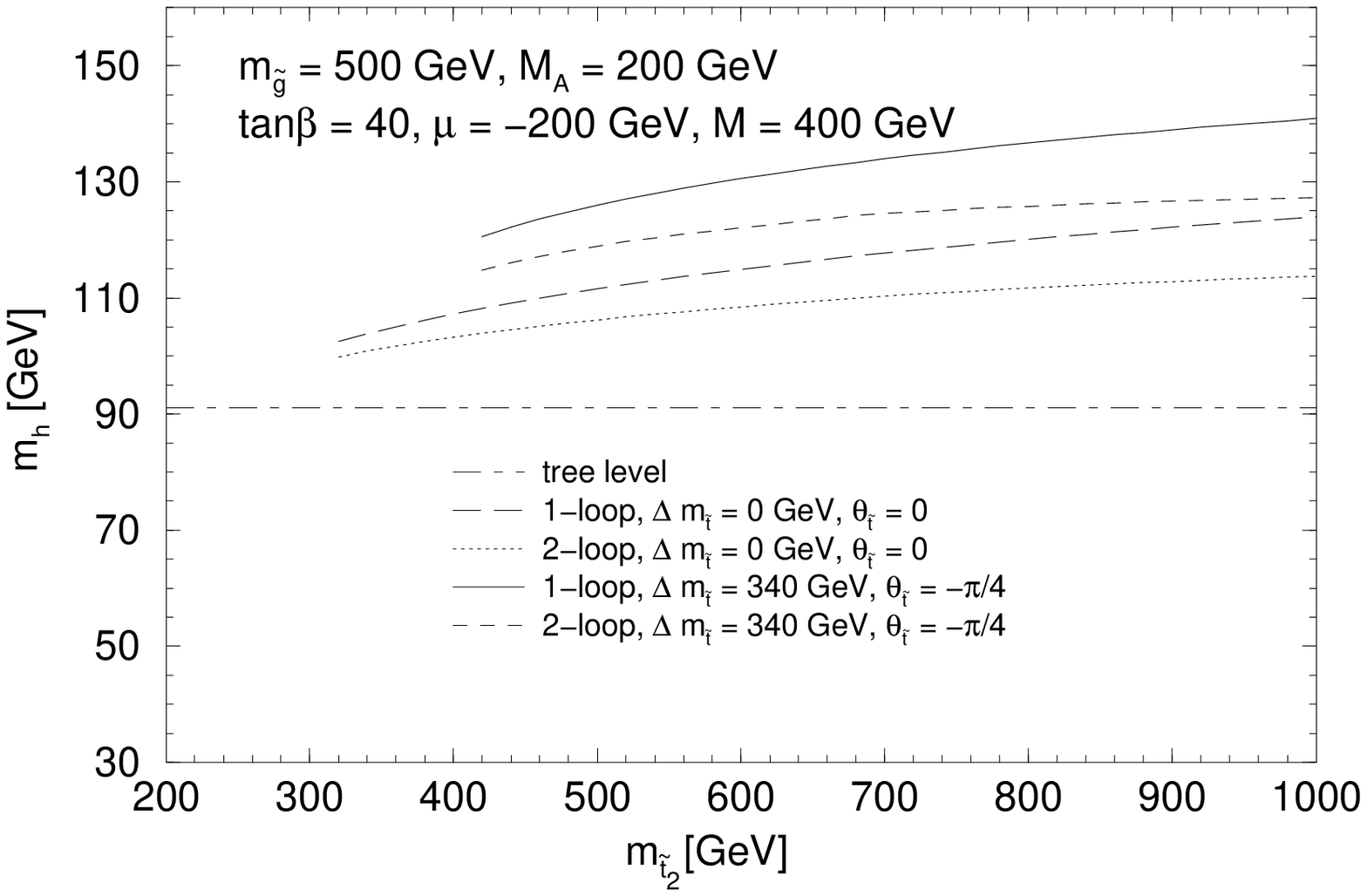,width=5.3cm,height=8cm,
                      bbllx=150pt,bblly=100pt,bburx=450pt,bbury=420pt}}
\hspace{7.5em}
\mbox{
\psfig{figure=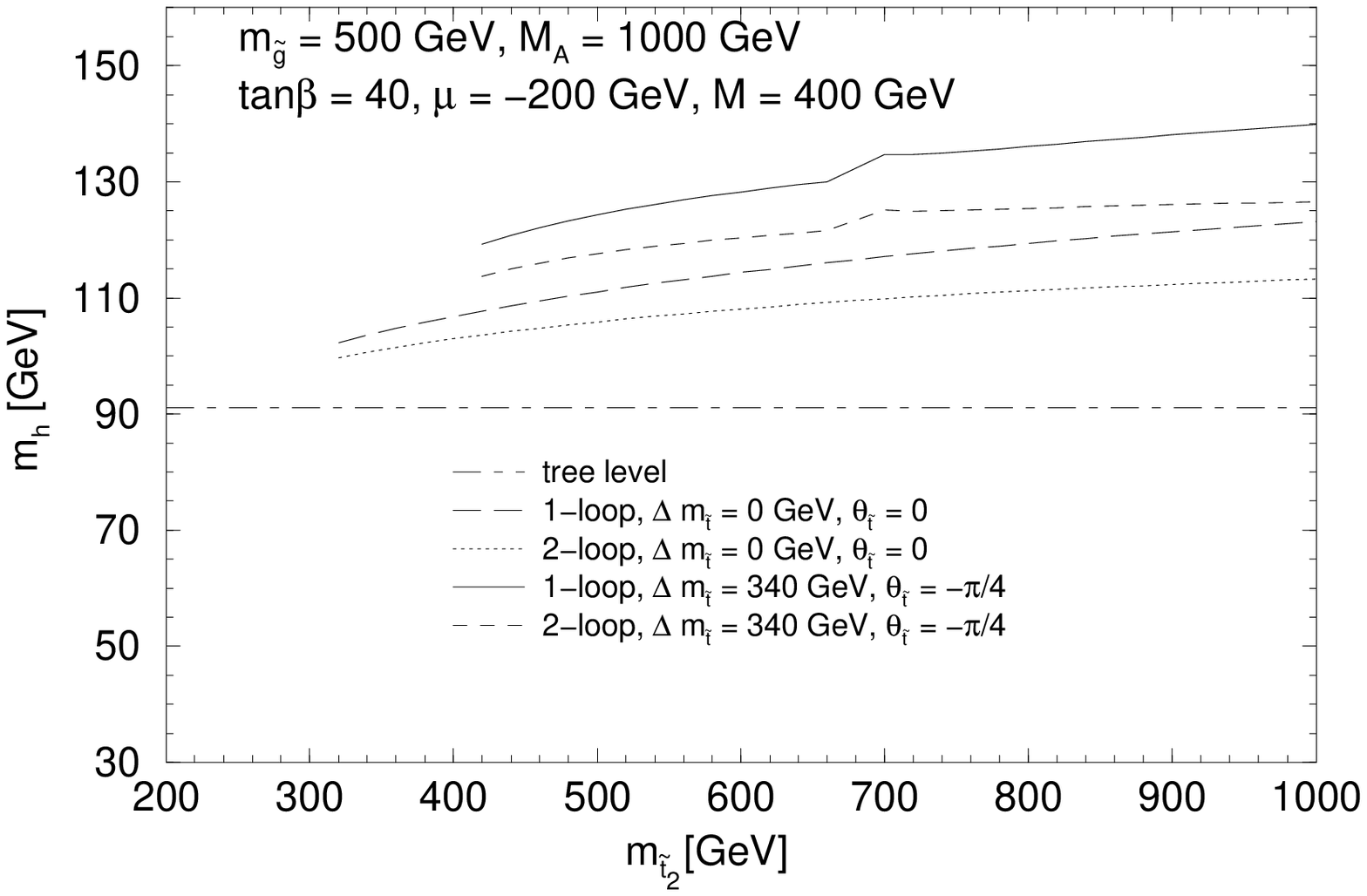,width=5.3cm,height=8cm,
                      bbllx=150pt,bblly=100pt,bburx=450pt,bbury=420pt}}
\end{center}
\caption[]{
The mass of the lightest Higgs boson in terms of the physical
parameters $\mste, \mstz$ and $\tst$, where 
$\delmst \equiv \mstz - \mste$. The scenarios 
$\delmst = 0 \gev, \tst = 0$ (no mixing) and
$\delmst = 340 \gev, \tst = -\pi/4$ (maximal mixing) are shown.
} 
\label{fig:mh_mst2}
\end{figure}

\bigskip
As pointed out in \citere{mhiggsletter2} it is desirable 
to express the predictions for the observable $\mh$ in terms
of other  
physical observables. This provides the possibility to directly
compare results 
obtained by different approaches making use of different
renormalization schemes. Therefore we show in \reffi{fig:mh_mst2}
the dependence of $\mh$ on the parameters $\mste, \mstz$ and
$\tst$, which, since we are working in the on-shell scheme, 
directly correspond to the physical ones.
We show $\mh$ as a function of $\mstz$ for 
$\De\mst = 0 \gev$ and $\tst = 0$ (no mixing) and for 
$\De\mst = 340 \gev$ and $\tst = -\pi/4$ (maximal mixing), where 
$\De\mst \equiv \mstz - \mste$. The choice of 
$\De\mst \approx 340 \gev$ corresponds to $\Mtlr/\msq \approx 2$ in
terms of the soft SUSY breaking parameters. The \twol\ results shown here
contain also the corrections beyond $\oaas$.
In these plots we have furthermore imposed the \rp\ constraint:
We have required that the contribution of the third generation
of scalar quarks to the $\rho$-parameter~\cite{drhosuqcdb,drhosuqcda} 
does not exceed the value of $1.3 \cdot 10^{-3}$, which corresponds
approximately to the resolution 
of $\De\rho$ when it is determined from experimental
data~\cite{delrhoexp}. 
For $\Tb = 1.6$ $\mh$ reaches about $76~(82) \gev$ for 
$\MA = 200~(1000) \gev, \mstz = 1 \tev$ and no mixing in the
$\Stop$-sector. In the maximal-mixing case the reached values of $\mh$
are $94~(101) \gev$. In the $\Tb = 40$ scenario, $\mh$ reaches 
$114~(127) \gev$ in the no-mixing (maximal-mixing) case 
for both values of $\MA$.
The peaks in the plots for $\MA = 1 \tev$ and maximal mixing in the
$\Stop$-sector around $\mstz = 660 \gev$ are due
to the threshold $\MA = \mste + \mstz$ in the
\onel\ contribution, originating from the stop-loop diagram in
the $A$~self-energy.

\subsection{Upper bound for $\mh$ as a function of $\Tb$}
\label{subsec:upperboundformh}

Since, as shown in \reffi{fig:mh_tb}, smaller values for $\mh$ are
obtained for small $\Tb$, this part of the parameter space can to a
large extend be covered at todays' colliders. 
The discovery limit for $\mh$ at LEP2 is expected to be slightly above 
$100 \gev$~\cite{lep2discpot}.
In this context it is of special interest
to know the maximally possible value for $\mh$ as a
function of $\Tb$ in the MSSM. To this end we have performed a
parameter scan, varying  
$\mgl, M, \mu, \MA$ and $\Mtlr$ for three values of $\mt$ and fixed
values of $\msq$ and $\Tb$. The maximal values for $\mh$, 
including also the Yukawa correction and the contribution from the
running top mass, were reached (in the case $\mt = 175 \gev$) for%
\footnote{
Due to threshold effects very high values for $\mh$ can occur. Since
this is regarded as an accidental effect, these isolated points of
parameter space are not considered here.
}
\BEA
\mgl  &\approx& 0.8\, \msq \non \\
M     &\approx& 0 \gev \non \\
\mu   &\approx& 0 \gev \non \\
\MA   &\approx& \KKKL \begin{array}{l}
                800 \gev\; \mbox{for small} \Tb \non \\
                360 \gev\; \mbox{for large} \Tb
                \end{array} \right. \non \\
\Mtlr &\approx& 2\, \msq 
\label{maxmhparameters}
\EEA
in all scenarios.

The value for $\MA$ in (\ref{maxmhparameters}) needs some further
explanation: as one can see in \reffi{fig:mh_MA}, due to a \onel\
threshold effect the value 
of $\mh$ can become very large for $\MA = 2\,\mt$. For large $\Tb$
this threshold effect results in a bigger $\mh$ value for $\MA$ in the
region around $2\,\mt$ than for larger values of $\MA$ (with
$\MA < 1500 \gev$, where we stopped our scan.) Of course the exact value 
$\MA = 2\,\mt = 2\cdot 175 \gev = 350 \gev$ would be a very specific
choice, giving a wrong impression of the possible size of $\mh$. 
(Exactly at the threshold also finite width effects for the $A$ boson
would have to be taken into account.) 
Therefore we have chosen the value $\MA = 360 \gev$ which is not
directly at the threshold, thus giving a more realistic impression
about the maximally possible values for $\mh$.
The choice $M = \mu = 0 \gev$ is experimentally excluded. We
nevertheless use
these values since the difference in $\mh$ to the case with
experimentally not 
excluded $M$ and $\mu$ is very small, typically below $0.5 \gev$.

\begin{figure}[ht!]
\begin{center}
\hspace{1em}
\mbox{
\psfig{figure=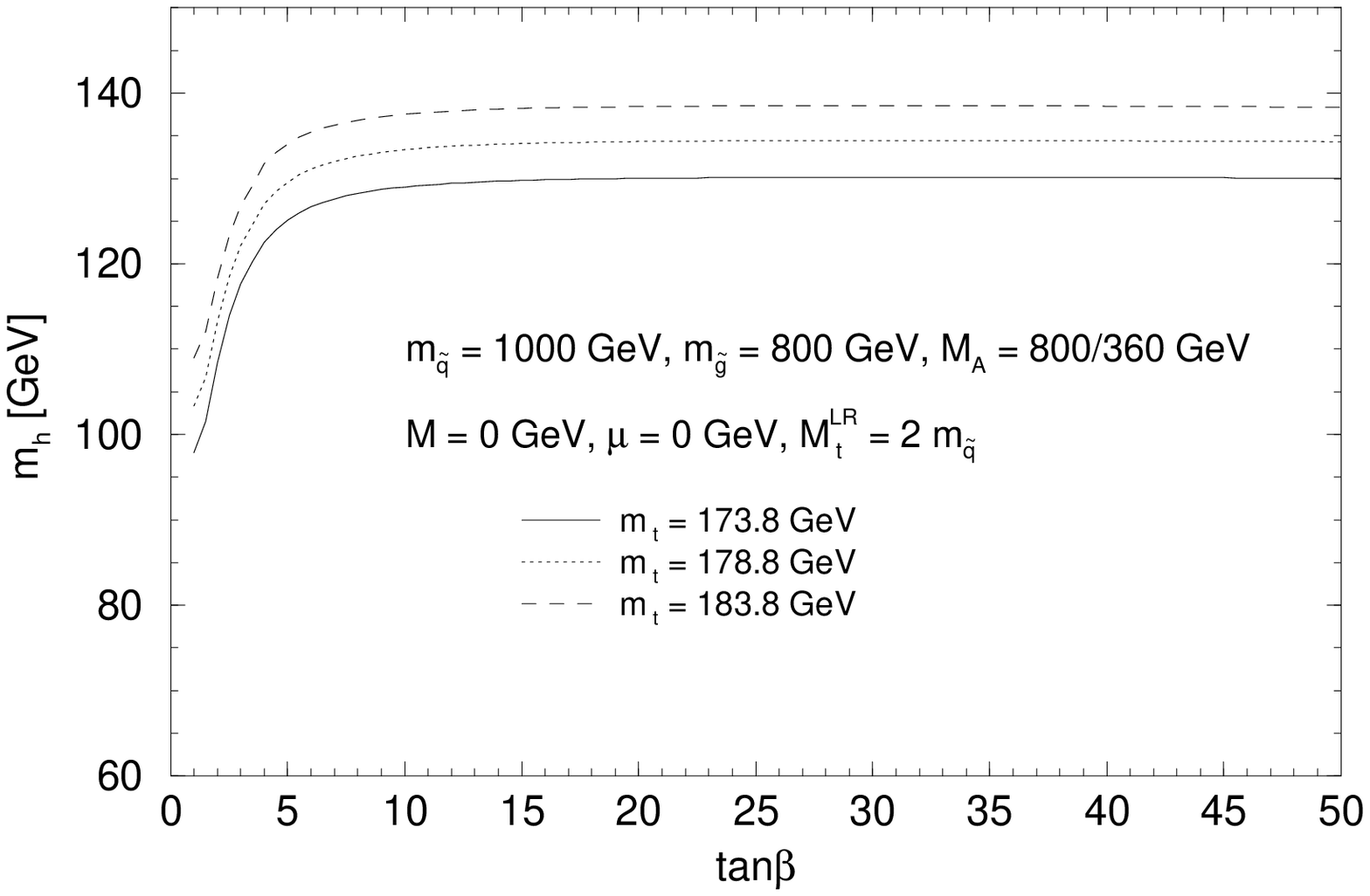,width=5.3cm,height=8cm,
                      bbllx=150pt,bblly=100pt,bburx=450pt,bbury=420pt}}
\hspace{7.5em}
\mbox{
\psfig{figure=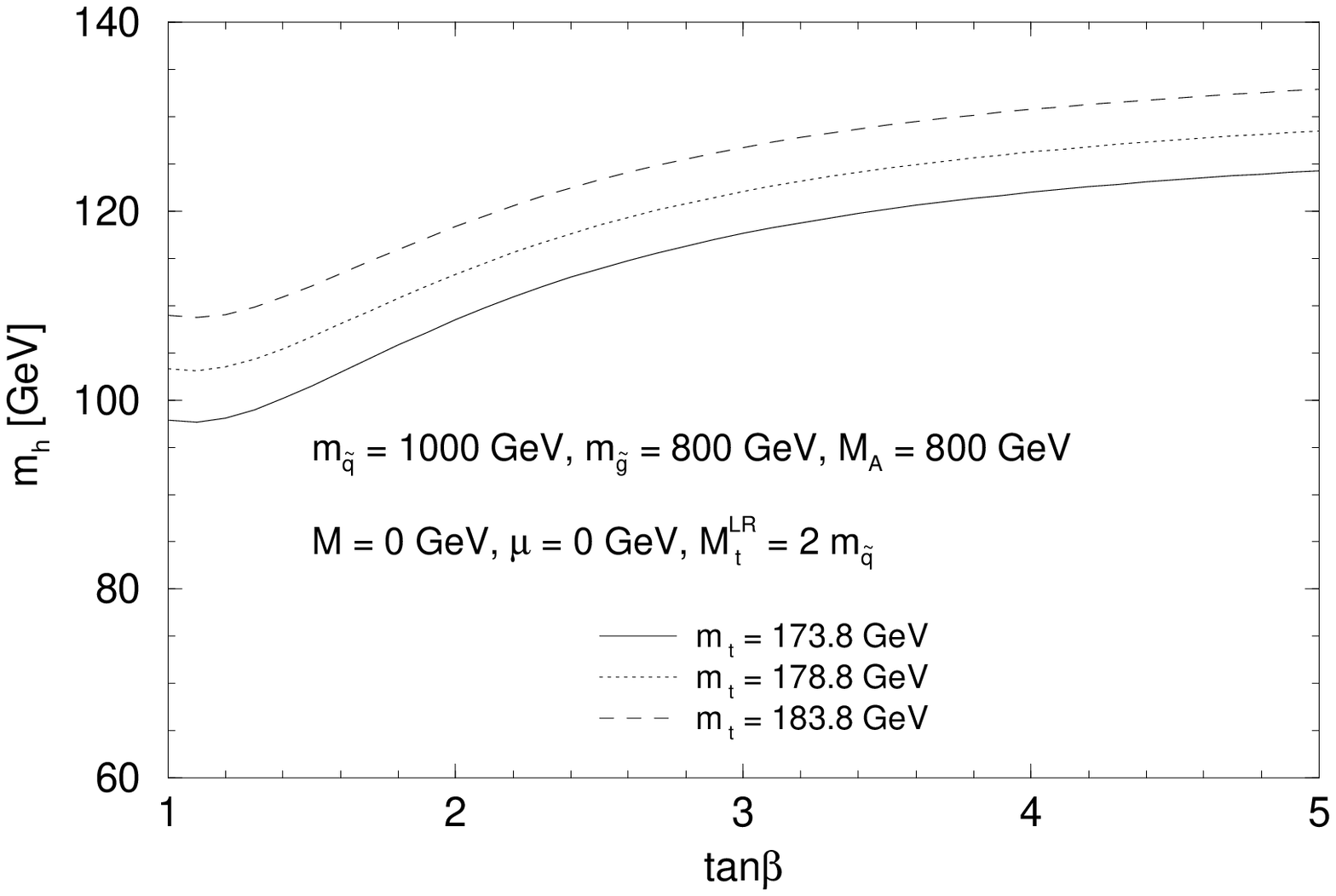,width=5.3cm,height=8cm,
                      bbllx=150pt,bblly=100pt,bburx=450pt,bbury=420pt}}
\end{center}

\vspace{1cm}

\begin{center}
\hspace{1em}
\mbox{
\psfig{figure=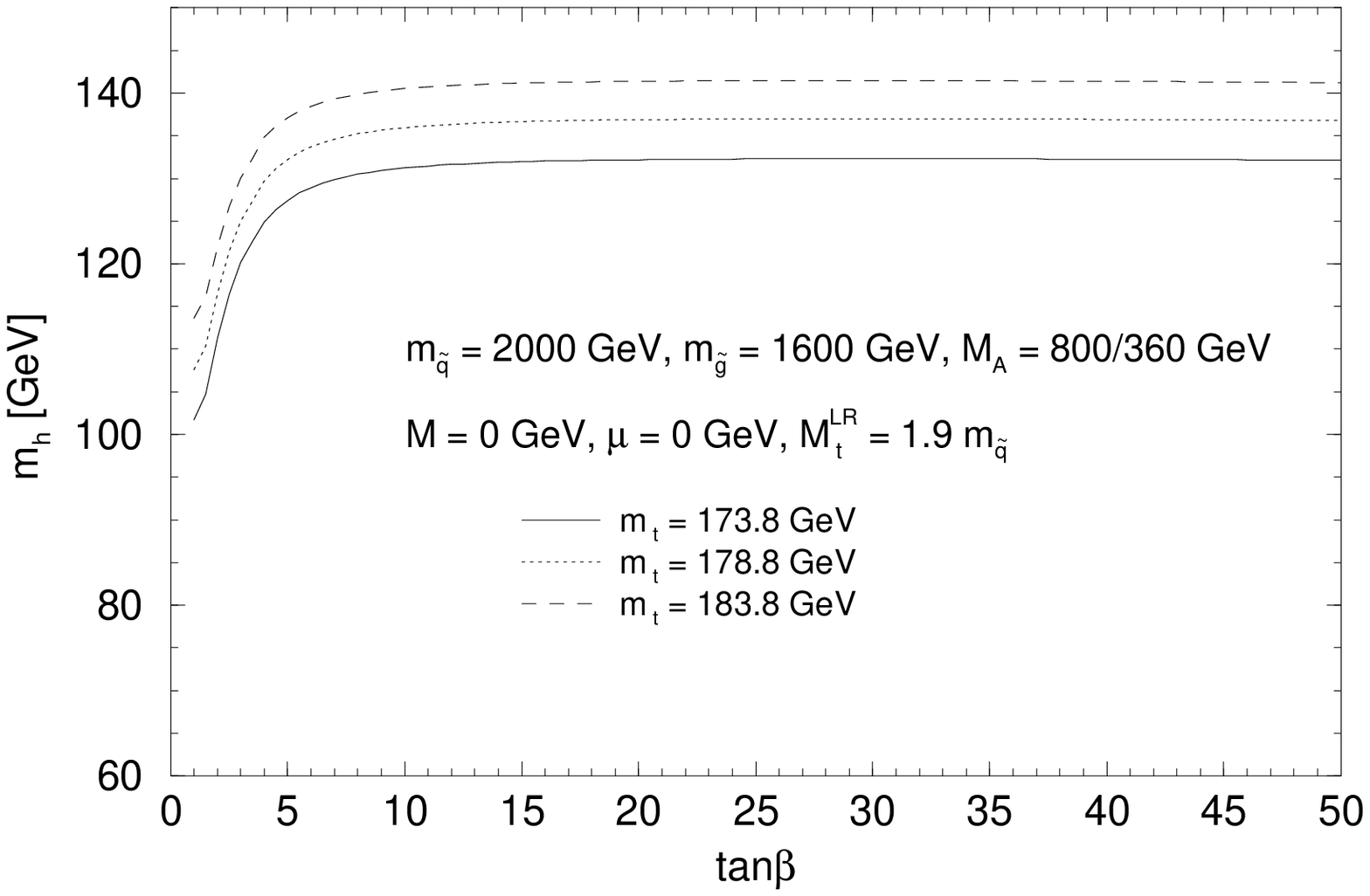,width=5.3cm,height=8cm,
                      bbllx=150pt,bblly=100pt,bburx=450pt,bbury=420pt}}
\hspace{7.5em}
\mbox{
\psfig{figure=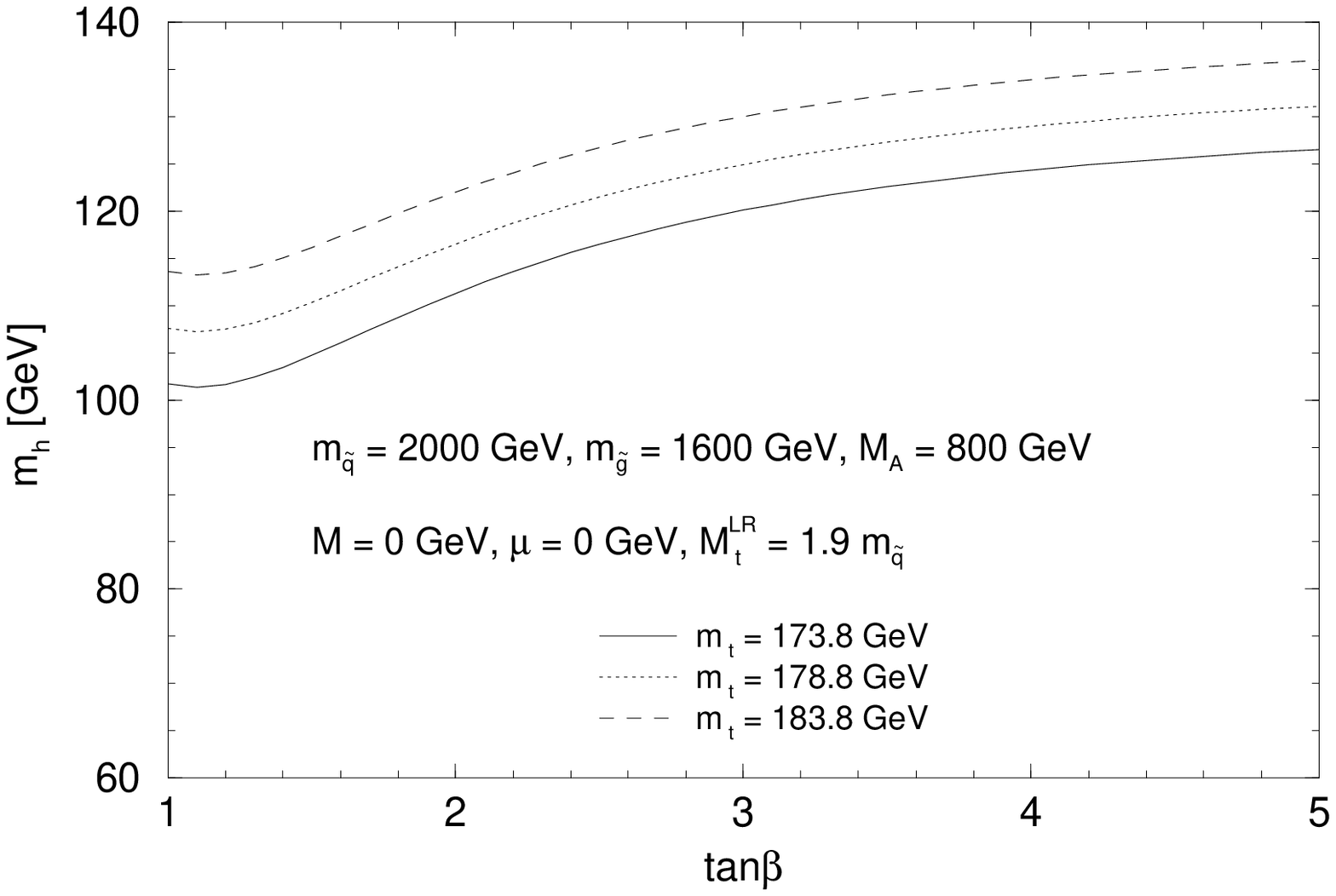,width=5.3cm,height=8cm,
                      bbllx=150pt,bblly=100pt,bburx=450pt,bbury=420pt}}
\end{center}
\caption[]{
The maximally possible value for $\mh$, including all refinement terms,
as a function of $\Tb$, depending 
on the unphysical parameters $\msq$ and $\Mtlr$. The other MSSM parameters
have been chosen according to~\refeqs{maxmhparameters}.
} 
\label{fig:mhmax_tb_UP}
\end{figure}

In \reffi{fig:mhmax_tb_UP} we show the maximal Higgs-boson mass value,
including also the corrections beyond $\oaas$, as a
function of $\Tb$; the other parameters are chosen according
to \refeqs{maxmhparameters}. For the top-quark mass the most recent
experimental value $\mt = 173.8 \gev$~\cite{mtexp} is chosen and, since $\mh$
grows with 
increasing $\mt$, the experimental value plus one and plus two
standard deviations ($\mt = 178.8, 183.8 \gev$)%
\footnote{
One should note, however, that the highest value for $\mt$ is
disfavored in the MSSM by internal consistency~\cite{nohighmt}.
}%
. The common squark mass
parameter is chosen to be $\msq = 1000 \gev$ as a high, and 
$\msq = 2000 \gev$ as a very high value. On the left side of
\reffi{fig:mhmax_tb_UP} we show the full $\Tb$ range ($\Tb \le 50$),
whereas on the right side we focus on the range especially
interesting for LEP2 and the upgraded Tevatron ($\Tb \le 5$).

In the $\Tb \le 5$ plot we have chosen $\MA = 800 \gev$. In the 
$\Tb \le 50$ plot, however, we have chosen $\MA = 800 \gev$ only for 
$\Tb \le 4$; for larger values we have switched to $\MA = 360 \gev$. For
the value $\Tb = 4$ one gets about the same maximal value
for $\mh$ for both choices of $\MA$.

\begin{table}[ht!]
\renewcommand{\arraystretch}{1.5}
\begin{center}
\begin{tabular}{|c||c|c|c|c|c|c|c|c|c|c|} 
\cline{2-11} \multicolumn{1}{c|}{} 
 & \multicolumn{2}{c|}{$\Tb = 1.6$} & \multicolumn{2}{c|}{$\Tb = 1.7$} &
   \multicolumn{2}{c|}{$\Tb = 1.8$} & \multicolumn{2}{c|}{$\Tb = 1.9$} & 
   \multicolumn{2}{c|}{$\Tb = 2.0$}  \\
\cline{2-11} \multicolumn{1}{c|}{}
 & \multicolumn{2}{c|}{$\msq$ =} & \multicolumn{2}{c|}{$\msq$ =} & 
   \multicolumn{2}{c|}{$\msq$ =} & \multicolumn{2}{c|}{$\msq$ =} & 
   \multicolumn{2}{c|}{$\msq$ =}\\ \cline{1-1}
$\mt$ & 1000 & 2000 & 1000 & 2000 & 1000 & 2000 & 1000 & 2000 
      & 1000 & 2000  \\ \hline \hline
173.8 & 103.0 & 106.1 & 104.4 & 107.4 & 105.8 & 108.8 & 107.2 & 110.1
      & 108.5 & 111.3 \\ \hline
178.8 & 108.1 & 111.6 & 109.4 & 112.9 & 110.8 & 114.2 & 112.1 & 115.4
      & 113.3 & 116.6 \\ \hline
183.8 & 113.4 & 117.4 & 114.7 & 118.6 & 116.0 & 119.8 & 117.2 & 120.9
      & 118.4 & 122.0 \\ \hline
\end{tabular}
\caption[]{Maximal values for $\mh$ for different choices of 
$\mt, \msq$ and $\Tb$. All other parameters have been chosen 
according to~\refeqs{maxmhparameters}. (All masses are in GeV.)}
\label{tab:mhmax_UP}
\end{center}
\end{table}

In the interesting region around $\Tb = 1.6$ the covered region of
the $\Tb$-parameter space depends strongly on the maximally
accessible energy of todays' colliders, see Tab.~\ref{tab:mhmax_UP}.
For an exclusion limit of $\mh > 107 \gev$, for instance, LEP2 covers 
$\Tb < 1.6$ completely only if $\mt$ is constrained to its present $1\,\si$
limit. On the other hand, taking a very conservative point of view and
choosing $\mt$ at the $2\,\si$ bound,
no limit on $\Tb$ can be set, even for $\msq = 1000 \gev$. 

One should keep in mind,
however, that the Higgs-boson masses depicted in \reffi{fig:mhmax_tb_UP}
are the maximally possible upper values,
i.e.\ for smaller mixing in the $\Stop$-sector the region 
$\Tb < 1.6$ can be covered by LEP2 for all other sets of parameters.
One can also see that a precise measurement of $\mt$ is
decisive in order to set stringent bounds on $\Tb$ in the MSSM.

In conclusion, our results confirm that for the scenario with 
$\Tb = 1.6$ the parameter space of the MSSM can be covered
to a very large extent. Only for maximal mixing, very large soft
SUSY breaking parameters in the $\Stop$-sector and
$\mt$ at its upper $(1-2)\si$ limit the light Higgs boson can escape
the detection at LEP2.
For increasing $\Tb$, however, the parameter space in which the Higgs boson
is not accessible at LEP2 increases rapidly.

Concerning the large $\Tb$ region, LEP2 and the upgraded Tevatron
can probe only the region of no mixing in the $\Stop$-sector. The 
LHC and a future linear $e^+e^-$-collider are needed in order to test
the parameter space with large $\Stop$-mixing.

\begin{figure}[ht!]
\begin{center}
\hspace{1em}
\mbox{
\psfig{figure=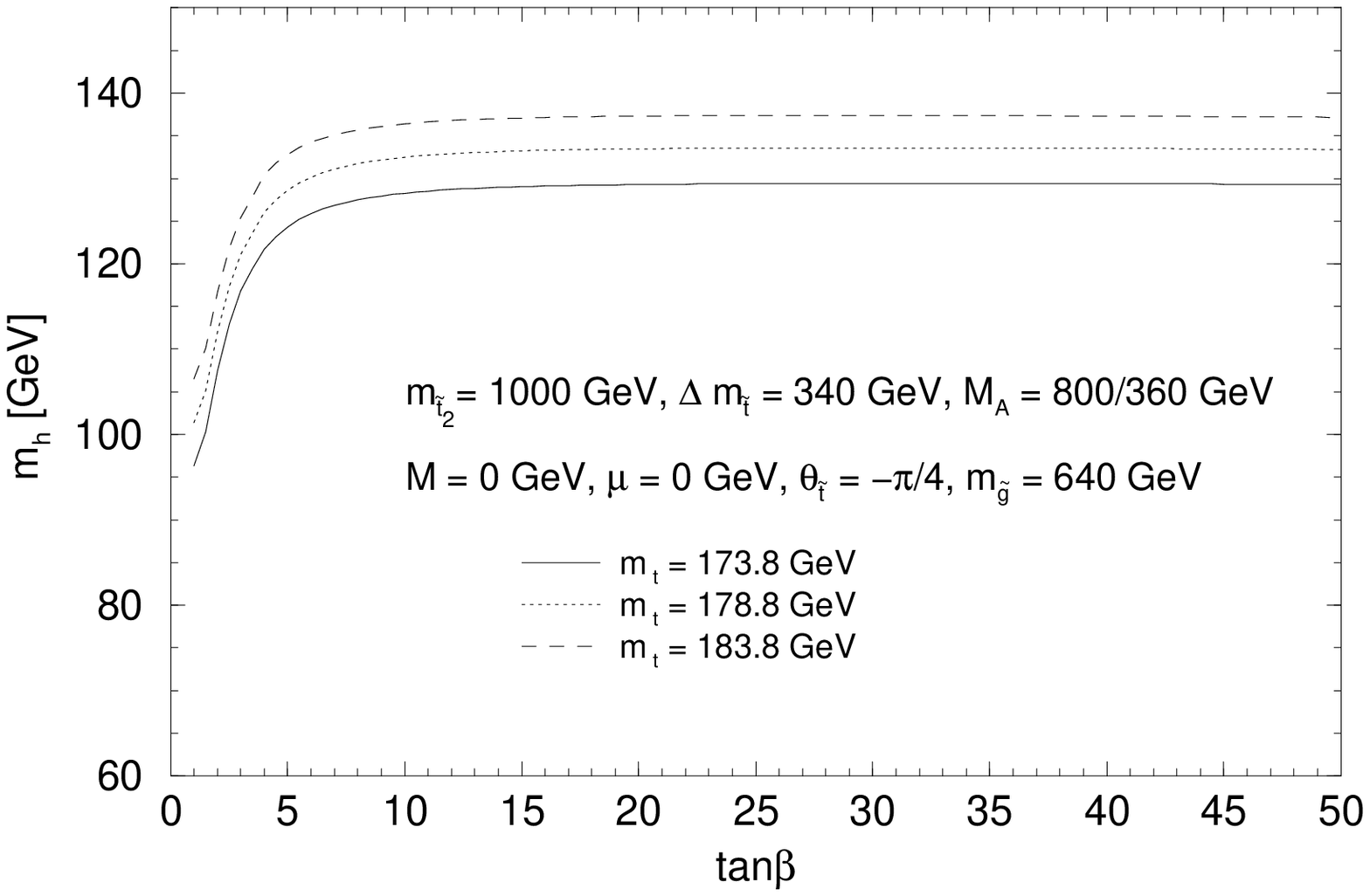,width=5.3cm,height=8cm,
                      bbllx=150pt,bblly=100pt,bburx=450pt,bbury=420pt}}
\hspace{7.5em}
\mbox{
\psfig{figure=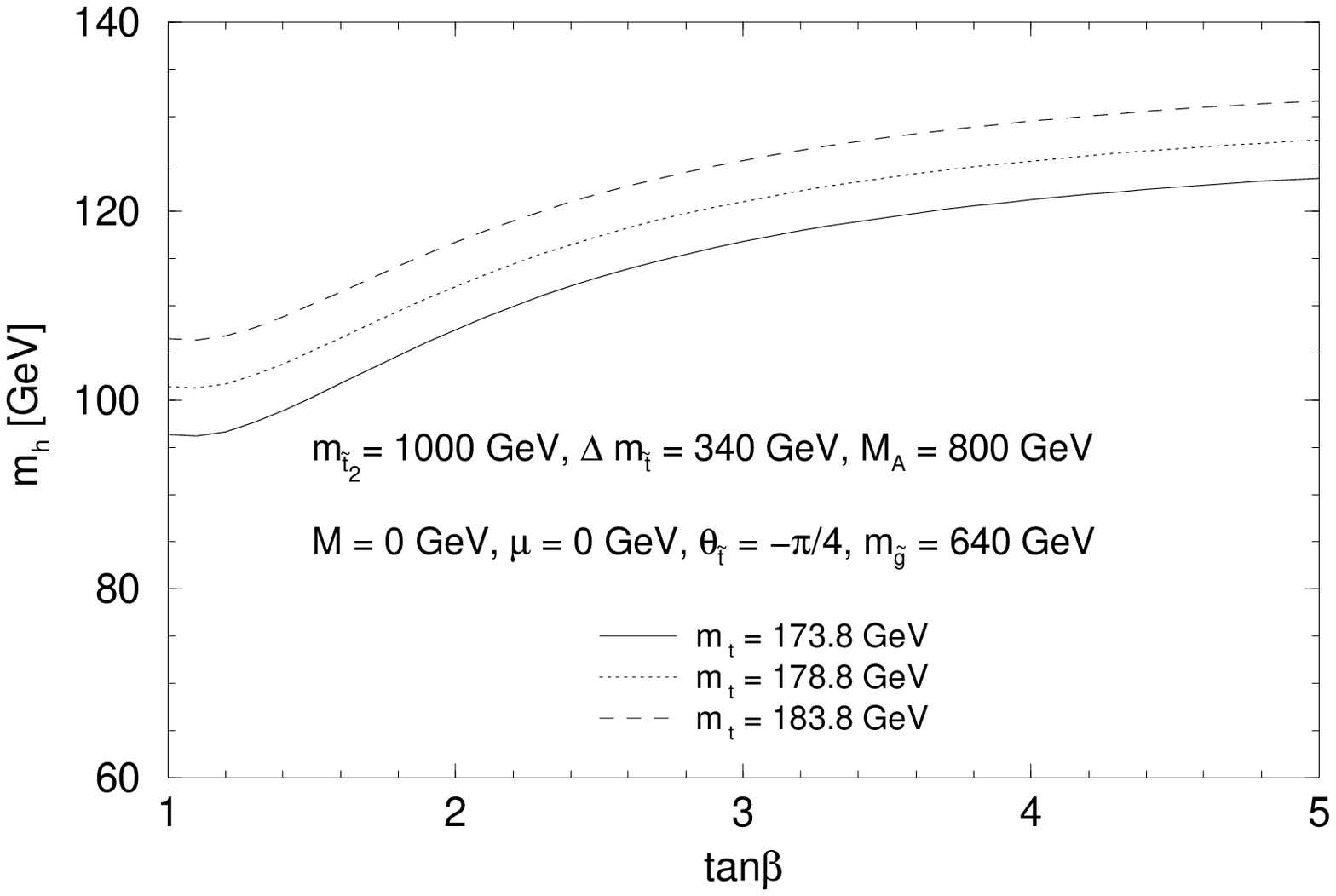,width=5.3cm,height=8cm,
                      bbllx=150pt,bblly=100pt,bburx=450pt,bbury=420pt}}
\end{center}

\vspace{1cm}

\begin{center}
\hspace{1em}
\mbox{
\psfig{figure=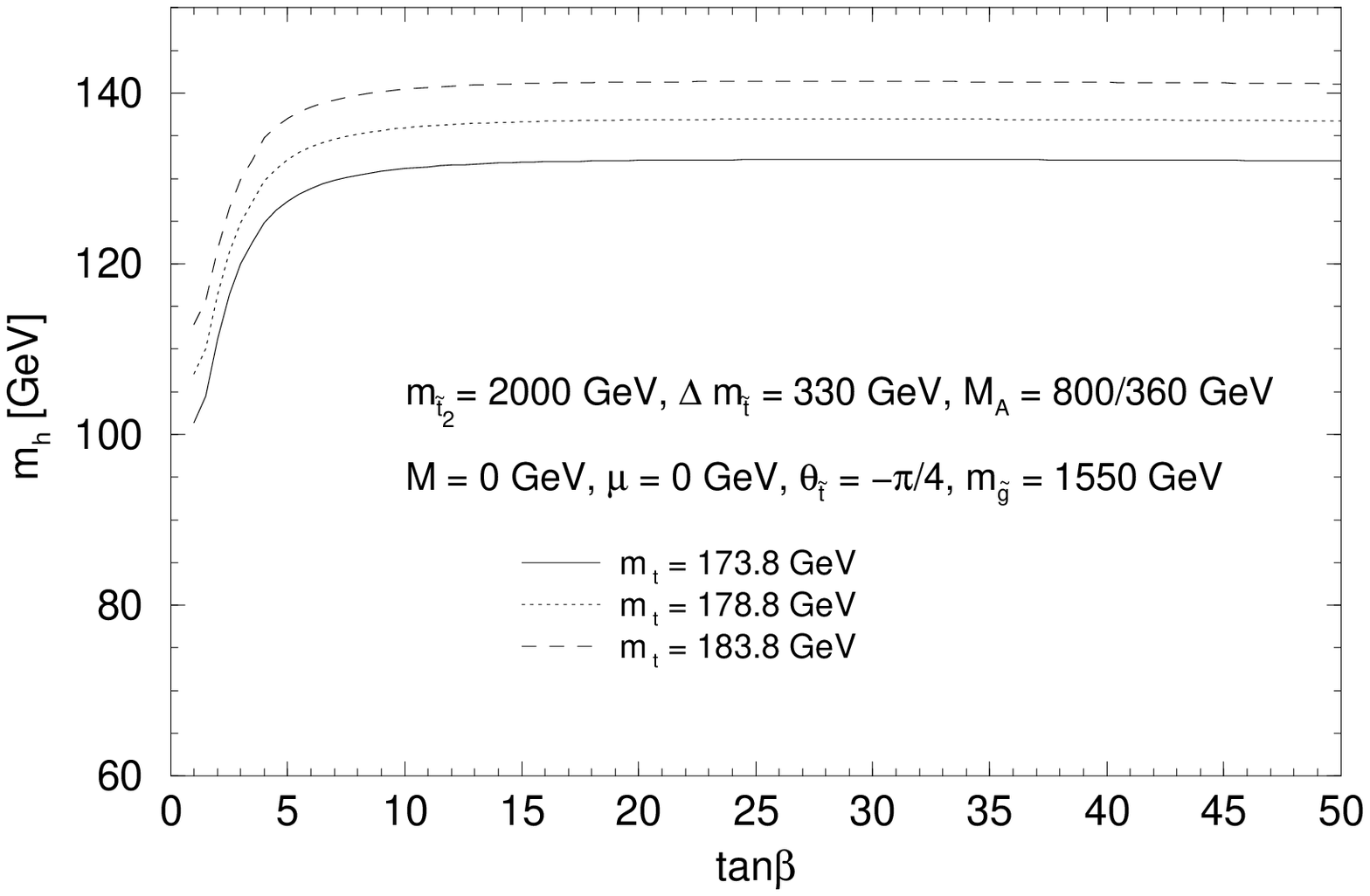,width=5.3cm,height=8cm,
                      bbllx=150pt,bblly=100pt,bburx=450pt,bbury=420pt}}
\hspace{7.5em}
\mbox{
\psfig{figure=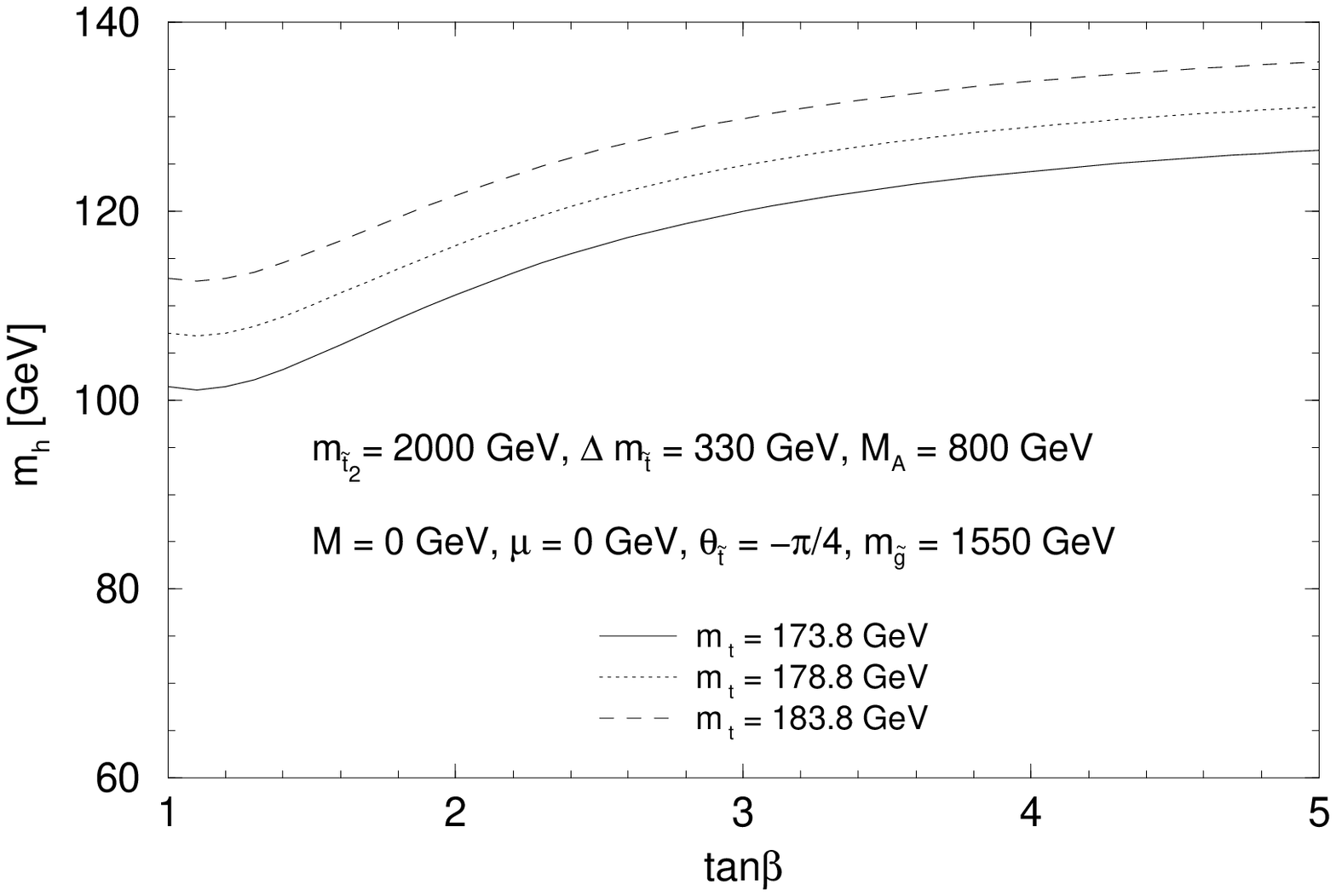,width=5.3cm,height=8cm,
                      bbllx=150pt,bblly=100pt,bburx=450pt,bbury=420pt}}
\end{center}
\caption[]{
The maximally possible value for $\mh$, including all refinement terms,
as a function of $\Tb$, depending 
on the physical parameters $\mstz, \delmst$ and $\tst$. The other MSSM
parameters have been chosen according to~\refeqs{maxmhparameters}.
} 
\label{fig:mhmax_tb_PP}
\end{figure}

\bigskip
In an analogous way we have also analyzed the maximal value for
$\mh$ as a function of the physical parameters, see
\reffi{fig:mhmax_tb_PP} and Tab.~\ref{tab:mhmax_PP}. We have chosen  
$\mstz = 1000,2000 \gev$, $\De\mst = 340 \gev$ and $\tst = -\pi/4$.
The other MSSM parameters are chosen according to~\refeqs{maxmhparameters}. 
\reffi{fig:mhmax_tb_PP} shows the result for the maximal range of
$\Tb$ and for the interesting range for LEP2, $\Tb \le 5$. The results
in \reffi{fig:mhmax_tb_PP} and Tab.~\ref{tab:mhmax_PP} are slightly
lower than the values obtained with the unphysical input parameters.
This is due to the fact that the values obtained for the squark
masses 
in the first scenario (for all other parameters chosen to be equal)
are always larger than for the latter case with 
physical input parameters. The analysis of the upper bound of $\mh$,
however, can be taken over 
directly from the case with unphysical input parameters.

\begin{table}[ht!]
\renewcommand{\arraystretch}{1.5}
\begin{center}
\begin{tabular}{|c||c|c|c|c|c|c|c|c|c|c|} 
\cline{2-11} \multicolumn{1}{c|}{} 
 & \multicolumn{2}{c|}{$\Tb = 1.6$} & \multicolumn{2}{c|}{$\Tb = 1.7$} &
   \multicolumn{2}{c|}{$\Tb = 1.8$} & \multicolumn{2}{c|}{$\Tb = 1.9$} & 
   \multicolumn{2}{c|}{$\Tb = 2.0$}  \\
\cline{2-11} \multicolumn{1}{c|}{}
 & \multicolumn{2}{c|}{$\mstz$ =} & \multicolumn{2}{c|}{$\mstz$ =} & 
   \multicolumn{2}{c|}{$\mstz$ =} & \multicolumn{2}{c|}{$\mstz$ =} & 
   \multicolumn{2}{c|}{$\mstz$ =}\\ \cline{1-1}
$\mt$ & 1000 & 2000 & 1000 & 2000 & 1000 & 2000 & 1000 & 2000 
      & 1000 & 2000  \\ \hline \hline
173.8 & 101.8 & 105.9 & 103.3 & 107.2 & 104.7 & 108.6 & 106.1 & 109.9
      & 107.5 & 111.2 \\ \hline
178.8 & 106.6 & 111.3 & 108.0 & 112.6 & 109.4 & 113.9 & 110.8 & 115.2
      & 112.1 & 116.4 \\ \hline
183.8 & 111.5 & 116.9 & 112.9 & 118.1 & 114.2 & 119.4 & 115.5 & 120.6
      & 116.7 & 121.7 \\ \hline
\end{tabular}
\caption[]{Maximal values for $\mh$ for different choices of 
$\mt, \mstz$ and $\Tb$. All other parameters have been chosen 
according to~\refeqs{maxmhparameters}. (All masses are in GeV.)} 
\label{tab:mhmax_PP}
\end{center}
\end{table}

\subsection{Numerical comparison with the RG approach}
\label{subsec:rgcomp}

\begin{figure}[ht!]
\begin{center}
\hspace{1em}
\mbox{
\psfig{figure=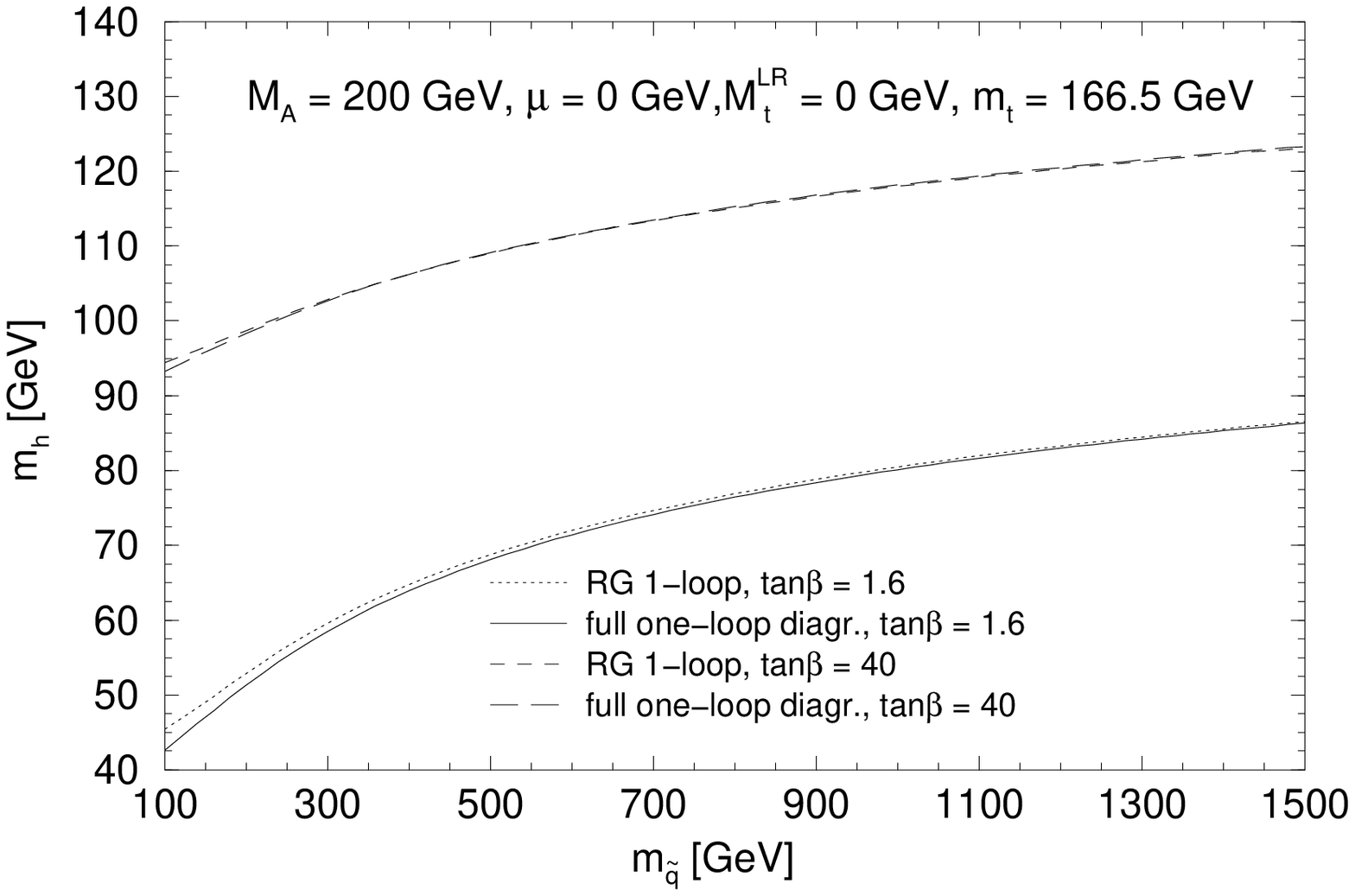,width=5.3cm,height=8cm,
                      bbllx=150pt,bblly=100pt,bburx=450pt,bbury=420pt}}
\hspace{7.5em}
\mbox{
\psfig{figure=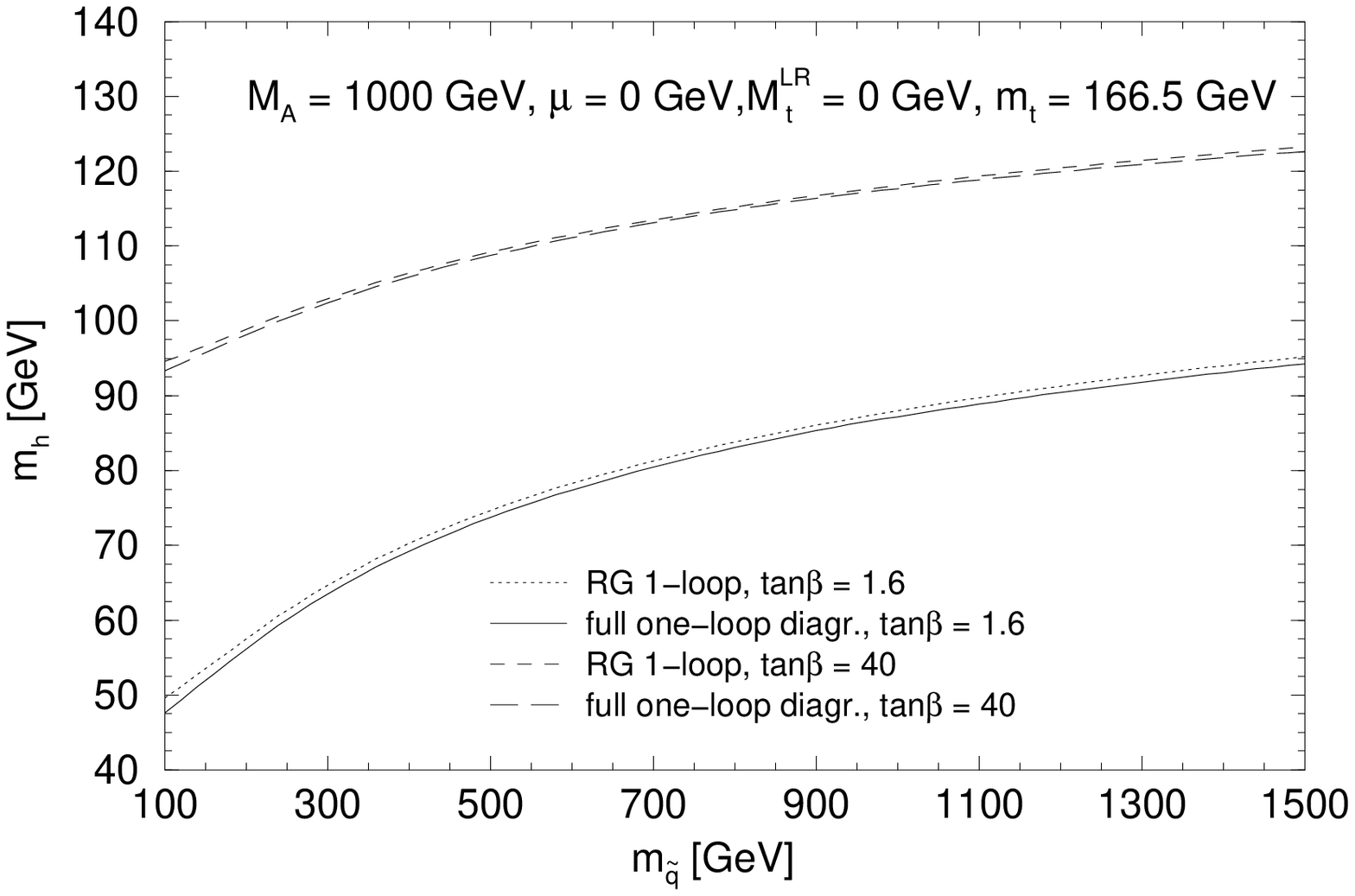,width=5.3cm,height=8cm,
                      bbllx=150pt,bblly=100pt,bburx=450pt,bbury=420pt}}
\end{center}

\caption[]{
Comparison between the \onel\ Feynman-diagrammatic calculations and the
results obtained by renormalization group methods~\cite{mhiggsRG2}.
The mass of the lightest Higgs boson is shown for the two scenarios with
$\Tb = 1.6$ and $\Tb = 40$ as a function of $\Mtlr/\msq$ for 
$\MA = 200,1000 \gev$.
} 
\label{fig:mh1loop_mq_nomix_RGVergleich}
\end{figure}

\begin{figure}[th!]
\begin{center}
\hspace{1em}
\mbox{
\psfig{figure=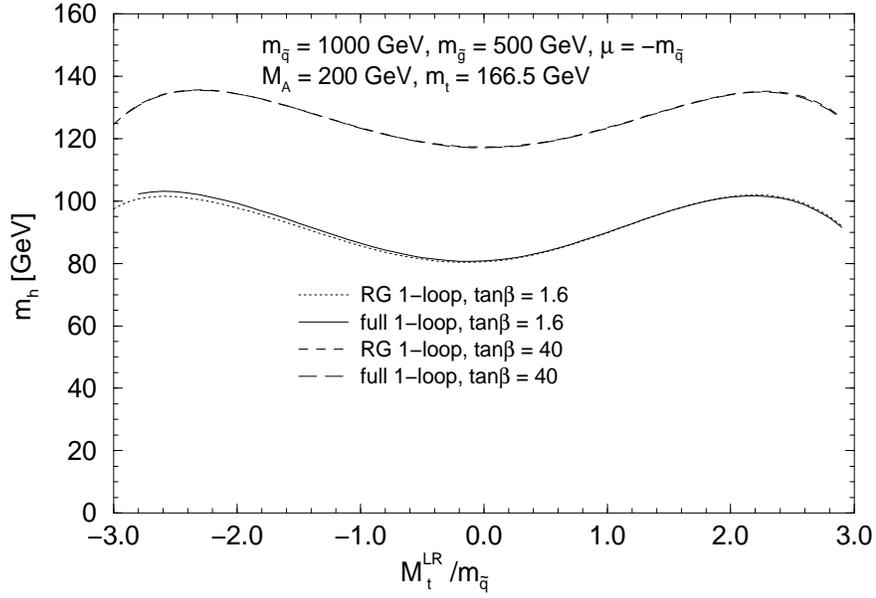,width=7.3cm,height=8cm,
                      bbllx=150pt,bblly=100pt,bburx=450pt,bbury=420pt}}
\end{center}

\vspace{.5cm}

\begin{center}
\hspace{1em}
\mbox{
\psfig{figure=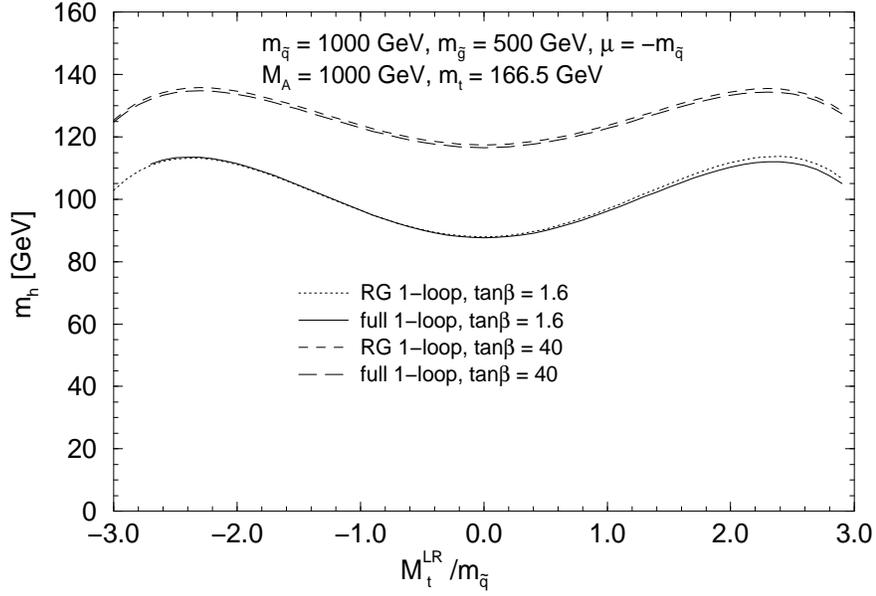,width=7.3cm,height=8cm,
                      bbllx=150pt,bblly=100pt,bburx=450pt,bbury=420pt}}
\end{center}
\caption[]{
Comparison between the \onel\ Feynman-diagrammatic calculations and the
results obtained by renormalization group methods~\cite{mhiggsRG2}.
The mass of the lightest Higgs boson is shown for the two scenarios with
$\Tb = 1.6$ and $\Tb = 40$ as a function of $\Mtlr/\msq$ for 
$\MA = 200,1000 \gev$. 
} 
\label{fig:mh1loop_mq_mix_RGVergleich}
\end{figure}

We now turn to the comparison of our diagrammatic results with the
predictions obtained via RG methods. 
For this comparison we made use of the {\tt FORTRAN} code
corresponding to~\citere{mhiggsRG1b}, except for the \onel\ results in
\reffis{fig:mh1loop_mq_nomix_RGVergleich} and
\ref{fig:mh1loop_mq_mix_RGVergleich}, where we 
used the code described in~\citere{mhiggsRG2}%
\footnote{
The RG results of \citere{mhiggsRG1b} and \citere{mhiggsRG2} 
agree within about $2 \gev$ with each other.
}%
.

\smallskip
We begin with the case of large values 
of $\MA$, for which the RG approach is most easily applicable and is
expected to work most accurately. In order to study different
contributions separately, we have first compared the diagrammatic
\onel\ on-shell result~\cite{mhiggsf1l} with the \onel\ leading log
result (without renormalization group improvement) given in
\citere{mhiggsRG2}. 
Since the available
code uses the running top mass $\mtms = \mtms(\mt) \approx 167.3 \gev$
we have also used this top mass for the full diagrammatic \onel\
calculation. 
In \reffi{fig:mh1loop_mq_nomix_RGVergleich} the lightest Higgs-boson
mass is shown in the
no-mixing scenario, i.e.\ $\Mtlr = 0 \gev$, whereas in
\reffi{fig:mh1loop_mq_mix_RGVergleich} $\mh$
is shown for increasing mixing in the $\Stop$-sector.
We found very good agreement, typically within 
$1 \gev$ for both mixing cases and low and high $\Tb$.
Only for very small values of $\msq$ a deviation up to $2 \gev$ arises.
For values of
$\MA$ below $100 \gev$ (which are not shown here) and large mixing
in the $\Stop$-sector deviations of about $5 \gev$ occur.

\begin{figure}[ht!]
\begin{center}
\hspace{1em}
\mbox{
\psfig{figure=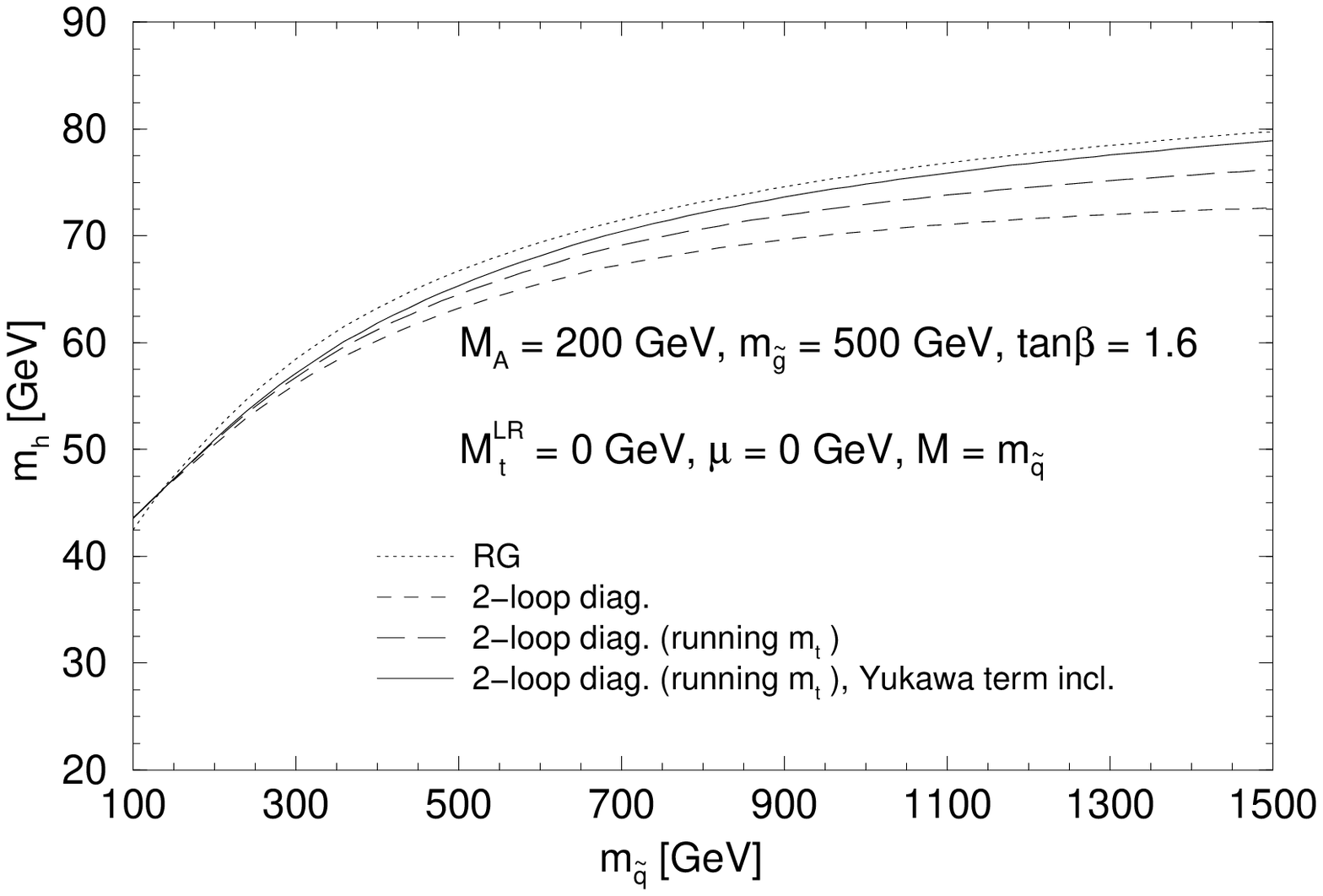,width=5.3cm,height=8cm,
                      bbllx=150pt,bblly=100pt,bburx=450pt,bbury=420pt}}
\hspace{7.5em}
\mbox{
\psfig{figure=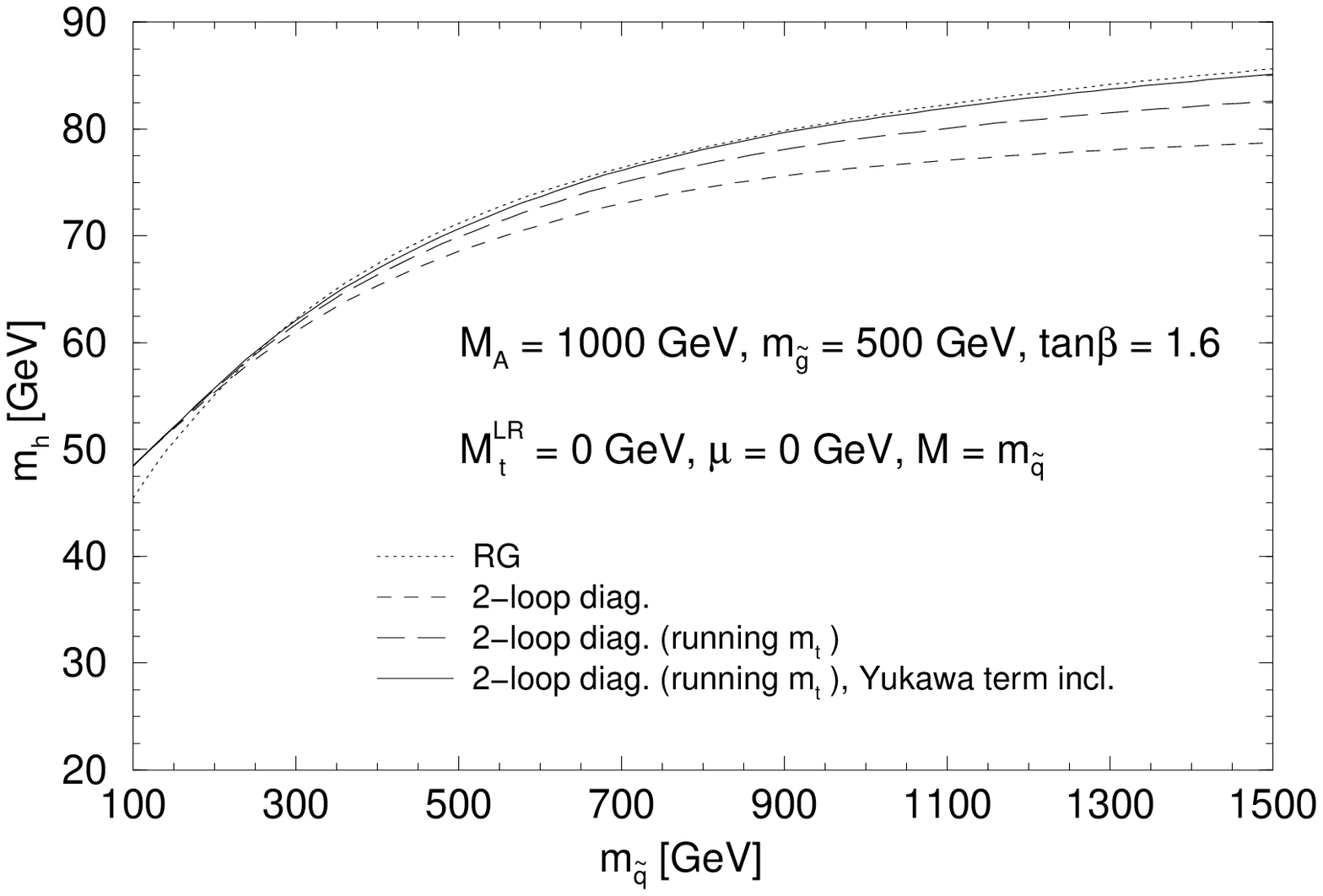,width=5.3cm,height=8cm,
                      bbllx=150pt,bblly=100pt,bburx=450pt,bbury=420pt}}
\end{center}

\vspace{.5cm}

\begin{center}
\hspace{1em}
\mbox{
\psfig{figure=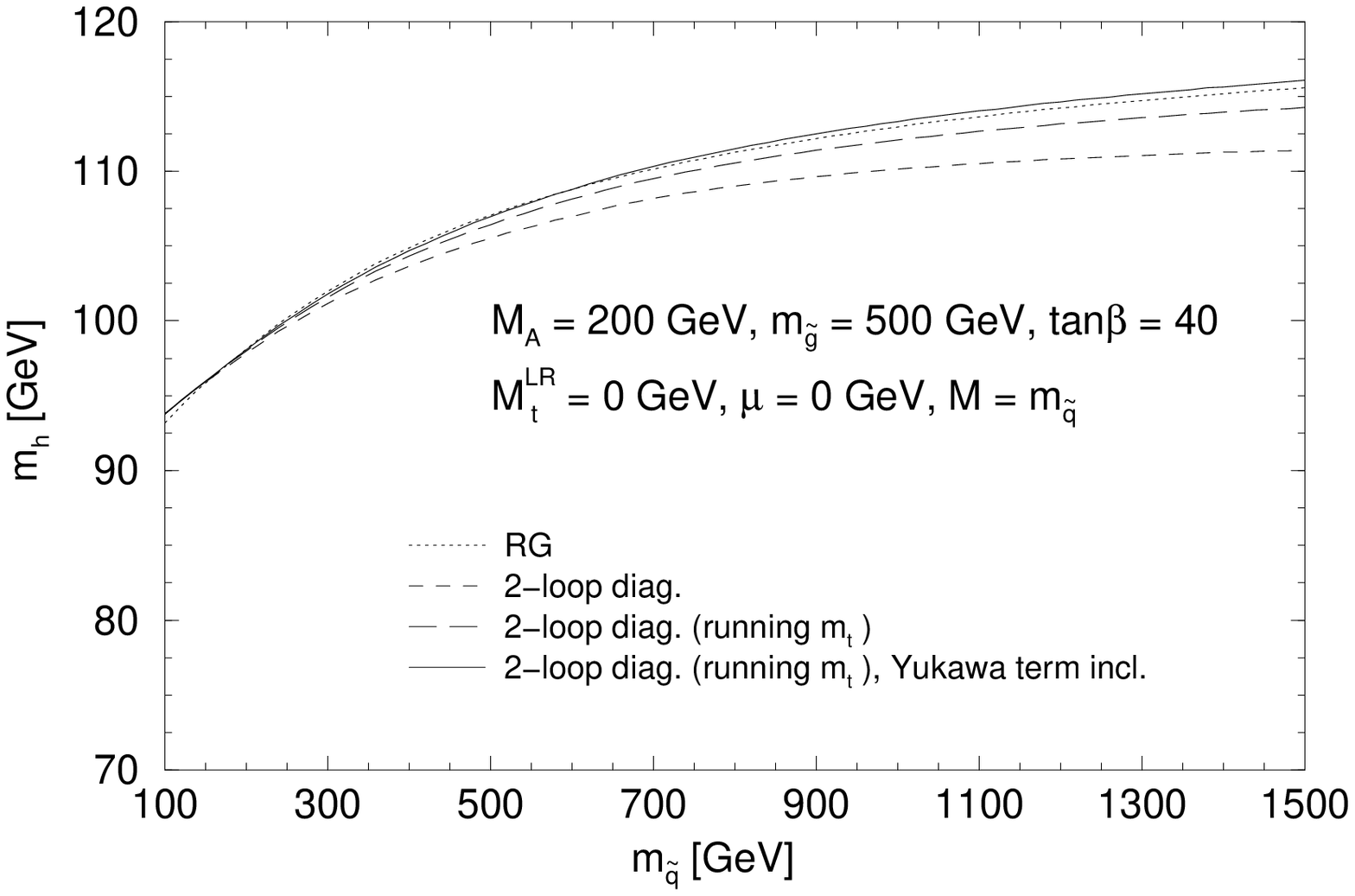,width=5.3cm,height=8cm,
                      bbllx=150pt,bblly=100pt,bburx=450pt,bbury=420pt}}
\hspace{7.5em}
\mbox{
\psfig{figure=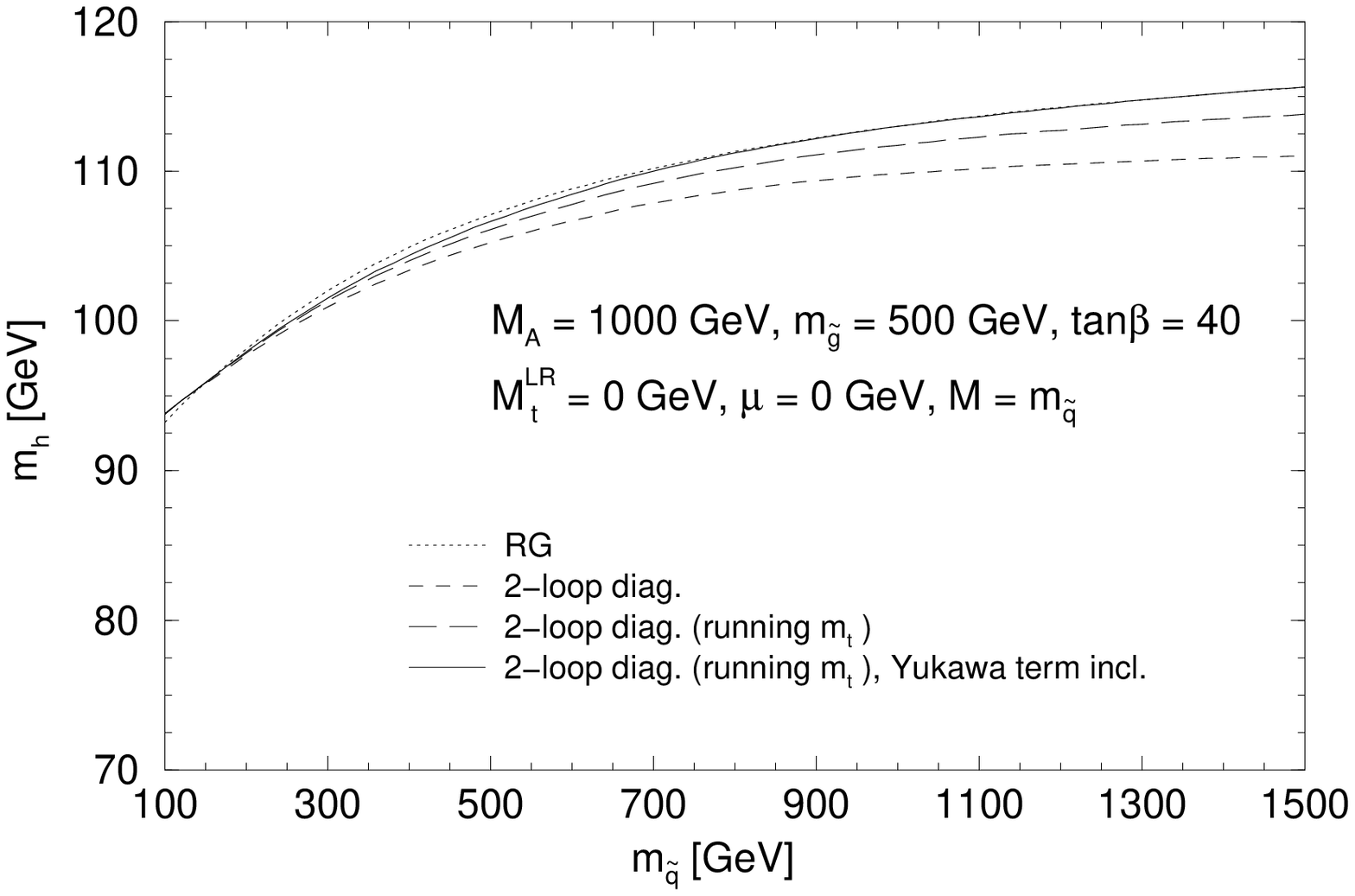,width=5.3cm,height=8cm,
                      bbllx=150pt,bblly=100pt,bburx=450pt,bbury=420pt}}
\end{center}
\caption[]{
Comparison between the Feynman-diagrammatic calculations and the
results obtained by renormalization group methods~\cite{mhiggsRG1b}.
The mass of the lightest Higgs boson is shown for the scenarios with
$\Tb = 1.6$ and $\Tb = 40$, $\MA = 200,1000 \gev$ for the case of
vanishing mixing in the $\Stop$-sector.
} 
\label{fig:mh_mq_RGVergleich}
\end{figure}

\smallskip
In the next step of comparison we analyzed the no-mixing case at the
\twol\ level: we have compared our diagrammatic result for the 
no-mixing case, including the Yukawa correction and the running top
mass effect, with the RG results obtained in \citere{mhiggsRG1b}.
We have adopted the scale $M$ (the $SU(2)$ gaugino mass parameter) as
$M = \msq$, in order to treat it in the same way as it has been done
in the RG approach.
As can be seen in \reffi{fig:mh_mq_RGVergleich}, after the inclusion
of the corrections beyond $\oaas$ the diagrammatic result for the
no-mixing case agrees very well with the RG result.
For the scenario with $\MA = 1000 \gev$ the deviation between the results
exceeds $2 \gev$ only for $\Tb = 1.6$ and $\msq < 150 \gev$. 
For $\MA = 200 \gev$ the deviation is in general slightly larger than
for $\MA = 1000 \gev$, but does not exceed $2 \gev$.

The RG results
do not contain the gluino mass as a parameter. Hence, varying $\mgl$,
which has been discussed in
Sec.~\ref{subsec:mhparameterdependence}, gives rise to an extra
deviation. In the no-mixing case this extra deviation does not exceed
$1 \gev$.  
Varying the other parameters $\mu$ and $M$ in general does not
lead to a sizable effect in the comparison with the corresponding RG
results (as long as $\Mtlr$ is taken as input and a variation of $\mu$
does not affect the $\Stop$-mixing.)

\begin{figure}[ht!]
\begin{center}
\hspace{1em}
\mbox{
\psfig{figure=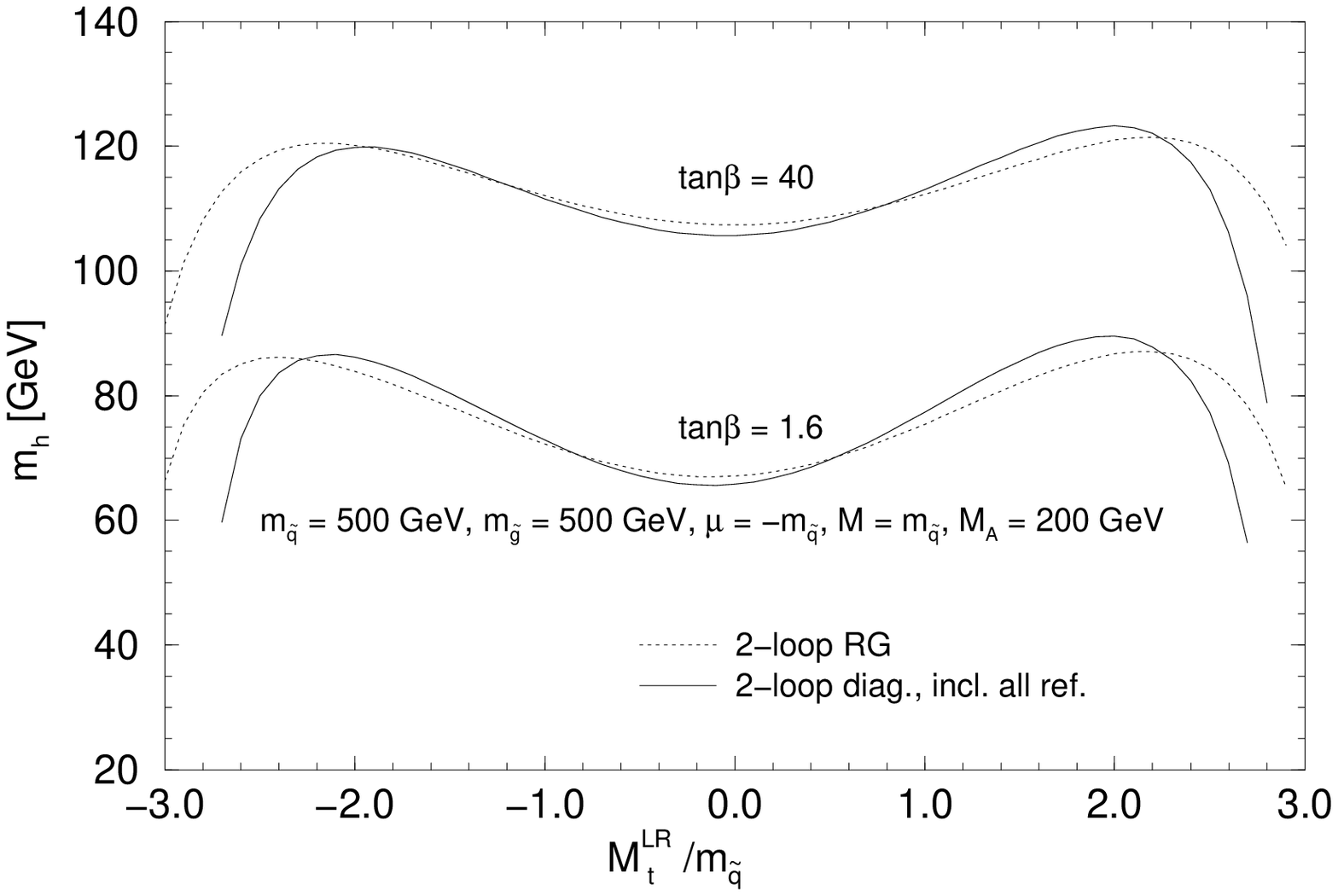,width=7.3cm,height=8cm,
                      bbllx=150pt,bblly=100pt,bburx=450pt,bbury=420pt}}
\end{center}

\vspace{.5cm}

\begin{center}
\hspace{1em}
\mbox{
\psfig{figure=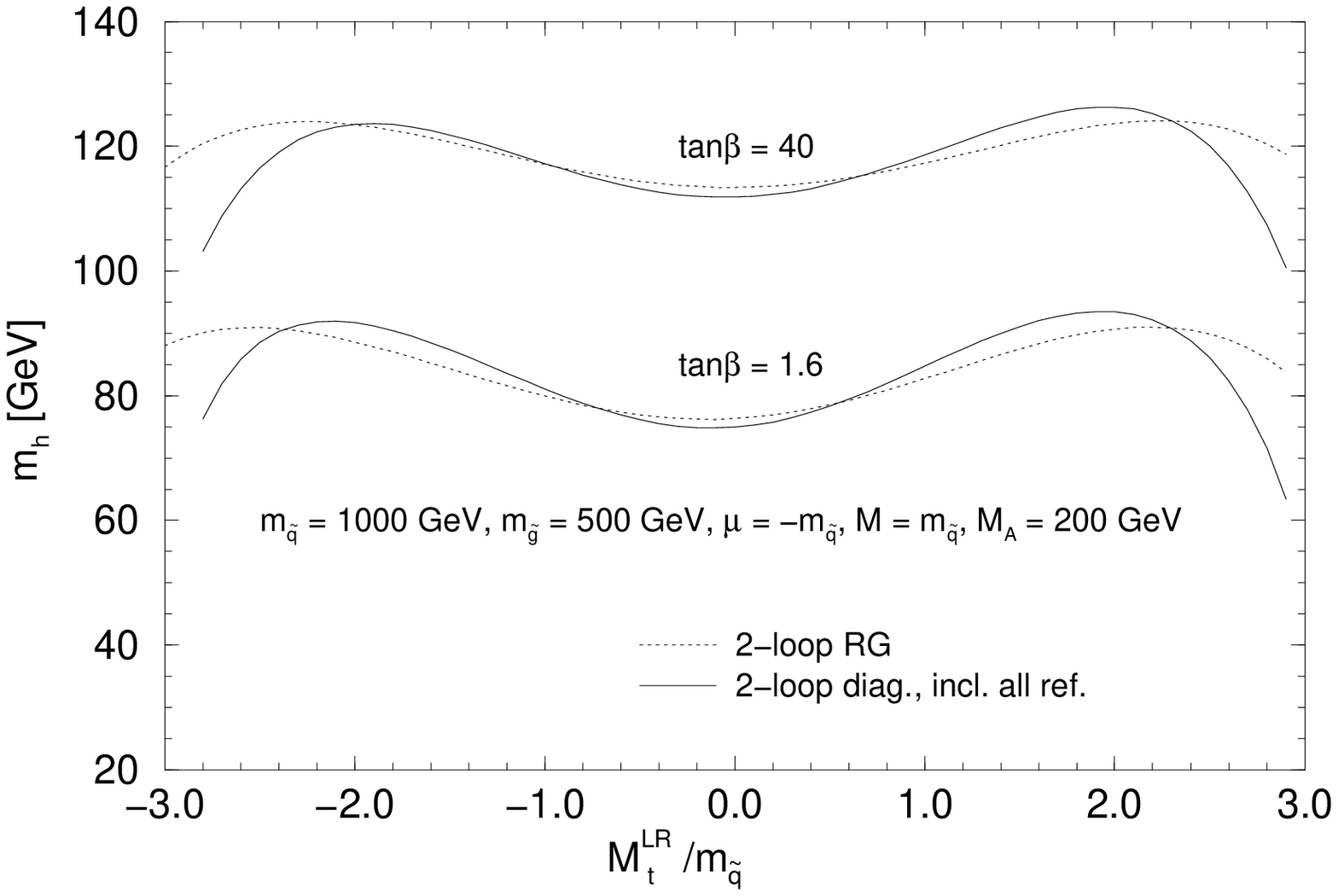,width=7.3cm,height=8cm,
                      bbllx=150pt,bblly=100pt,bburx=450pt,bbury=420pt}}
\end{center}
\caption[]{
Comparison between the Feynman-diagrammatic calculations and the
results obtained by renormalization group methods~\cite{mhiggsRG1b}.
The mass of the lightest Higgs boson is shown for the two scenarios with
$\Tb = 1.6$ and $\Tb = 40$ as a function of $\Mtlr/\msq$ for
$\msq = 200,1000 \gev$ and $\MA = 200 \gev$. 
} 
\label{fig:mh_MtLRdivmq_MA200_RGVergleich}
\end{figure}

\begin{figure}[ht!]
\begin{center}
\hspace{1em}
\mbox{
\psfig{figure=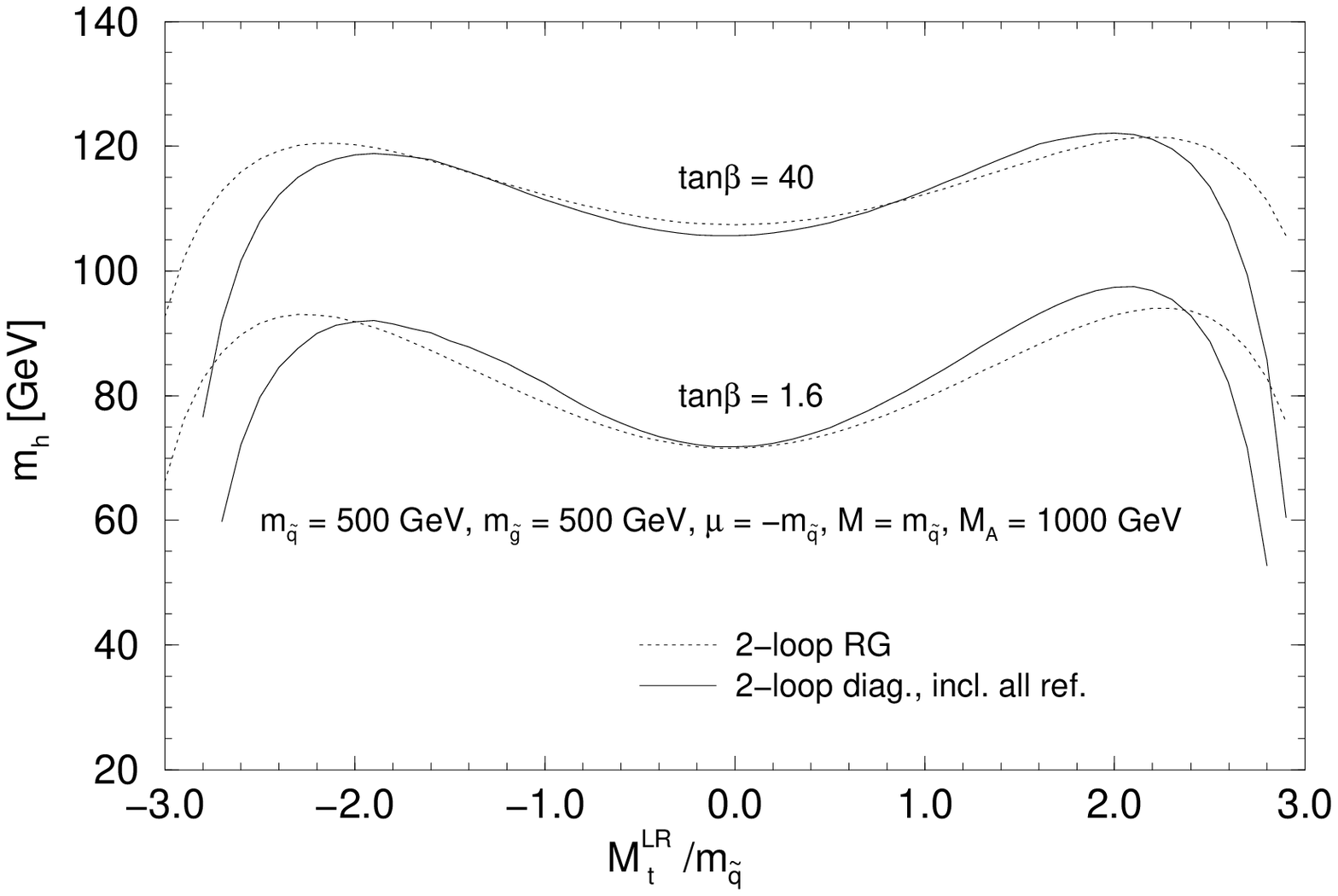,width=7.3cm,height=8cm,
                      bbllx=150pt,bblly=100pt,bburx=450pt,bbury=420pt}}
\end{center}

\vspace{.5cm}

\begin{center}
\hspace{1em}
\mbox{
\psfig{figure=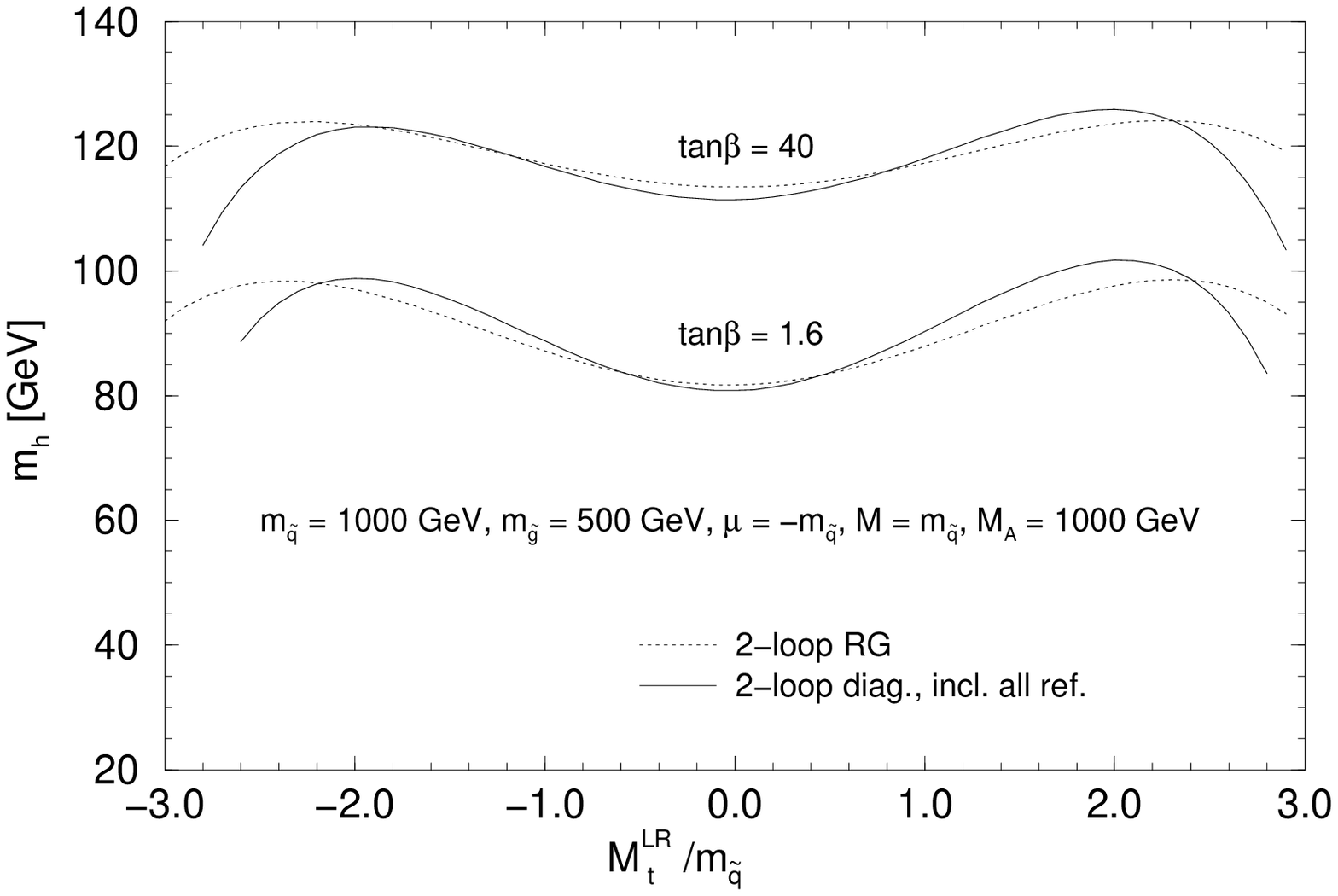,width=7.3cm,height=8cm,
                      bbllx=150pt,bblly=100pt,bburx=450pt,bbury=420pt}}
\end{center}
\caption[]{
Comparison between the Feynman-diagrammatic calculations and the
results obtained by renormalization group methods~\cite{mhiggsRG1b}.
The mass of the lightest Higgs boson is shown for the two scenarios with
$\Tb = 1.6$ and $\Tb = 40$  as a function of $\Mtlr/\msq$ for
$\msq = 200,1000 \gev$ and $\MA = 1000 \gev$. 
} 
\label{fig:mh_MtLRdivmq_MA1000_RGVergleich}
\end{figure}

\bigskip
Finally we consider the situation where mixing in the $\Stop$~sector is
taken into account. 
In \reffi{fig:mh_MtLRdivmq_MA200_RGVergleich} our diagrammatic result,
including the Yukawa correction and the running top mass effect,
is compared with the RG results~\cite{mhiggsRG1b}
as a function of
$\Mtlr/\msq$ for the cases $\Tb = 1.6$ and $\Tb = 40$, and for 
$\msq = 500,1000 \gev$ and $\MA = 200 \gev$.
The $\MA = 1000 \gev$ scenario is depicted in
\reffi{fig:mh_MtLRdivmq_MA1000_RGVergleich} for the same set of
parameters. The point $\Mtlr/\msq = 0$
corresponds to the plots shown in \reffi{fig:mh_mq_RGVergleich}, except 
that the parameter $\mu$ is set to $\mu = -\msq$ here.
For larger
$\Stop$-mixing, sizable deviations between the diagrammatic and the RG
results occur. They can reach $5 \gev$ for moderate mixing and become
very large for $|\Mtlr/\msq| \gsim 2.5$. 
As already mentioned above, the maximal
value for $\mh$ in the diagrammatic approach is reached for
$\Mtlr/\msq \approx \pm 2$, whereas the RG results have a maximum at 
$\Mtlr/\msq \approx \pm 2.4$, i.e.\ at the \onel\ value. 
This holds for all combinations of $\Tb, \msq$ and $\MA$.
In the case of positive $\Mtlr$,
the maximal values for $\mh$ reached in the diagrammatic calculation are
up to $5~(3)\gev$ larger than the ones of the RG method for 
$\Tb = 1.6~(40)$.
The dependence on $\Mtlr$ is asymmetric;
for negative $\Mtlr$ about the same maximal values are reached in the two
approaches.

The diagrammatic result varies with $\mgl$ as shown
in \reffi{fig:mh_mgl}. In the case of mixing in the $\Stop$-sector
this leads in general to a 
larger effect than in the no-mixing case and shifts the diagrammatic
result relative to the RG result within $\pm 2 \gev$.

\bigskip
Up to now we have compared the results of our diagrammatic on-shell
calculation and the RG methods in terms of the (unphysical) soft SUSY
breaking parameters
of the $\Stop$ mass matrix $\MstL$, $\MstR$ and $\Mtlr$, 
since the available numerical codes for the RG 
results~\cite{mhiggsRG1b,mhiggsRG2} are given in terms of these
parameters. 
However, since the two approaches rely on different
renormalization schemes, the meaning of these non-observable
parameters is not precisely the same in the two approaches
starting from \twol\ order.
Indeed we have checked that
assuming fixed values for the physical parameters $\mste$,
$\mstz$, and $\tst$ and deriving the corresponding values of the 
parameters $\MstL$, $\MstR$ and $\Mtlr$ in the on-shell scheme as well
as in the $\overline{{\rm MS}}$ scheme, sizable differences occur between the
values of the mixing parameter $\Mtlr$ in the two schemes.
On the other hand the
parameters $\MstL$, $\MstR$ are approximately equal in both schemes.
Thus, part of the different shape of the curves in
\reffi{fig:mh_MtLRdivmq_MA200_RGVergleich} and  
\reffi{fig:mh_MtLRdivmq_MA1000_RGVergleich} 
may be attributed to a different meaning of the parameter $\Mtlr$ in
the on-shell scheme and in the RG calculation.

\begin{figure}[ht!]
\begin{center}
\hspace{1em}
\mbox{
\psfig{figure=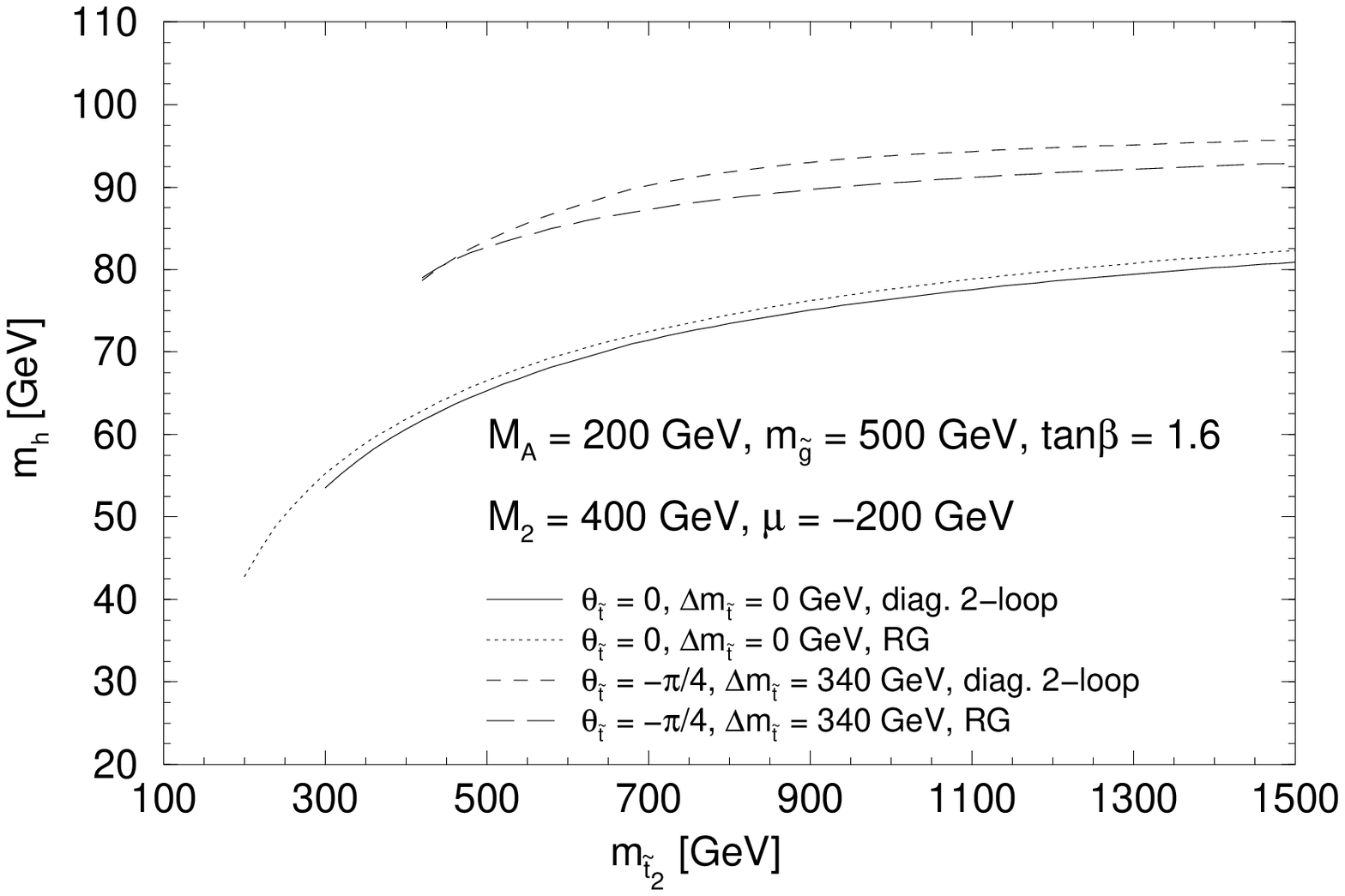,width=5.3cm,height=8cm,
                      bbllx=150pt,bblly=100pt,bburx=450pt,bbury=420pt}}
\hspace{7.5em}
\mbox{
\psfig{figure=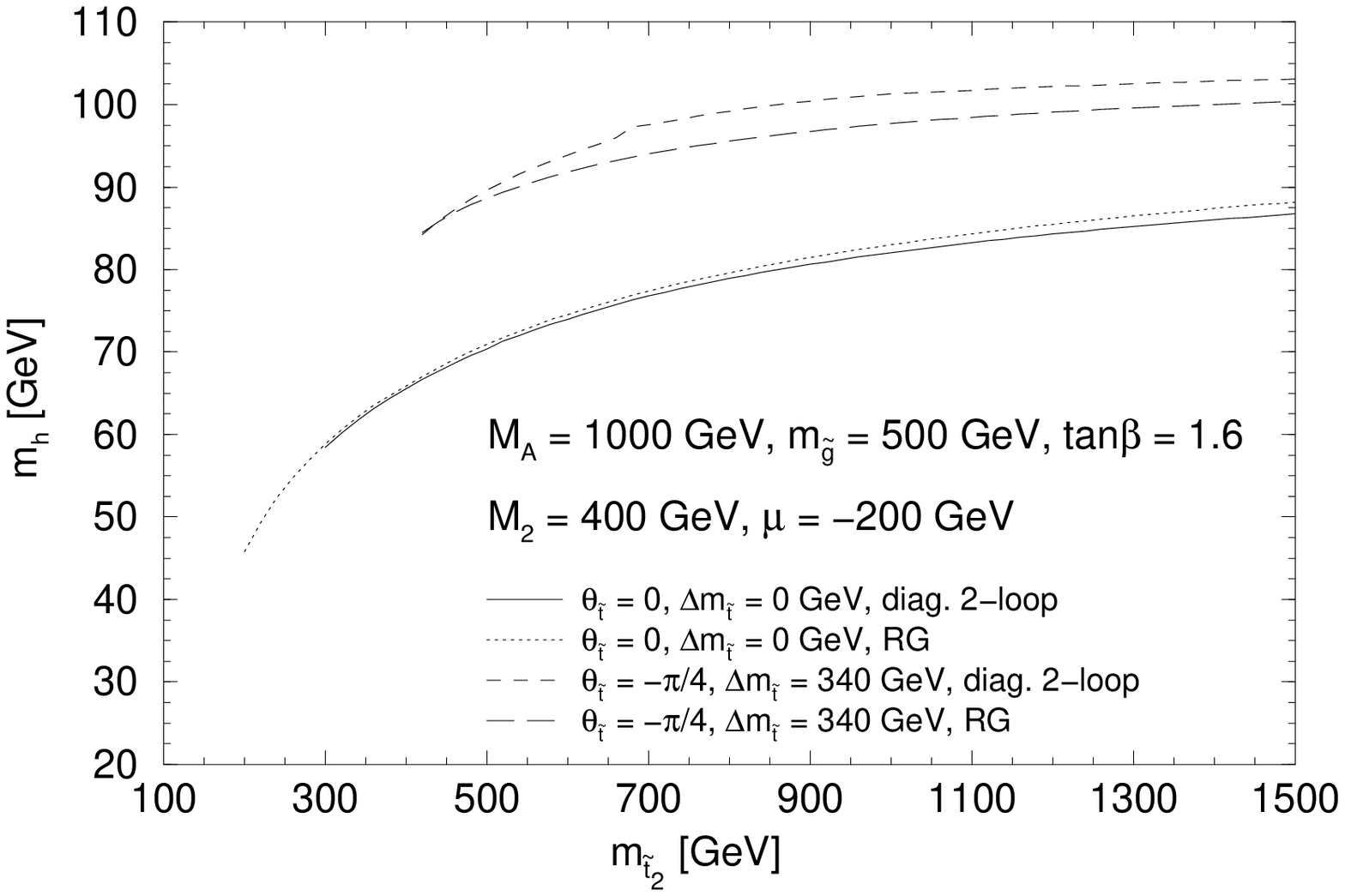,width=5.3cm,height=8cm,
                      bbllx=150pt,bblly=100pt,bburx=450pt,bbury=420pt}}
\end{center}


\begin{center}
\hspace{1em}
\mbox{
\psfig{figure=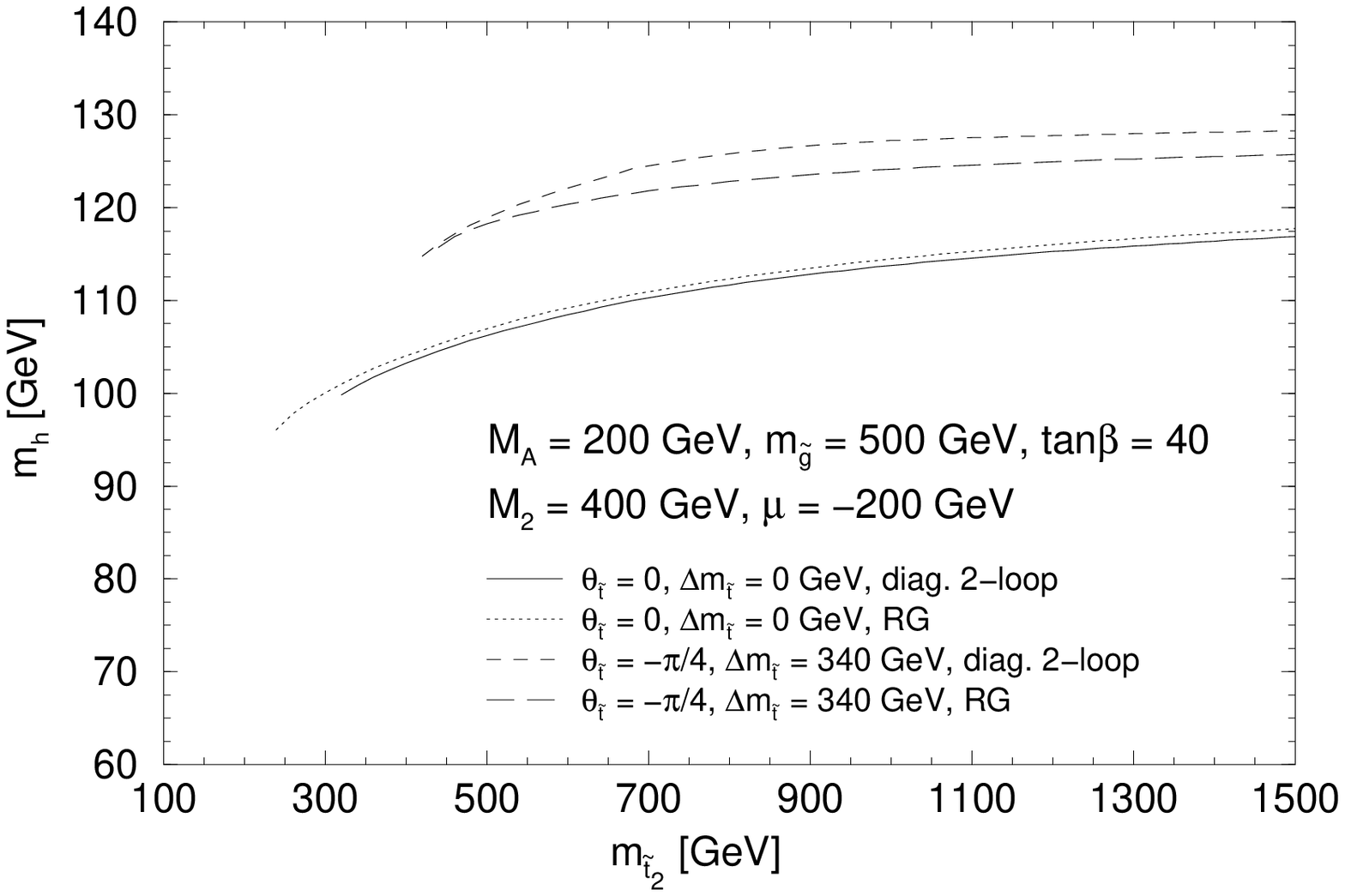,width=5.3cm,height=8cm,
                      bbllx=150pt,bblly=100pt,bburx=450pt,bbury=420pt}}
\hspace{7.5em}
\mbox{
\psfig{figure=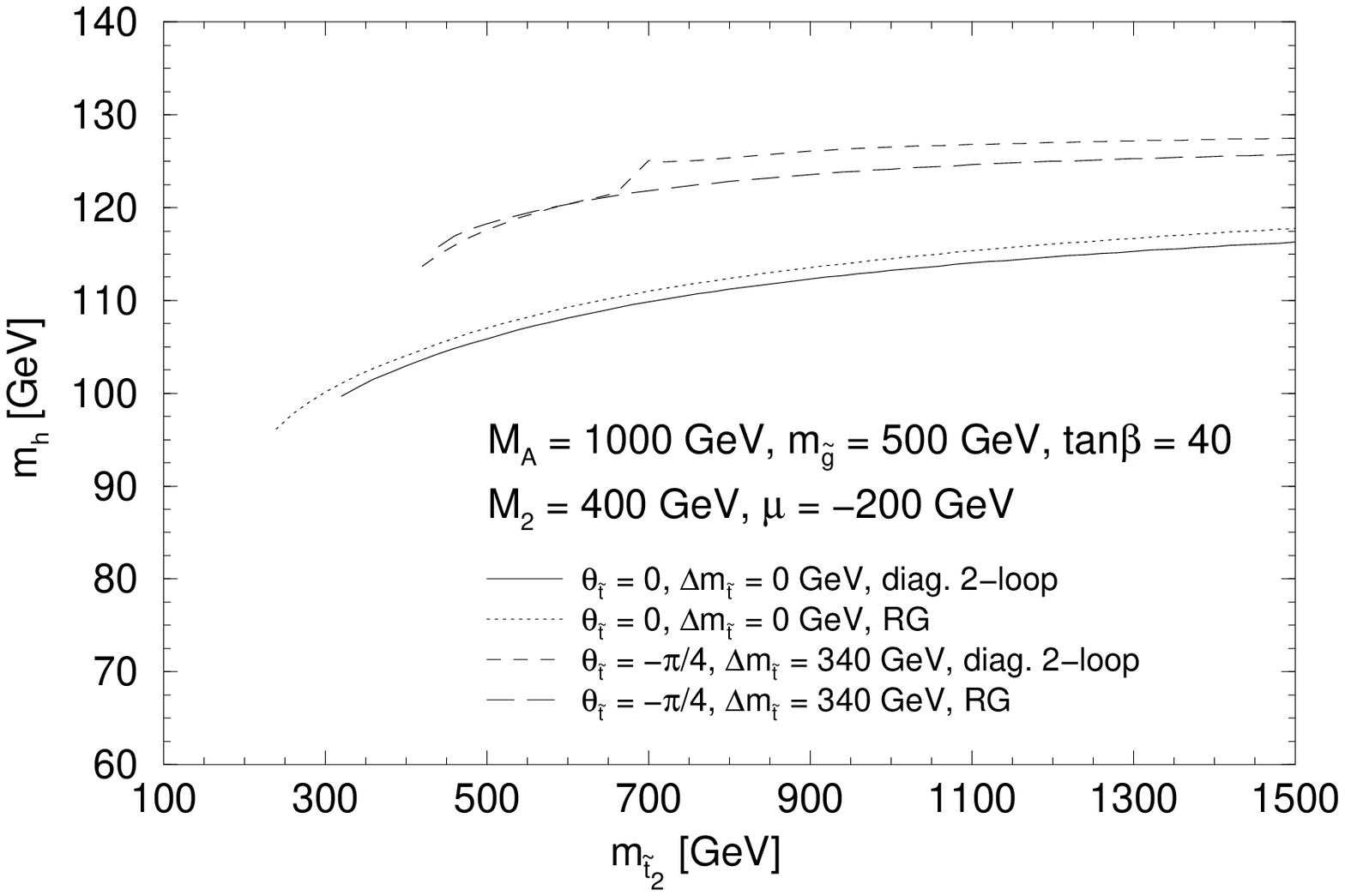,width=5.3cm,height=8cm,
                      bbllx=150pt,bblly=100pt,bburx=450pt,bbury=420pt}}
\end{center}
\caption[]{
Comparison between the Feynman-diagrammatic calculations and the
results obtained by renormalization group methods~\cite{mhiggsRG1b} in
terms of physical parameters.
The mass of the lightest Higgs boson is shown for the two scenarios with
$\Tb = 1.6$ and $\Tb = 40$ and for $\MA = 200,1000 \gev$ as a function
of the heavier physical  $\Stop$ mass $\mstz$.
For the curves with $\tst = 0$ a mass difference $\delmst = 0 \gev$ is
taken, whereas for $\tst = -\pi/4$ we choose $\delmst = 340 \gev$, for
which the maximal Higgs-boson masses are achieved.
} 
\label{fig:mh_mst2_RGVergleich}
\end{figure}

In order to avoid this problem in comparing results
obtained by different approaches making use 
of different renormalization schemes, we find it preferable to compare
predictions for physical observables in terms of other observables
(instead of unphysical parameters).
Therefore we switch from the set of
unphysical parameters to a set of physical parameters: 
\BE
\MstL, \MstR, \Mtlr~~\to~~\mstz, \delmst (\equiv \mstz - \mste), \tst~.
\EE
In \reffi{fig:mh_mst2_RGVergleich} we compare 
the results for the lightest Higgs-boson mass, obtained by the
Feynman-diagrammatic method and by the RG method, in terms of this new
set of parameters: $\mh$ is shown as a function of $\mstz$ with the mass
difference $\delmst \equiv \mstz - \mste$ and the mixing angle $\tst$ as
further input parameters.
In the context of the RG approach the running $\Stop$-masses, derived 
from the $\Stop$ mass matrix, are considered as an approximation for
the physical masses.
In our approach, on the other hand,
since we are working in the on-shell scheme, the
$\Stop$-masses and the mixing angle directly correspond to
physical parameters. 
In \reffi{fig:mh_mst2_RGVergleich} we have furthermore implemented
the same $\De\rho$ constraints on the range of the third generation
scalar quark masses as in \reffi{fig:mh_mst2}.

Similarly to the comparison shown in
\reffi{fig:mh_MtLRdivmq_MA200_RGVergleich} and 
\ref{fig:mh_MtLRdivmq_MA1000_RGVergleich},
very good agreement is found in \reffi{fig:mh_mst2_RGVergleich} between
the results of the two approaches in the case of vanishing
$\Stop$-mixing. The deviation is typically less than $1 \gev$ and
never exceeds $2 \gev$.  
Using the physical parameters as input, the 
maximal-mixing scenario is realized by setting 
$\tst = -\pi/4$ and $\De\mst \approx 340 \gev$ (i.e.\ the $\Stop$-masses
obtained for $\Mtlr/\msq \approx 2$ have a mass difference of about
$340 \gev$.) In this scenario again (as in
\reffis{fig:mh_MtLRdivmq_MA200_RGVergleich} and 
\ref{fig:mh_MtLRdivmq_MA1000_RGVergleich})
the diagrammatic result yields values for $\mh$ which are higher by
about $5 \gev$.
The peaks in the plots for $\MA = 1 \tev$ and maximal mixing in the
$\Stop$-sector around $\mstz = 660 \gev$ are again due to the threshold
$\MA = \mste + \mstz$ in the
\onel\ contribution, originating from the stop-loop diagram in
the $A$~self-energy.

\section{Conclusions}

Using the Feynman diagrammatic method we have calculated
the leading $\oaas$ 
corrections to the masses of the neutral $\cp$-even Higgs bosons in
the MSSM. 
The \twol\ result has been implemented into the prediction based
on the complete diagrammatic \onel\ on-shell result. Two further
corrections beyond $\oaas$ have
been added in order to incorporate leading electroweak \twol\ and
higher-order QCD contributions.
The results have been obtained using the on-shell scheme, which
means a renormalization of all sectors of the MSSM at \onel\ order and
of the Higgs-boson sector at \twol\ order.
In our \twol\ calculation we have imposed no restrictions on 
the parameters of the Higgs and scalar top sector of the model.
Thus the results are valid for arbitrary values of the relevant 
MSSM parameters. 
The complete result has been implemented into the {\tt FORTRAN}
program \fh~\cite{feynhiggs} which is available via its WWW page\\
{\tt http://www-itp.physik.uni-karlsruhe.de/feynhiggs}~.\\
In this way we provide the at present
most precise prediction for $\mh$ and $\mH$ based on Feynman-diagrammatic
calculations.

\bigskip
The \twol\ corrections lead to a large reduction of
the one-loop on-shell result. 
We have performed a detailed analysis of the dependence of $\mh$ on
the various MSSM parameters. 
Concerning the scalar top sector the analysis has been carried out 
in terms of the (unphysical) soft SUSY breaking parameters 
$\MstL$, $\MstR$ 
and $\Mtlr$ as well as in terms of the physical parameters
$\mstz$, $\delmst \equiv \mstz - \mste$ and $\tst$.

A scan over the parameters $\mu, M, \mgl, \MA$ and $\Mtlr$ has been
performed in order to determine the maximally possible value for $\mh$
as a function of $\Tb$. 
Our results show that
for the scenario with $\Tb = 1.6$ 
the parameter space of the MSSM can be covered
almost completely. Only for maximal mixing, very large soft
SUSY breaking parameters in the $\Stop$-sector and
$\mt$ at its upper experimental limit the light Higgs boson can escape
the detection at LEP2 in this scenario.
Concerning the large $\Tb$ region, LEP2 and the upgraded Tevatron
can probe only the region of no mixing in the $\Stop$-sector.

\bigskip
We have compared our results, obtained by a Feynman diagrammatic
calculation (where also the corrections beyond $\oaas$ have been
included), with the results obtained via RG methods.
Concerning the \onel\ contributions we find very good agreement
between these two approaches. The same is valid for the \twol\
corrections in the case of vanishing mixing in the $\Stop$-sector.
On the other hand, in the case of non-vanishing mixing sizable
deviations between the two approaches occur. For moderate mixing they
reach up to $5 \gev$, 
for $|\Mtlr/\msq| \gsim 2.5$ they can be very large.
In the diagrammatic approach the maximal
value for $\mh$ is reached for
$\Mtlr/\msq \approx \pm 2$, whereas the RG results have a maximum at 
$\Mtlr/\msq \approx \pm 2.4$, i.e.\ at the \onel\ value. 
This holds for all combinations of $\Tb, \msq$ and $\MA$.
The fact that the parameter $\mgl$ is absent in the RG results can give
rise to an additional deviation between the two approaches of about
$\pm 2 \gev$.

\bigskip
We have furthermore discussed the issue of how 
results obtained via different approaches using 
different renormalization schemes can be 
readily compared to each other
also when corrections beyond \onel\ order are incorporated.  
For this purpose it is adequate to express the prediction for the
Higgs-boson masses in terms of other physical observables, i.e.\ 
the physical masses and mixing angles of the model.

Accordingly, we have compared the results obtained by our diagrammatic
\twol\ calculation with those obtained by 
RG methods in terms of the physical observables 
$\mstz$, $\delmst \equiv \mstz - \mste$ and $\tst$.
As for the comparison in terms of the unphysical parameters, we have
found good agreement for 
the case of vanishing mixing in the $\Stop$-sector. For large
splitting between the 
$\Stop$-masses, however, the Higgs-boson masses obtained by the
Feynman diagrammatic calculation are about $5 \gev$ larger than the
ones calculated in the RG approach.

\bigskip
\subsection*{Acknowledgements}
W.H. gratefully acknowledges support by the Volkswagenstiftung.



\end{document}